\documentclass[acmsmall, screen, authorversion, nonacm, a4paper]{acmart}
\pdfoutput=1

\author{Danielle Marshall}
\orcid{0000-0002-4284-3757}
\affiliation{
  \department{School of Computing} 
  \institution{University of Kent}
  \country{United Kingdom} 
}
\email{dm635@kent.ac.uk} 

\author{Dominic Orchard}
\orcid{0000-0002-7058-7842}
\affiliation{
  \department{School of Computing}
  \institution{University of Kent}
  \country{United Kingdom}
}
\email{D.A.Orchard@kent.ac.uk}
\affiliation{
  \department{Department of Computer Science and Technology}
  \institution{University of Cambridge}
  \country{United Kingdom}
}
\email{dao29@cam.ac.uk}

\authorsaddresses{}

\newcommand{\SYSTEMdrule}[4][]{{\displaystyle\frac{\begin{array}{l}#2\end{array}}{#3}\quad\SYSTEMdrulename{#4}}}

\newcommand{\SYSTEMpremise}[1]{ #1 \\}
\newenvironment{SYSTEMdefnblock}[3][]{ \framebox{\mbox{#2}} \quad #3 \\[0pt]}{}

\newcommand{\SYSTEMnt}[1]{\mathit{#1}}
\newcommand{\SYSTEMmv}[1]{\mathit{#1}}
\newcommand{\SYSTEMkw}[1]{\mathbf{#1}}
\newcommand{\SYSTEMsym}[1]{#1}

\newcommand{\SYSTEMdrulename}[1]{\textsc{#1}}

\IfFileExists{xcolor.sty}{
	\RequirePackage{xcolor}
}{
	\RequirePackage{color}
}

 \usepackage{stmaryrd}
 \definecolor{coeffectColor}{HTML}{0750d0}
 \definecolor{effectColor}{HTML}{d64800}
 \definecolor{uniqueColor}{HTML}{c20232}
 \definecolor{borrowColor}{HTML}{226622}
 \definecolor{couniqueColor}{HTML}{7e0080}

\newcommand{\SYSTEMdruleTyvarName}[0]{\SYSTEMdrulename{Tyvar}}
\newcommand{\SYSTEMdruleTyvar}[1]{\SYSTEMdrule[#1]{%
}{
    \SYSTEMsym{0}  \cdot  \Gamma   ,   \SYSTEMmv{x}  :  \SYSTEMnt{A}    \vdash  \SYSTEMmv{x}  :  \SYSTEMnt{A} }{%
{\SYSTEMdruleTyvarName}{}%
}}

\newcommand{\SYSTEMdruleTyabsName}[0]{\SYSTEMdrulename{Tyabs}}
\newcommand{\SYSTEMdruleTyabs}[1]{\SYSTEMdrule[#1]{%
\SYSTEMpremise{  \Gamma ,   \SYSTEMmv{x}  :  \SYSTEMnt{A}    \vdash  \SYSTEMnt{t}  :  \SYSTEMnt{B} }%
}{
 \Gamma  \vdash   \lambda  \SYSTEMmv{x}  .  \SYSTEMnt{t}   :  \SYSTEMnt{A}  \multimap  \SYSTEMnt{B} }{%
{\SYSTEMdruleTyabsName}{}%
}}

\newcommand{\SYSTEMdruleTyappName}[0]{\SYSTEMdrulename{Tyapp}}
\newcommand{\SYSTEMdruleTyapp}[1]{\SYSTEMdrule[#1]{%
\SYSTEMpremise{ \begin{array}{cc}   \Gamma_{{\mathrm{1}}}  \vdash  \SYSTEMnt{t_{{\mathrm{1}}}}  :  \SYSTEMnt{A}  \multimap  \SYSTEMnt{B}   \;\, & \;\,   \Gamma_{{\mathrm{2}}}  \vdash  \SYSTEMnt{t_{{\mathrm{2}}}}  :  \SYSTEMnt{A}   \end{array} }%
}{
  \Gamma_{{\mathrm{1}}}  +  \Gamma_{{\mathrm{2}}}   \vdash  \SYSTEMnt{t_{{\mathrm{1}}}} \, \SYSTEMnt{t_{{\mathrm{2}}}}  :  \SYSTEMnt{B} }{%
{\SYSTEMdruleTyappName}{}%
}}

\newcommand{\SYSTEMdruleTyprName}[0]{\SYSTEMdrulename{Typr}}
\newcommand{\SYSTEMdruleTypr}[1]{\SYSTEMdrule[#1]{%
\SYSTEMpremise{ \begin{array}{cc}   \Gamma  \vdash  \SYSTEMnt{t}  :  \SYSTEMnt{A}   \;\, & \;\,   \neg \mathsf{resourceAllocator}(  \SYSTEMnt{t}  )   \end{array} }%
}{
  r  \cdot  \Gamma   \vdash  \SYSTEMsym{[}  \SYSTEMnt{t}  \SYSTEMsym{]}  :   {\textcolor{coeffectColor}{\Box_{ r } } }  \SYSTEMnt{A}  }{%
{\SYSTEMdruleTyprName}{}%
}}

\newcommand{\SYSTEMdruleTyelimName}[0]{\SYSTEMdrulename{Tyelim}}
\newcommand{\SYSTEMdruleTyelim}[1]{\SYSTEMdrule[#1]{%
\SYSTEMpremise{ \begin{array}{cc}   \Gamma_{{\mathrm{1}}}  \vdash  \SYSTEMnt{t_{{\mathrm{1}}}}  :   {\textcolor{coeffectColor}{\Box_{ r } } }  \SYSTEMnt{A}    \;\, & \;\,    \Gamma_{{\mathrm{2}}} ,   \textcolor{coeffectColor}{ \SYSTEMmv{x}  : [  \SYSTEMnt{A}  ]_{ r } }    \vdash  \SYSTEMnt{t_{{\mathrm{2}}}}  :  \SYSTEMnt{B}   \end{array} }%
}{
  \Gamma_{{\mathrm{1}}}  +  \Gamma_{{\mathrm{2}}}   \vdash  \textbf{let} \, \SYSTEMsym{[}  \SYSTEMmv{x}  \SYSTEMsym{]}  \SYSTEMsym{=}  \SYSTEMnt{t_{{\mathrm{1}}}} \, \textbf{in} \, \SYSTEMnt{t_{{\mathrm{2}}}}  :  \SYSTEMnt{B} }{%
{\SYSTEMdruleTyelimName}{}%
}}

\newcommand{\SYSTEMdruleTyderName}[0]{\SYSTEMdrulename{Tyder}}
\newcommand{\SYSTEMdruleTyder}[1]{\SYSTEMdrule[#1]{%
\SYSTEMpremise{  \Gamma ,   \SYSTEMmv{x}  :  \SYSTEMnt{A}    \vdash  \SYSTEMnt{t}  :  \SYSTEMnt{B} }%
}{
  \Gamma ,   \textcolor{coeffectColor}{ \SYSTEMmv{x}  : [  \SYSTEMnt{A}  ]_{ \SYSTEMsym{1} } }    \vdash  \SYSTEMnt{t}  :  \SYSTEMnt{B} }{%
{\SYSTEMdruleTyderName}{}%
}}

\newcommand{\SYSTEMdruleTyapproxName}[0]{\SYSTEMdrulename{Tyapprox}}
\newcommand{\SYSTEMdruleTyapprox}[1]{\SYSTEMdrule[#1]{%
\SYSTEMpremise{ \begin{array}{cc}     \Gamma ,   \textcolor{coeffectColor}{ \SYSTEMmv{x}  : [  \SYSTEMnt{A}  ]_{ r } }    ,  \Gamma'   \vdash  \SYSTEMnt{t}  :  \SYSTEMnt{B}   \;\, & \;\,   r  \sqsubseteq  s   \end{array} }%
}{
   \Gamma ,   \textcolor{coeffectColor}{ \SYSTEMmv{x}  : [  \SYSTEMnt{A}  ]_{ s } }    ,  \Gamma'   \vdash  \SYSTEMnt{t}  :  \SYSTEMnt{B} }{%
{\SYSTEMdruleTyapproxName}{}%
}}

\newcommand{\SYSTEMdruleTypairIntroName}[0]{\SYSTEMdrulename{TypairIntro}}
\newcommand{\SYSTEMdruleTypairIntro}[1]{\SYSTEMdrule[#1]{%
\SYSTEMpremise{ \begin{array}{cc}   \Gamma_{{\mathrm{1}}}  \vdash  \SYSTEMnt{t_{{\mathrm{1}}}}  :  \SYSTEMnt{A}   \;\, & \;\,   \Gamma_{{\mathrm{2}}}  \vdash  \SYSTEMnt{t_{{\mathrm{2}}}}  :  \SYSTEMnt{B}   \end{array} }%
}{
  \Gamma_{{\mathrm{1}}}  +  \Gamma_{{\mathrm{2}}}   \vdash  \SYSTEMsym{(}  \SYSTEMnt{t_{{\mathrm{1}}}}  \SYSTEMsym{,}  \SYSTEMnt{t_{{\mathrm{2}}}}  \SYSTEMsym{)}  :   \SYSTEMnt{A}  \otimes  \SYSTEMnt{B}  }{%
{\SYSTEMdruleTypairIntroName}{}%
}}

\newcommand{\SYSTEMdruleTypairElimName}[0]{\SYSTEMdrulename{TypairElim}}
\newcommand{\SYSTEMdruleTypairElim}[1]{\SYSTEMdrule[#1]{%
\SYSTEMpremise{ \begin{array}{cc}   \Gamma_{{\mathrm{1}}}  \vdash  \SYSTEMnt{t_{{\mathrm{1}}}}  :   \SYSTEMnt{A}  \otimes  \SYSTEMnt{B}    \;\, & \;\,     \Gamma_{{\mathrm{2}}} ,   \SYSTEMmv{x}  :  \SYSTEMnt{A}   ,   \SYSTEMmv{y}  :  \SYSTEMnt{B}    \vdash  \SYSTEMnt{t_{{\mathrm{2}}}}  :  \SYSTEMnt{C}   \end{array} }%
}{
  \Gamma_{{\mathrm{1}}}  +  \Gamma_{{\mathrm{2}}}   \vdash  \textbf{let} \, \SYSTEMsym{(x,}  \SYSTEMsym{y)}  \SYSTEMsym{=}  \SYSTEMnt{t_{{\mathrm{1}}}} \, \textbf{in} \, \SYSTEMnt{t_{{\mathrm{2}}}}  :  \SYSTEMnt{C} }{%
{\SYSTEMdruleTypairElimName}{}%
}}

\newcommand{\SYSTEMdruleTyunitIntroName}[0]{\SYSTEMdrulename{TyunitIntro}}
\newcommand{\SYSTEMdruleTyunitIntro}[1]{\SYSTEMdrule[#1]{%
}{
  \SYSTEMsym{0}  \cdot  \Gamma   \vdash   ()   :   \mathsf{unit}  }{%
{\SYSTEMdruleTyunitIntroName}{}%
}}

\newcommand{\SYSTEMdruleTyunitElimName}[0]{\SYSTEMdrulename{TyunitElim}}
\newcommand{\SYSTEMdruleTyunitElim}[1]{\SYSTEMdrule[#1]{%
\SYSTEMpremise{ \begin{array}{cc}   \Gamma_{{\mathrm{1}}}  \vdash  \SYSTEMnt{t_{{\mathrm{1}}}}  :   \mathsf{unit}    \;\, & \;\,   \Gamma_{{\mathrm{2}}}  \vdash  \SYSTEMnt{t_{{\mathrm{2}}}}  :  \SYSTEMnt{B}   \end{array} }%
}{
  \Gamma_{{\mathrm{1}}}  +  \Gamma_{{\mathrm{2}}}   \vdash   \textbf{let}\ () =  \SYSTEMnt{t_{{\mathrm{1}}}} \ \textbf{in}\  \SYSTEMnt{t_{{\mathrm{2}}}}   :  \SYSTEMnt{B} }{%
{\SYSTEMdruleTyunitElimName}{}%
}}

\newcommand{\SYSTEMdruleTyreturnGenName}[0]{\SYSTEMdrulename{TyreturnGen}}
\newcommand{\SYSTEMdruleTyreturnGen}[1]{\SYSTEMdrule[#1]{%
\SYSTEMpremise{ \Gamma  \vdash  \SYSTEMnt{t}  :   {\textcolor{uniqueColor}{\ast} }{ \SYSTEMnt{A} }  }%
}{
 \Gamma  \vdash  \SYSTEMkw{share} \, \SYSTEMnt{t}  :   {\textcolor{coeffectColor}{\Box_{ r } } }  \SYSTEMnt{A}  }{%
{\SYSTEMdruleTyreturnGenName}{}%
}}

\newcommand{\SYSTEMdruleTybindGenFreshName}[0]{\SYSTEMdrulename{TybindGenFresh}}
\newcommand{\SYSTEMdruleTybindGenFresh}[1]{\SYSTEMdrule[#1]{%
\SYSTEMpremise{ \begin{array}{cc}   \SYSTEMsym{1}  \sqsubseteq  r   \;\, & \;\,   \mathsf{cloneable} (  \SYSTEMnt{A}  )   \end{array} }%
\SYSTEMpremise{ \begin{array}{cc}    \Gamma_{{\mathrm{1}}} ,   \overline{ \SYSTEMnt{id} }    \vdash  \SYSTEMnt{t_{{\mathrm{1}}}}  :   {\textcolor{coeffectColor}{\Box_{ r } } }  \SYSTEMnt{A}    \;\, & \;\,    \Gamma_{{\mathrm{2}}} ,   \SYSTEMmv{x}  :   \exists \overline{ \SYSTEMnt{id'} } .   {\textcolor{uniqueColor}{\ast} }{  (   \SYSTEMnt{A}  [ \overline{ \SYSTEMnt{id'} } / \overline{ \SYSTEMnt{id} } ]   )  }      \vdash  \SYSTEMnt{t_{{\mathrm{2}}}}  :  \SYSTEMnt{B}   \end{array} }%
}{
  \SYSTEMsym{(}   \Gamma_{{\mathrm{1}}}  +  \Gamma_{{\mathrm{2}}}   \SYSTEMsym{)} ,   \overline{ \SYSTEMnt{id} }    \vdash  \SYSTEMkw{clone} \, \SYSTEMnt{t_{{\mathrm{1}}}} \, \textbf{as} \, \SYSTEMmv{x} \, \textbf{in} \, \SYSTEMnt{t_{{\mathrm{2}}}}  :  \SYSTEMnt{B} }{%
{\SYSTEMdruleTybindGenFreshName}{}%
}}

\newcommand{\SYSTEMdruleTypackName}[0]{\SYSTEMdrulename{Typack}}

\newcommand{\SYSTEMdruleTyunpackName}[0]{\SYSTEMdrulename{Tyunpack}}

\newcommand{\SYSTEMdruleTywithBorrowName}[0]{\SYSTEMdrulename{TywithBorrow}}
\newcommand{\SYSTEMdruleTywithBorrow}[1]{\SYSTEMdrule[#1]{%
\SYSTEMpremise{ \begin{array}{cc}   \Gamma_{{\mathrm{1}}}  \vdash  \SYSTEMnt{t_{{\mathrm{1}}}}  :   {\textcolor{uniqueColor}{\ast} }{ \SYSTEMnt{A} }    \;\, & \;\,   \Gamma_{{\mathrm{2}}}  \vdash  \SYSTEMnt{t_{{\mathrm{2}}}}  :    {\textcolor{borrowColor}{\&_{ \SYSTEMsym{1} } } }  \SYSTEMnt{A}    \multimap    {\textcolor{borrowColor}{\&_{ \SYSTEMsym{1} } } }  \SYSTEMnt{B}     \end{array} }%
}{
  \Gamma_{{\mathrm{1}}}  +  \Gamma_{{\mathrm{2}}}   \vdash   \textbf{withBorrow}\  \SYSTEMnt{t_{{\mathrm{1}}}} \  \SYSTEMnt{t_{{\mathrm{2}}}}   :   {\textcolor{uniqueColor}{\ast} }{ \SYSTEMnt{B} }  }{%
{\SYSTEMdruleTywithBorrowName}{}%
}}

\newcommand{\SYSTEMdruleTysplitName}[0]{\SYSTEMdrulename{Tysplit}}
\newcommand{\SYSTEMdruleTysplit}[1]{\SYSTEMdrule[#1]{%
\SYSTEMpremise{ \Gamma  \vdash  \SYSTEMnt{t}  :   {\textcolor{borrowColor}{\&_{ p } } }  \SYSTEMnt{A}  }%
}{
 \Gamma  \vdash   \textbf{split}\  \SYSTEMnt{t}   :      {\textcolor{borrowColor}{\&_{   \frac{ p }{ \SYSTEMsym{2} }   } } }  \SYSTEMnt{A}    \otimes    {\textcolor{borrowColor}{\&_{   \frac{ p }{ \SYSTEMsym{2} }   } } }  \SYSTEMnt{A}     }{%
{\SYSTEMdruleTysplitName}{}%
}}

\newcommand{\SYSTEMdruleTyjoinName}[0]{\SYSTEMdrulename{Tyjoin}}
\newcommand{\SYSTEMdruleTyjoin}[1]{\SYSTEMdrule[#1]{%
\SYSTEMpremise{ \begin{array}{cc}    \begin{array}{cc}   \Gamma_{{\mathrm{1}}}  \vdash  \SYSTEMnt{t_{{\mathrm{1}}}}  :   {\textcolor{borrowColor}{\&_{ p } } }  \SYSTEMnt{A}    \;\, & \;\,   \Gamma_{{\mathrm{2}}}  \vdash  \SYSTEMnt{t_{{\mathrm{2}}}}  :   {\textcolor{borrowColor}{\&_{ q } } }  \SYSTEMnt{A}    \end{array}    \;\, & \;\,     p  +  q    \leq  \SYSTEMsym{1}   \end{array} }%
}{
  \Gamma_{{\mathrm{1}}}  +  \Gamma_{{\mathrm{2}}}   \vdash   \textbf{join}\  \SYSTEMnt{t_{{\mathrm{1}}}} \  \SYSTEMnt{t_{{\mathrm{2}}}}   :   {\textcolor{borrowColor}{\&_{   p  +  q   } } }  \SYSTEMnt{A}  }{%
{\SYSTEMdruleTyjoinName}{}%
}}

\newcommand{\SYSTEMdruleTypushName}[0]{\SYSTEMdrulename{Typush}}
\newcommand{\SYSTEMdruleTypush}[1]{\SYSTEMdrule[#1]{%
\SYSTEMpremise{ \Gamma  \vdash  \SYSTEMnt{t}  :   {\textcolor{borrowColor}{\&_{ p } } }   (   \SYSTEMnt{A}  \otimes  \SYSTEMnt{B}   )   }%
}{
 \Gamma  \vdash  \SYSTEMkw{push} \, \SYSTEMnt{t}  :    (   {\textcolor{borrowColor}{\&_{ p } } }  \SYSTEMnt{A}   )   \otimes   (   {\textcolor{borrowColor}{\&_{ p } } }  \SYSTEMnt{B}   )   }{%
{\SYSTEMdruleTypushName}{}%
}}

\newcommand{\SYSTEMdruleTypullName}[0]{\SYSTEMdrulename{Typull}}
\newcommand{\SYSTEMdruleTypull}[1]{\SYSTEMdrule[#1]{%
\SYSTEMpremise{ \Gamma  \vdash  \SYSTEMnt{t}  :    (   {\textcolor{borrowColor}{\&_{ p } } }  \SYSTEMnt{A}   )   \otimes   (   {\textcolor{borrowColor}{\&_{ p } } }  \SYSTEMnt{B}   )   }%
}{
 \Gamma  \vdash  \SYSTEMkw{pull} \, \SYSTEMnt{t}  :   {\textcolor{borrowColor}{\&_{ p } } }   (   \SYSTEMnt{A}  \otimes  \SYSTEMnt{B}   )   }{%
{\SYSTEMdruleTypullName}{}%
}}

\newcommand{\SYSTEMdruleRTynecName}[0]{\SYSTEMdrulename{RTynec}}
\newcommand{\SYSTEMdruleRTynec}[1]{\SYSTEMdrule[#1]{%
\SYSTEMpremise{ \gamma  \vdash  \SYSTEMnt{t}  :  \SYSTEMnt{A} }%
}{
    \SYSTEMsym{0}  \cdot  \Gamma    ,  \gamma   \vdash  \ast  \SYSTEMnt{t}  :   {\textcolor{borrowColor}{\&_{ p } } }  \SYSTEMnt{A}  }{%
{\SYSTEMdruleRTynecName}{}%
}}

\newcommand{\SYSTEMdruleRTyresName}[0]{\SYSTEMdrulename{RTyres}}
\newcommand{\SYSTEMdruleRTyres}[1]{\SYSTEMdrule[#1]{%
}{
    \SYSTEMsym{0}  \cdot  \Gamma   ,   \SYSTEMmv{ref}  :   \SYSTEMnt{Res} _{ \SYSTEMnt{id} } \  \SYSTEMnt{A}     \vdash  \SYSTEMmv{ref}  :   \SYSTEMnt{Res} _{ \SYSTEMnt{id} } \  \SYSTEMnt{A}  }{%
{\SYSTEMdruleRTyresName}{}%
}}

\newcommand{\SYSTEMdruleRTyunborrowName}[0]{\SYSTEMdrulename{RTyunborrow}}
\newcommand{\SYSTEMdruleRTyunborrow}[1]{\SYSTEMdrule[#1]{%
\SYSTEMpremise{ \Gamma  \vdash  \SYSTEMnt{t}  :   {\textcolor{borrowColor}{\&_{ \SYSTEMsym{1} } } }  \SYSTEMnt{A}  }%
}{
 \Gamma  \vdash   \textbf{unborrow}\  \SYSTEMnt{t}   :   {\textcolor{uniqueColor}{\ast} }{ \SYSTEMnt{A} }  }{%
{\SYSTEMdruleRTyunborrowName}{}%
}}

\newcommand{\SYSTEMdruleHeapexistentialBetaName}[0]{\SYSTEMdrulename{HeapexistentialBeta}}
\newcommand{\SYSTEMdruleHeapexistentialBeta}[1]{\SYSTEMdrule[#1]{%
\SYSTEMpremise{ \SYSTEMmv{y}  \# \{ H  ,  \SYSTEMnt{v}  ,  \SYSTEMnt{t} \} }%
}{
 H  \vdash   \textbf{unpack}\ \langle{  \SYSTEMnt{id}  ,  \SYSTEMmv{x}  }\rangle
     =   \textbf{pack}\ \langle{  \SYSTEMnt{id'}  ,  \SYSTEMnt{v}  }\rangle  \ \textbf{in}\  \SYSTEMnt{t}   \,\leadsto_{ s } \,   H ,   \SYSTEMmv{y}  \textcolor{coeffectColor}{\mapsto_{ r } }  \SYSTEMnt{v}    \vdash   \SYSTEMnt{t}  [  \SYSTEMmv{y}  /  \SYSTEMmv{x}  ]  }{%
{\SYSTEMdruleHeapexistentialBetaName}{}%
}}

\newcommand{\SYSTEMdruleHeapvarName}[0]{\SYSTEMdrulename{Heapvar}}
\newcommand{\SYSTEMdruleHeapvar}[1]{\SYSTEMdrule[#1]{%
\SYSTEMpremise{ \exists  r'  .\    s  +  r'   \sqsubseteq  r  }%
}{
  H ,   \SYSTEMmv{x}  \textcolor{coeffectColor}{\mapsto_{ r } }  \SYSTEMnt{v}    \vdash  \SYSTEMmv{x}  \,\leadsto_{ s } \,   H ,   \SYSTEMmv{x}  \textcolor{coeffectColor}{\mapsto_{ r } }  \SYSTEMnt{v}    \vdash   \SYSTEMnt{v} _{      }  }{%
{\SYSTEMdruleHeapvarName}{}%
}}

\newcommand{\SYSTEMdruleHeapbetaName}[0]{\SYSTEMdrulename{Heapbeta}}
\newcommand{\SYSTEMdruleHeapbeta}[1]{\SYSTEMdrule[#1]{%
\SYSTEMpremise{ \SYSTEMmv{y}  \# \{ H  ,  \SYSTEMnt{v}  ,  \SYSTEMnt{t} \} }%
}{
 H  \vdash  \SYSTEMsym{(}   \lambda  \SYSTEMmv{x}  .  \SYSTEMnt{t}   \SYSTEMsym{)} \, \SYSTEMnt{v}  \,\leadsto_{ s } \,   H ,   \SYSTEMmv{y}  \textcolor{coeffectColor}{\mapsto_{ s } }  \SYSTEMnt{v}    \vdash   \SYSTEMnt{t}  [  \SYSTEMmv{y}  /  \SYSTEMmv{x}  ]  }{%
{\SYSTEMdruleHeapbetaName}{}%
}}

\newcommand{\SYSTEMdruleHeapappLName}[0]{\SYSTEMdrulename{HeapappL}}
\newcommand{\SYSTEMdruleHeapappL}[1]{\SYSTEMdrule[#1]{%
\SYSTEMpremise{ H  \vdash  \SYSTEMnt{t_{{\mathrm{1}}}}  \,\leadsto_{ s } \,  H'  \vdash  \SYSTEMnt{t'_{{\mathrm{1}}}} }%
}{
 H  \vdash  \SYSTEMnt{t_{{\mathrm{1}}}} \, \SYSTEMnt{t_{{\mathrm{2}}}}  \,\leadsto_{ s } \,  H'  \vdash  \SYSTEMnt{t'_{{\mathrm{1}}}} \, \SYSTEMnt{t_{{\mathrm{2}}}} }{%
{\SYSTEMdruleHeapappLName}{}%
}}

\newcommand{\SYSTEMdruleHeapappRName}[0]{\SYSTEMdrulename{HeapappR}}
\newcommand{\SYSTEMdruleHeapappR}[1]{\SYSTEMdrule[#1]{%
\SYSTEMpremise{ H  \vdash  \SYSTEMnt{t_{{\mathrm{2}}}}  \,\leadsto_{ s } \,  H'  \vdash  \SYSTEMnt{t'_{{\mathrm{2}}}} }%
}{
 H  \vdash  \SYSTEMnt{v} \, \SYSTEMnt{t_{{\mathrm{2}}}}  \,\leadsto_{ s } \,  H'  \vdash  \SYSTEMnt{v} \, \SYSTEMnt{t'_{{\mathrm{2}}}} }{%
{\SYSTEMdruleHeapappRName}{}%
}}

\newcommand{\SYSTEMdruleHeapbetaBoxName}[0]{\SYSTEMdrulename{HeapbetaBox}}
\newcommand{\SYSTEMdruleHeapbetaBox}[1]{\SYSTEMdrule[#1]{%
\SYSTEMpremise{ \SYSTEMmv{y}  \# \{ H  ,  \SYSTEMnt{v}  ,  \SYSTEMnt{t} \} }%
}{
 H  \vdash  \textbf{let} \, \SYSTEMsym{[}  \SYSTEMmv{x}  \SYSTEMsym{]}  \SYSTEMsym{=}   [ { \SYSTEMnt{v} } ]_{\color{coeffectColor}{ r } }  \, \textbf{in} \, \SYSTEMnt{t}  \,\leadsto_{ s } \,   H ,   \SYSTEMmv{y}  \textcolor{coeffectColor}{\mapsto_{  (   s  *  r   )  } }  \SYSTEMnt{v}    \vdash   \SYSTEMnt{t}  [  \SYSTEMmv{y}  /  \SYSTEMmv{x}  ]  }{%
{\SYSTEMdruleHeapbetaBoxName}{}%
}}

\newcommand{\SYSTEMdruleHeapcongBoxElimName}[0]{\SYSTEMdrulename{HeapcongBoxElim}}
\newcommand{\SYSTEMdruleHeapcongBoxElim}[1]{\SYSTEMdrule[#1]{%
\SYSTEMpremise{ H  \vdash  \SYSTEMnt{t_{{\mathrm{1}}}}  \,\leadsto_{ s } \,  H'  \vdash  \SYSTEMnt{t'_{{\mathrm{1}}}} }%
}{
 H  \vdash  \textbf{let} \, \SYSTEMsym{[}  \SYSTEMmv{x}  \SYSTEMsym{]}  \SYSTEMsym{=}  \SYSTEMnt{t_{{\mathrm{1}}}} \, \textbf{in} \, \SYSTEMnt{t_{{\mathrm{2}}}}  \,\leadsto_{ s } \,  H'  \vdash  \textbf{let} \, \SYSTEMsym{[}  \SYSTEMmv{x}  \SYSTEMsym{]}  \SYSTEMsym{=}  \SYSTEMnt{t'_{{\mathrm{1}}}} \, \textbf{in} \, \SYSTEMnt{t_{{\mathrm{2}}}} }{%
{\SYSTEMdruleHeapcongBoxElimName}{}%
}}

\newcommand{\SYSTEMdruleHeapcongPromotionName}[0]{\SYSTEMdrulename{HeapcongPromotion}}
\newcommand{\SYSTEMdruleHeapcongPromotion}[1]{\SYSTEMdrule[#1]{%
\SYSTEMpremise{ H  \vdash  \SYSTEMnt{t}  \,\leadsto_{   s  *  r   } \,  H'  \vdash  \SYSTEMnt{t'} }%
}{
 H  \vdash   [ { \SYSTEMnt{t} } ]_{\color{coeffectColor}{ r } }   \,\leadsto_{ s } \,  H'  \vdash   [ { \SYSTEMnt{t'} } ]_{\color{coeffectColor}{ r } }  }{%
{\SYSTEMdruleHeapcongPromotionName}{}%
}}

\newcommand{\SYSTEMdruleHeapcongPairLName}[0]{\SYSTEMdrulename{HeapcongPairL}}
\newcommand{\SYSTEMdruleHeapcongPairL}[1]{\SYSTEMdrule[#1]{%
\SYSTEMpremise{ H  \vdash  \SYSTEMnt{t_{{\mathrm{1}}}}  \,\leadsto_{ s } \,  H'  \vdash  \SYSTEMnt{t'_{{\mathrm{1}}}} }%
}{
 H  \vdash  \SYSTEMsym{(}  \SYSTEMnt{t_{{\mathrm{1}}}}  \SYSTEMsym{,}  \SYSTEMnt{t_{{\mathrm{2}}}}  \SYSTEMsym{)}  \,\leadsto_{ s } \,  H'  \vdash  \SYSTEMsym{(}  \SYSTEMnt{t'_{{\mathrm{1}}}}  \SYSTEMsym{,}  \SYSTEMnt{t_{{\mathrm{2}}}}  \SYSTEMsym{)} }{%
{\SYSTEMdruleHeapcongPairLName}{}%
}}

\newcommand{\SYSTEMdruleHeapcongPairRName}[0]{\SYSTEMdrulename{HeapcongPairR}}
\newcommand{\SYSTEMdruleHeapcongPairR}[1]{\SYSTEMdrule[#1]{%
\SYSTEMpremise{ H  \vdash  \SYSTEMnt{t_{{\mathrm{2}}}}  \,\leadsto_{ s } \,  H'  \vdash  \SYSTEMnt{t'_{{\mathrm{2}}}} }%
}{
 H  \vdash  \SYSTEMsym{(}  \SYSTEMnt{v}  \SYSTEMsym{,}  \SYSTEMnt{t_{{\mathrm{2}}}}  \SYSTEMsym{)}  \,\leadsto_{ s } \,  H'  \vdash  \SYSTEMsym{(}  \SYSTEMnt{v}  \SYSTEMsym{,}  \SYSTEMnt{t'_{{\mathrm{2}}}}  \SYSTEMsym{)} }{%
{\SYSTEMdruleHeapcongPairRName}{}%
}}

\newcommand{\SYSTEMdruleHeappairBetaName}[0]{\SYSTEMdrulename{HeappairBeta}}
\newcommand{\SYSTEMdruleHeappairBeta}[1]{\SYSTEMdrule[#1]{%
\SYSTEMpremise{ \begin{array}{cc}   \SYSTEMmv{x'}  \# \{ H  ,  \SYSTEMnt{v_{{\mathrm{1}}}}  ,  \SYSTEMnt{v_{{\mathrm{2}}}}  ,  \SYSTEMnt{t} \}   \;\, & \;\,   \SYSTEMmv{y'}  \# \{ H  ,  \SYSTEMnt{v_{{\mathrm{1}}}}  ,  \SYSTEMnt{v_{{\mathrm{2}}}}  ,  \SYSTEMnt{t} \}   \end{array} }%
}{
 H  \vdash  \textbf{let} \, \SYSTEMsym{(x,}  \SYSTEMsym{y)}  \SYSTEMsym{=}  \SYSTEMsym{(}  \SYSTEMnt{v_{{\mathrm{1}}}}  \SYSTEMsym{,}  \SYSTEMnt{v_{{\mathrm{2}}}}  \SYSTEMsym{)} \, \textbf{in} \, \SYSTEMnt{t}  \,\leadsto_{ s } \,     H ,   \SYSTEMmv{x'}  \textcolor{coeffectColor}{\mapsto_{ s } }  \SYSTEMnt{v_{{\mathrm{1}}}}    ,   \SYSTEMmv{y'}  \textcolor{coeffectColor}{\mapsto_{ s } }  \SYSTEMnt{v_{{\mathrm{2}}}}    \vdash    \SYSTEMnt{t}  [  \SYSTEMmv{y'}  /  \SYSTEMmv{y}  ]   [  \SYSTEMmv{x'}  /  \SYSTEMmv{x}  ]  }{%
{\SYSTEMdruleHeappairBetaName}{}%
}}

\newcommand{\SYSTEMdruleHeapcongPairElimName}[0]{\SYSTEMdrulename{HeapcongPairElim}}
\newcommand{\SYSTEMdruleHeapcongPairElim}[1]{\SYSTEMdrule[#1]{%
\SYSTEMpremise{ H  \vdash  \SYSTEMnt{t_{{\mathrm{1}}}}  \,\leadsto_{ s } \,  H'  \vdash  \SYSTEMnt{t'_{{\mathrm{1}}}} }%
}{
 H  \vdash  \textbf{let} \, \SYSTEMsym{(x,}  \SYSTEMsym{y)}  \SYSTEMsym{=}  \SYSTEMnt{t_{{\mathrm{1}}}} \, \textbf{in} \, \SYSTEMnt{t_{{\mathrm{2}}}}  \,\leadsto_{ s } \,  H'  \vdash  \textbf{let} \, \SYSTEMsym{(x,}  \SYSTEMsym{y)}  \SYSTEMsym{=}  \SYSTEMnt{t'_{{\mathrm{1}}}} \, \textbf{in} \, \SYSTEMnt{t_{{\mathrm{2}}}} }{%
{\SYSTEMdruleHeapcongPairElimName}{}%
}}

\newcommand{\SYSTEMdruleHeapcongShareName}[0]{\SYSTEMdrulename{HeapcongShare}}
\newcommand{\SYSTEMdruleHeapcongShare}[1]{\SYSTEMdrule[#1]{%
\SYSTEMpremise{ H  \vdash  \SYSTEMnt{t}  \,\leadsto_{ s } \,  H'  \vdash  \SYSTEMnt{t'} }%
}{
 H  \vdash  \SYSTEMkw{share} \, \SYSTEMnt{t}  \,\leadsto_{ s } \,  H'  \vdash  \SYSTEMkw{share} \, \SYSTEMnt{t'} }{%
{\SYSTEMdruleHeapcongShareName}{}%
}}

\newcommand{\SYSTEMdruleHeapshareName}[0]{\SYSTEMdrulename{Heapshare}}
\newcommand{\SYSTEMdruleHeapshare}[1]{\SYSTEMdrule[#1]{%
\SYSTEMpremise{   \mathsf{dom}  (  H  )   \equiv    \mathsf{refs}  (  \SYSTEMnt{v}  )  }%
}{
   H  ,  H'    \vdash  \SYSTEMkw{share} \, \SYSTEMsym{(}  \ast  \SYSTEMnt{v}  \SYSTEMsym{)}  \,\leadsto_{ s } \,    \SYSTEMsym{(}   { {[  H  ]}_{\textcolor{uniqueColor}{ \SYSTEMsym{0} } } }   \SYSTEMsym{)}  ,  H'    \vdash  \SYSTEMsym{[}  \SYSTEMnt{v}  \SYSTEMsym{]} }{%
{\SYSTEMdruleHeapshareName}{}%
}}

\newcommand{\SYSTEMdruleHeapsplitRefName}[0]{\SYSTEMdrulename{HeapsplitRef}}
\newcommand{\SYSTEMdruleHeapsplitRef}[1]{\SYSTEMdrule[#1]{%
\SYSTEMpremise{ \begin{array}{cc}   \SYSTEMmv{ref_{{\mathrm{1}}}}  \#  H   \;\, & \;\,   \SYSTEMmv{ref_{{\mathrm{2}}}}  \#  H   \end{array} }%
}{
   H ,   \SYSTEMmv{ref}  \textcolor{uniqueColor}{\mapsto_{ p } }  \SYSTEMnt{id}   ,   \SYSTEMnt{id}  \mapsto  \SYSTEMnt{v}    \vdash   \textbf{split}\  \SYSTEMsym{(}  \ast  \SYSTEMmv{ref}  \SYSTEMsym{)}   \,\leadsto_{ s } \,     H ,   \SYSTEMmv{ref_{{\mathrm{1}}}}  \textcolor{uniqueColor}{\mapsto_{  \frac{ p }{ \SYSTEMsym{2} }  } }  \SYSTEMnt{id}   ,   \SYSTEMmv{ref_{{\mathrm{2}}}}  \textcolor{uniqueColor}{\mapsto_{  \frac{ p }{ \SYSTEMsym{2} }  } }  \SYSTEMnt{id}   ,   \SYSTEMnt{id}  \mapsto  \SYSTEMnt{v}    \vdash  \SYSTEMsym{(}  \ast  \SYSTEMmv{ref_{{\mathrm{1}}}}  \SYSTEMsym{,}  \ast  \SYSTEMmv{ref_{{\mathrm{2}}}}  \SYSTEMsym{)} }{%
{\SYSTEMdruleHeapsplitRefName}{}%
}}

\newcommand{\SYSTEMdruleHeapsplitPairName}[0]{\SYSTEMdrulename{HeapsplitPair}}
\newcommand{\SYSTEMdruleHeapsplitPair}[1]{\SYSTEMdrule[#1]{%
\SYSTEMpremise{ H  \vdash   \textbf{split}\  \SYSTEMsym{(}  \ast  \SYSTEMnt{v}  \SYSTEMsym{)}   \,\leadsto_{ s } \,  H'  \vdash  \SYSTEMsym{(}  \ast  \SYSTEMnt{v_{{\mathrm{1}}}}  \SYSTEMsym{,}  \ast  \SYSTEMnt{v_{{\mathrm{2}}}}  \SYSTEMsym{)} }%
\SYSTEMpremise{ H'  \vdash   \textbf{split}\  \SYSTEMsym{(}  \ast  \SYSTEMnt{w}  \SYSTEMsym{)}   \,\leadsto_{ s } \,  H''  \vdash  \SYSTEMsym{(}  \ast  \SYSTEMnt{w_{{\mathrm{1}}}}  \SYSTEMsym{,}  \ast  \SYSTEMnt{w_{{\mathrm{2}}}}  \SYSTEMsym{)} }%
}{
 H  \vdash   \textbf{split}\  \SYSTEMsym{(}  \ast  \SYSTEMsym{(}  \SYSTEMnt{v}  \SYSTEMsym{,}  \SYSTEMnt{w}  \SYSTEMsym{)}  \SYSTEMsym{)}   \,\leadsto_{ s } \,  H''  \vdash  \SYSTEMsym{(}  \ast  \SYSTEMsym{(}  \SYSTEMnt{v_{{\mathrm{1}}}}  \SYSTEMsym{,}  \SYSTEMnt{w_{{\mathrm{1}}}}  \SYSTEMsym{)}  \SYSTEMsym{,}  \ast  \SYSTEMsym{(}  \SYSTEMnt{v_{{\mathrm{2}}}}  \SYSTEMsym{,}  \SYSTEMnt{w_{{\mathrm{2}}}}  \SYSTEMsym{)}  \SYSTEMsym{)} }{%
{\SYSTEMdruleHeapsplitPairName}{}%
}}

\newcommand{\SYSTEMdruleHeapjoinPairName}[0]{\SYSTEMdrulename{HeapjoinPair}}
\newcommand{\SYSTEMdruleHeapjoinPair}[1]{\SYSTEMdrule[#1]{%
\SYSTEMpremise{ H  \vdash   \textbf{join}\  \SYSTEMsym{(}  \ast  \SYSTEMnt{v_{{\mathrm{1}}}}  \SYSTEMsym{)} \  \SYSTEMsym{(}  \ast  \SYSTEMnt{v_{{\mathrm{2}}}}  \SYSTEMsym{)}   \,\leadsto_{ s } \,  H'  \vdash  \ast  \SYSTEMnt{v} }%
\SYSTEMpremise{ H'  \vdash   \textbf{join}\  \SYSTEMsym{(}  \ast  \SYSTEMnt{w_{{\mathrm{1}}}}  \SYSTEMsym{)} \  \SYSTEMsym{(}  \ast  \SYSTEMnt{w_{{\mathrm{2}}}}  \SYSTEMsym{)}   \,\leadsto_{ s } \,  H''  \vdash  \ast  \SYSTEMnt{w} }%
}{
 H  \vdash   \textbf{join}\  \SYSTEMsym{(}  \ast  \SYSTEMsym{(}  \SYSTEMnt{v_{{\mathrm{1}}}}  \SYSTEMsym{,}  \SYSTEMnt{w_{{\mathrm{1}}}}  \SYSTEMsym{)}  \SYSTEMsym{)} \  \SYSTEMsym{(}  \ast  \SYSTEMsym{(}  \SYSTEMnt{v_{{\mathrm{2}}}}  \SYSTEMsym{,}  \SYSTEMnt{w_{{\mathrm{2}}}}  \SYSTEMsym{)}  \SYSTEMsym{)}   \,\leadsto_{ s } \,  H''  \vdash  \ast  \SYSTEMsym{(}  \SYSTEMnt{v}  \SYSTEMsym{,}  \SYSTEMnt{w}  \SYSTEMsym{)} }{%
{\SYSTEMdruleHeapjoinPairName}{}%
}}

\newcommand{\SYSTEMdruleHeapjoinRefName}[0]{\SYSTEMdrulename{HeapjoinRef}}
\newcommand{\SYSTEMdruleHeapjoinRef}[1]{\SYSTEMdrule[#1]{%
\SYSTEMpremise{ \SYSTEMmv{ref}  \#  H }%
}{
    H ,   \SYSTEMmv{ref_{{\mathrm{1}}}}  \textcolor{uniqueColor}{\mapsto_{ p } }  \SYSTEMnt{id}   ,   \SYSTEMmv{ref_{{\mathrm{2}}}}  \textcolor{uniqueColor}{\mapsto_{ q } }  \SYSTEMnt{id}   ,   \SYSTEMnt{id}  \mapsto  \SYSTEMnt{v}    \vdash   \textbf{join}\  \SYSTEMsym{(}  \ast  \SYSTEMmv{ref_{{\mathrm{1}}}}  \SYSTEMsym{)} \  \SYSTEMsym{(}  \ast  \SYSTEMmv{ref_{{\mathrm{2}}}}  \SYSTEMsym{)}   \,\leadsto_{ s } \,    H ,   \SYSTEMmv{ref}  \textcolor{uniqueColor}{\mapsto_{  (   p  +  q   )  } }  \SYSTEMnt{id}   ,   \SYSTEMnt{id}  \mapsto  \SYSTEMnt{v}    \vdash  \ast  \SYSTEMmv{ref} }{%
{\SYSTEMdruleHeapjoinRefName}{}%
}}

\newcommand{\SYSTEMdruleHeapwithBorrowName}[0]{\SYSTEMdrulename{HeapwithBorrow}}
\newcommand{\SYSTEMdruleHeapwithBorrow}[1]{\SYSTEMdrule[#1]{%
\SYSTEMpremise{ \SYSTEMmv{y}  \# \{ H  ,  \SYSTEMnt{v}  ,  \SYSTEMnt{t} \} }%
}{
 H  \vdash   \textbf{withBorrow}\  \SYSTEMsym{(}   \lambda  \SYSTEMmv{x}  .  \SYSTEMnt{t}   \SYSTEMsym{)} \  \SYSTEMsym{(}  \ast  \SYSTEMnt{v}  \SYSTEMsym{)}   \,\leadsto_{ s } \,   H ,   \SYSTEMmv{y}  \textcolor{coeffectColor}{\mapsto_{ s } }  \SYSTEMsym{(}  \ast  \SYSTEMnt{v}  \SYSTEMsym{)}    \vdash   \textbf{unborrow}\    \SYSTEMnt{t}  [  \SYSTEMmv{y}  /  \SYSTEMmv{x}  ]    }{%
{\SYSTEMdruleHeapwithBorrowName}{}%
}}

\newcommand{\SYSTEMdruleHeapunborrowBorrowName}[0]{\SYSTEMdrulename{HeapunborrowBorrow}}
\newcommand{\SYSTEMdruleHeapunborrowBorrow}[1]{\SYSTEMdrule[#1]{%
}{
 H  \vdash   \textbf{unborrow}\  \SYSTEMsym{(}  \ast  \SYSTEMnt{v}  \SYSTEMsym{)}   \,\leadsto_{ s } \,  H  \vdash  \ast  \SYSTEMnt{v} }{%
{\SYSTEMdruleHeapunborrowBorrowName}{}%
}}

\newcommand{\SYSTEMdruleHeapcongUnborrowName}[0]{\SYSTEMdrulename{HeapcongUnborrow}}
\newcommand{\SYSTEMdruleHeapcongUnborrow}[1]{\SYSTEMdrule[#1]{%
\SYSTEMpremise{ H  \vdash  \SYSTEMnt{t}  \,\leadsto_{ s } \,  H'  \vdash  \SYSTEMnt{t'} }%
}{
 H  \vdash   \textbf{unborrow}\  \SYSTEMnt{t}   \,\leadsto_{ s } \,  H'  \vdash   \textbf{unborrow}\  \SYSTEMnt{t'}  }{%
{\SYSTEMdruleHeapcongUnborrowName}{}%
}}

\newcommand{\SYSTEMdruleHeapnewArrayName}[0]{\SYSTEMdrulename{HeapnewArray}}
\newcommand{\SYSTEMdruleHeapnewArray}[1]{\SYSTEMdrule[#1]{%
\SYSTEMpremise{ \begin{array}{cc}   \SYSTEMmv{ref}  \#  H   \;\, & \;\,   \SYSTEMnt{id}  \#  H   \end{array} }%
}{
 H  \vdash   \textbf{newArray}  \, \SYSTEMmv{n}  \,\leadsto_{ s } \,    H ,   \SYSTEMmv{ref}  \textcolor{uniqueColor}{\mapsto_{ \SYSTEMsym{1} } }  \SYSTEMnt{id}   ,   \SYSTEMnt{id}  \mapsto   \mathsf{init}     \vdash   \textbf{pack}\ \langle{  \SYSTEMnt{id}  ,  \ast  \SYSTEMmv{ref}  }\rangle  }{%
{\SYSTEMdruleHeapnewArrayName}{}%
}}

\newcommand{\SYSTEMdruleHeapreadArrayName}[0]{\SYSTEMdrulename{HeapreadArray}}
\newcommand{\SYSTEMdruleHeapreadArray}[1]{\SYSTEMdrule[#1]{%
}{
   H ,   \SYSTEMmv{ref}  \textcolor{uniqueColor}{\mapsto_{ p } }  \SYSTEMnt{id}   ,   \SYSTEMnt{id}  \mapsto   \textbf{arr}  [  \SYSTEMmv{i}  ] =  \SYSTEMnt{v}     \vdash   \textbf{readArray}  \, \SYSTEMsym{(}  \ast  \SYSTEMmv{ref}  \SYSTEMsym{)} \, \SYSTEMmv{i}  \,\leadsto_{ s } \,    H ,   \SYSTEMmv{ref}  \textcolor{uniqueColor}{\mapsto_{ p } }  \SYSTEMnt{id}   ,   \SYSTEMnt{id}  \mapsto   \textbf{arr}  [  \SYSTEMmv{i}  ] =  \SYSTEMnt{v}     \vdash  \SYSTEMsym{(}  \SYSTEMnt{v}  \SYSTEMsym{,}  \ast  \SYSTEMmv{ref}  \SYSTEMsym{)} }{%
{\SYSTEMdruleHeapreadArrayName}{}%
}}

\newcommand{\SYSTEMdruleHeapwriteArrayName}[0]{\SYSTEMdrulename{HeapwriteArray}}
\newcommand{\SYSTEMdruleHeapwriteArray}[1]{\SYSTEMdrule[#1]{%
}{
   H ,   \SYSTEMmv{ref}  \textcolor{uniqueColor}{\mapsto_{ p } }  \SYSTEMnt{id}   ,   \SYSTEMnt{id}  \mapsto  \textbf{arr}    \vdash   \textbf{writeArray}  \, \SYSTEMsym{(}  \ast  \SYSTEMmv{ref}  \SYSTEMsym{)} \, \SYSTEMmv{i} \, \SYSTEMnt{v}  \,\leadsto_{ s } \,    H ,   \SYSTEMmv{ref}  \textcolor{uniqueColor}{\mapsto_{ p } }  \SYSTEMnt{id}   ,   \SYSTEMnt{id}  \mapsto   \textbf{arr}  [  \SYSTEMmv{i}  ] =  \SYSTEMnt{v}     \vdash  \ast  \SYSTEMmv{ref} }{%
{\SYSTEMdruleHeapwriteArrayName}{}%
}}

\newcommand{\SYSTEMdruleHeapdeleteArrayName}[0]{\SYSTEMdrulename{HeapdeleteArray}}
\newcommand{\SYSTEMdruleHeapdeleteArray}[1]{\SYSTEMdrule[#1]{%
}{
   H ,   \SYSTEMmv{ref}  \textcolor{uniqueColor}{\mapsto_{ p } }  \SYSTEMnt{id}   ,   \SYSTEMnt{id}  \mapsto  \textbf{arr}    \vdash   \textbf{deleteArray}  \, \SYSTEMsym{(}  \ast  \SYSTEMmv{ref}  \SYSTEMsym{)}  \,\leadsto_{ s } \,  H  \vdash   ()  }{%
{\SYSTEMdruleHeapdeleteArrayName}{}%
}}

\newcommand{\SYSTEMdruleHeapcopyBetaName}[0]{\SYSTEMdrulename{HeapcopyBeta}}
\newcommand{\SYSTEMdruleHeapcopyBeta}[1]{\SYSTEMdrule[#1]{%
\SYSTEMpremise{ \begin{array}{cc}    \begin{array}{cc}      \mathsf{dom}  (  H'  )   \equiv    \mathsf{refs}  (  \SYSTEMnt{v}  )     \;\, & \;\,     (  H''  , \theta , \overline{  \SYSTEMnt{id}  } )   =   \mathsf{copy}(  H'  )     \end{array}    \;\, & \;\,   \SYSTEMmv{y}  \# \{ H  ,  \SYSTEMnt{v}  ,  \SYSTEMnt{t} \}   \end{array} }%
}{
   H  ,  H'    \vdash  \SYSTEMkw{clone} \,  [ { \SYSTEMnt{v} } ]_{\color{coeffectColor}{ r } }  \, \textbf{as} \, \SYSTEMmv{x} \, \textbf{in} \, \SYSTEMnt{t_{{\mathrm{2}}}}  \,\leadsto_{ s } \,       H  ,  H'    ,  H''   ,   \SYSTEMmv{y}  \textcolor{coeffectColor}{\mapsto_{ s } }   \textbf{pack}\ \langle{ \overline{ \SYSTEMnt{id} } ,   \ast  \SYSTEMsym{(}   \theta( \SYSTEMnt{v} )   \SYSTEMsym{)}   }\rangle     \vdash   \SYSTEMnt{t_{{\mathrm{2}}}}  [  \SYSTEMmv{y}  /  \SYSTEMmv{x}  ]  }{%
{\SYSTEMdruleHeapcopyBetaName}{}%
}}

\newcommand{\SYSTEMdruleHeappushUniqueName}[0]{\SYSTEMdrulename{HeappushUnique}}
\newcommand{\SYSTEMdruleHeappushUnique}[1]{\SYSTEMdrule[#1]{%
}{
 H  \vdash  \SYSTEMkw{push} \, \SYSTEMsym{(}  \ast  \SYSTEMsym{(}  \SYSTEMnt{v_{{\mathrm{1}}}}  \SYSTEMsym{,}  \SYSTEMnt{v_{{\mathrm{2}}}}  \SYSTEMsym{)}  \SYSTEMsym{)}  \,\leadsto_{ s } \,  H  \vdash  \SYSTEMsym{(}  \ast  \SYSTEMnt{v_{{\mathrm{1}}}}  \SYSTEMsym{,}  \ast  \SYSTEMnt{v_{{\mathrm{2}}}}  \SYSTEMsym{)} }{%
{\SYSTEMdruleHeappushUniqueName}{}%
}}

\newcommand{\SYSTEMdruleHeappullUniqueName}[0]{\SYSTEMdrulename{HeappullUnique}}
\newcommand{\SYSTEMdruleHeappullUnique}[1]{\SYSTEMdrule[#1]{%
}{
 H  \vdash  \SYSTEMkw{pull} \, \SYSTEMsym{(}  \ast  \SYSTEMnt{v_{{\mathrm{1}}}}  \SYSTEMsym{,}  \ast  \SYSTEMnt{v_{{\mathrm{2}}}}  \SYSTEMsym{)}  \,\leadsto_{ s } \,  H  \vdash  \ast  \SYSTEMsym{(}  \SYSTEMnt{v_{{\mathrm{1}}}}  \SYSTEMsym{,}  \SYSTEMnt{v_{{\mathrm{2}}}}  \SYSTEMsym{)} }{%
{\SYSTEMdruleHeappullUniqueName}{}%
}}

\newcommand{\SYSTEMdruleHeapMultireflName}[0]{\SYSTEMdrulename{HeapMultirefl}}
\newcommand{\SYSTEMdruleHeapMultirefl}[1]{\SYSTEMdrule[#1]{%
}{
 H  \vdash  \SYSTEMnt{t} \ \Rightarrow_{ s } \  H  \vdash  \SYSTEMnt{t} }{%
{\SYSTEMdruleHeapMultireflName}{}%
}}

\newcommand{\SYSTEMdruleHeapMultiextName}[0]{\SYSTEMdrulename{HeapMultiext}}
\newcommand{\SYSTEMdruleHeapMultiext}[1]{\SYSTEMdrule[#1]{%
\SYSTEMpremise{ \begin{array}{cc}   H  \vdash  \SYSTEMnt{t_{{\mathrm{1}}}}  \,\leadsto_{ s } \,  H'  \vdash  \SYSTEMnt{t_{{\mathrm{2}}}}   \;\, & \;\,   H'  \vdash  \SYSTEMnt{t_{{\mathrm{2}}}} \ \Rightarrow_{ s } \  H''  \vdash  \SYSTEMnt{t_{{\mathrm{3}}}}   \end{array} }%
}{
 H  \vdash  \SYSTEMnt{t_{{\mathrm{1}}}} \ \Rightarrow_{ s } \  H''  \vdash  \SYSTEMnt{t_{{\mathrm{3}}}} }{%
{\SYSTEMdruleHeapMultiextName}{}%
}}

\newcommand{\SYSTEMdruleHeapCtxtCompatbaseName}[0]{\SYSTEMdrulename{HeapCtxtCompatbase}}
\newcommand{\SYSTEMdruleHeapCtxtCompatbase}[1]{\SYSTEMdrule[#1]{%
}{
  \emptyset   \bowtie   \emptyset  }{%
{\SYSTEMdruleHeapCtxtCompatbaseName}{}%
}}

\newcommand{\SYSTEMdruleHeapCtxtCompatgcArrName}[0]{\SYSTEMdrulename{HeapCtxtCompatgcArr}}
\newcommand{\SYSTEMdruleHeapCtxtCompatgcArr}[1]{\SYSTEMdrule[#1]{%
\SYSTEMpremise{ H  \bowtie   \emptyset  }%
}{
  H ,   \SYSTEMmv{ref}  \textcolor{uniqueColor}{\mapsto_{ p } }  \SYSTEMnt{id}    \bowtie   \emptyset  }{%
{\SYSTEMdruleHeapCtxtCompatgcArrName}{}%
}}

\newcommand{\SYSTEMdruleHeapCtxtCompatextResName}[0]{\SYSTEMdrulename{HeapCtxtCompatextRes}}
\newcommand{\SYSTEMdruleHeapCtxtCompatextRes}[1]{\SYSTEMdrule[#1]{%
\SYSTEMpremise{ \begin{array}{cc}    H ,   \SYSTEMnt{id}  \mapsto  v_r    \bowtie   \Gamma  +  \gamma    \;\, & \;\,   \gamma  \vdash   v_r   :   \SYSTEMnt{Res} _{ \SYSTEMnt{id} } \  \SYSTEMnt{A}    \end{array} }%
}{
   H ,   \SYSTEMmv{ref}  \textcolor{uniqueColor}{\mapsto_{ p } }  \SYSTEMnt{id}   ,   \SYSTEMnt{id}  \mapsto  v_r    \bowtie  \SYSTEMsym{(}   \Gamma ,   \SYSTEMmv{ref}  :   \SYSTEMnt{Res} _{ \SYSTEMnt{id} } \  \SYSTEMnt{A}     \SYSTEMsym{)} }{%
{\SYSTEMdruleHeapCtxtCompatextResName}{}%
}}

\newcommand{\SYSTEMdruleHeapCtxtCompatextGrName}[0]{\SYSTEMdrulename{HeapCtxtCompatextGr}}
\newcommand{\SYSTEMdruleHeapCtxtCompatextGr}[1]{\SYSTEMdrule[#1]{%
\SYSTEMpremise{ \begin{array}{cc}    \begin{array}{cc}    \begin{array}{cc}   H  \bowtie    \Gamma  +   s  \cdot  \Gamma'      \;\, & \;\,   \SYSTEMmv{x}  \not\in \mathsf{dom}( H )   \end{array}    \;\, & \;\,   \Gamma'  \vdash  \SYSTEMnt{v}  :  \SYSTEMnt{A}   \end{array}    \;\, & \;\,   \exists  r'  .\    s  +  r'   \equiv  r    \end{array} }%
}{
 \SYSTEMsym{(}   H ,   \SYSTEMmv{x}  \textcolor{coeffectColor}{\mapsto_{ r } }  \SYSTEMnt{v}    \SYSTEMsym{)}  \bowtie  \SYSTEMsym{(}   \Gamma ,   \textcolor{coeffectColor}{ \SYSTEMmv{x}  : [  \SYSTEMnt{A}  ]_{ s } }    \SYSTEMsym{)} }{%
{\SYSTEMdruleHeapCtxtCompatextGrName}{}%
}}

\newcommand{\SYSTEMdruleHeapCtxtCompatextLinName}[0]{\SYSTEMdrulename{HeapCtxtCompatextLin}}
\newcommand{\SYSTEMdruleHeapCtxtCompatextLin}[1]{\SYSTEMdrule[#1]{%
\SYSTEMpremise{ \begin{array}{cc}    \begin{array}{cc}    \begin{array}{cc}   H  \bowtie    \Gamma  +  \Gamma'     \;\, & \;\,   \SYSTEMmv{x}  \not\in \mathsf{dom}( H )   \end{array}    \;\, & \;\,   \Gamma'  \vdash  \SYSTEMnt{v}  :  \SYSTEMnt{A}   \end{array}    \;\, & \;\,   \exists  r'  .\    \SYSTEMsym{1}  +  r'   \equiv  r    \end{array} }%
}{
 \SYSTEMsym{(}   H ,   \SYSTEMmv{x}  \textcolor{coeffectColor}{\mapsto_{ r } }  \SYSTEMnt{v}    \SYSTEMsym{)}  \bowtie  \SYSTEMsym{(}   \Gamma ,   \SYSTEMmv{x}  :  \SYSTEMnt{A}    \SYSTEMsym{)} }{%
{\SYSTEMdruleHeapCtxtCompatextLinName}{}%
}}

\providecommand{\SYSTEMdruleTyabsName}{}
\providecommand{\SYSTEMdruleTyvarName}{}
\providecommand{\SYSTEMdruleTyappName}{}
\providecommand{\SYSTEMdruleTyprName}{}
\providecommand{\SYSTEMdruleTyelimName}{}
\providecommand{\SYSTEMdruleTyderName}{}

\providecommand{\SYSTEMdruleTyapproxName}{}
\providecommand{\SYSTEMdruleTypairIntroName}{}
\providecommand{\SYSTEMdruleTypairElimName}{}
\providecommand{\SYSTEMdruleTyunitIntroName}{}
\providecommand{\SYSTEMdruleTyunitElimName}{}

\providecommand{\SYSTEMdruleTyreturnGenName}{}

\providecommand{\SYSTEMdruleTybindGenFreshName}{}

\providecommand{\SYSTEMdruleTywithBorrowName}{}
\providecommand{\SYSTEMdruleTysplitName}{}
\providecommand{\SYSTEMdruleTyjoinName}{}

\providecommand{\SYSTEMdruleTypushName}{}
\providecommand{\SYSTEMdruleTypullName}{}

\providecommand{\SYSTEMdruleTypackName}{}
\providecommand{\SYSTEMdruleTyunpackName}{}

\providecommand{\SYSTEMdruleRTynecName}{}
\providecommand{\SYSTEMdruleRTyresName}{}
\providecommand{\SYSTEMdruleRTyresName}{}

\providecommand{\SYSTEMdruleRTyunborrowName}{}

\providecommand{\SYSTEMdruleHeapCtxtCompatbaseName}{}
\providecommand{\SYSTEMdruleHeapCtxtCompatextGrName}{}
\providecommand{\SYSTEMdruleHeapCtxtCompatextLinName}{}
\providecommand{\SYSTEMdruleHeapCtxtCompatextResName}{}
\providecommand{\SYSTEMdruleHeapCtxtCompatextResName}{}

\renewcommand{\SYSTEMdruleTyabsName}{$\textsc{abs}$}
\renewcommand{\SYSTEMdruleTyappName}{$\textsc{app}$}
\renewcommand{\SYSTEMdruleTyvarName}{$\textsc{var}$}
\renewcommand{\SYSTEMdruleTyprName}{$\textsc{pr}$}
\renewcommand{\SYSTEMdruleTyelimName}{$\textsc{elim}$}
\renewcommand{\SYSTEMdruleTyderName}{$\textsc{der}$}

\renewcommand{\SYSTEMdruleTyapproxName}{$\textsc{approx}$}
\renewcommand{\SYSTEMdruleTypairIntroName}{$\otimes_I$}
\renewcommand{\SYSTEMdruleTypairElimName}{$\otimes_E$}
\renewcommand{\SYSTEMdruleTyunitIntroName}{$1_I$}
\renewcommand{\SYSTEMdruleTyunitElimName}{$1_E$}

\renewcommand{\SYSTEMdruleTybindGenFreshName}{$\textsc{clone}$}

\renewcommand{\SYSTEMdruleTyreturnGenName}{$\textsc{share}$}

\renewcommand{\SYSTEMdruleTywithBorrowName}{$\textsc{with}\with$}
\renewcommand{\SYSTEMdruleTysplitName}{$\textsc{split}$}
\renewcommand{\SYSTEMdruleTyjoinName}{$\textsc{join}$}

\renewcommand{\SYSTEMdruleTypushName}{$\textsc{push}$}
\renewcommand{\SYSTEMdruleTypullName}{$\textsc{pull}$}

\renewcommand{\SYSTEMdruleTypackName}{$\textsc{pack}$}
\renewcommand{\SYSTEMdruleTyunpackName}{$\textsc{unpack}$}

\renewcommand{\SYSTEMdruleRTynecName}{$\textsc{nec}$}
\renewcommand{\SYSTEMdruleRTyresName}{$\textsc{ref}$}

\renewcommand{\SYSTEMdruleRTyunborrowName}{$\textsc{unborrow}$}

\renewcommand{\SYSTEMdruleHeapMultireflName}{$\textsc{refl}$}
\renewcommand{\SYSTEMdruleHeapMultiextName}{$\textsc{ext}$}

\renewcommand{\SYSTEMdruleHeapvarName}{$\leadsto_{{\textsc{var}}}$}
\renewcommand{\SYSTEMdruleHeapbetaName}{$\leadsto_\beta$}
\renewcommand{\SYSTEMdruleHeapappLName}{$\leadsto_{\textsc{appL}}$}
\renewcommand{\SYSTEMdruleHeapappRName}{$\leadsto_{\textsc{appR}}$}
\renewcommand{\SYSTEMdruleHeapbetaBoxName}{$\leadsto_{\square\beta}$}
\renewcommand{\SYSTEMdruleHeapcongBoxElimName}{$\leadsto_{\textsc{let}\square}$}
\renewcommand{\SYSTEMdruleHeapcongPromotionName}{$\leadsto_{\square}$}
\renewcommand{\SYSTEMdruleHeapcongPairElimName}{$\leadsto_{\textsc{let}\otimes}$}
\renewcommand{\SYSTEMdruleHeapcongPairLName}{$\leadsto_{\otimes\textsc{L}}$}
\renewcommand{\SYSTEMdruleHeapcongPairRName}{$\leadsto_{\otimes\textsc{R}}$}
\renewcommand{\SYSTEMdruleHeappairBetaName}{$\leadsto_{\otimes\beta}$}

\renewcommand{\SYSTEMdruleHeapcongShareName}{$\leadsto_\textsc{share}$}
\renewcommand{\SYSTEMdruleHeapshareName}{$\leadsto_{\textsc{share}\beta}$}

\renewcommand{\SYSTEMdruleHeapsplitRefName}{$\leadsto_{\textsc{split}\textsc{Ref}}$}

\renewcommand{\SYSTEMdruleHeapsplitPairName}{$\leadsto_{\textsc{split}\otimes}$}

\renewcommand{\SYSTEMdruleHeapjoinRefName}{$\leadsto_{\textsc{join}\textsc{Ref}}$}

\renewcommand{\SYSTEMdruleHeapjoinPairName}{$\leadsto_{\textsc{join}\otimes}$}

\renewcommand{\SYSTEMdruleHeapwithBorrowName}{$\leadsto_{\textsc{with}\with}$}
\renewcommand{\SYSTEMdruleHeapunborrowBorrowName}{$\leadsto_{\textsc{un}\with}$}
\renewcommand{\SYSTEMdruleHeapcongUnborrowName}{$\leadsto_\textsc{unborrow}$}

\renewcommand{\SYSTEMdruleHeapnewArrayName}{$\leadsto_{\textsc{newArray}}$}
\renewcommand{\SYSTEMdruleHeapreadArrayName}{$\leadsto_{\textsc{readArray}}$}
\renewcommand{\SYSTEMdruleHeapwriteArrayName}{$\leadsto_{\textsc{writeArray}}$}
\renewcommand{\SYSTEMdruleHeapdeleteArrayName}{$\leadsto_{\textsc{deleteArray}}$}

\renewcommand{\SYSTEMdruleHeapcopyBetaName}{$\leadsto_{\textsc{clone}\beta}$}

\renewcommand{\SYSTEMdruleHeappushUniqueName}{$\leadsto_{\textsc{push}\ast}$}

\renewcommand{\SYSTEMdruleHeappullUniqueName}{$\leadsto_{\textsc{pull}\ast}$}

\renewcommand{\SYSTEMdruleHeapexistentialBetaName}{$\leadsto_{\exists\beta}$}

\renewcommand{\SYSTEMdruleHeapCtxtCompatbaseName}{\textsc{empty}}
\renewcommand{\SYSTEMdruleHeapCtxtCompatextGrName}{\textsc{ext}}
\renewcommand{\SYSTEMdruleHeapCtxtCompatextLinName}{\textsc{extLin}}
\renewcommand{\SYSTEMdruleHeapCtxtCompatextResName}{\textsc{extArrRef}}
\renewcommand{\SYSTEMdruleHeapCtxtCompatextResName}{\textsc{extRes}}

\providecommand{\SYSTEMdruleHeapCtxtCompatgcArrName}{}
\renewcommand{\SYSTEMdruleHeapCtxtCompatgcArrName}{\textsc{GCArr}}

\renewcommand{\SYSTEMdrulename}[1]{\!\!\!\textsc{#1}}

\usepackage{mathpartir}
\usepackage{cmll}
\usepackage{enumitem}
\usepackage{resizegather}

\usepackage{listings}
\usepackage{pifont}% http://ctan.org/pkg/pifont
\newcommand{\xmark}{\ding{55}}%

\lstdefinelanguage{Granule}{%
  mathescape=true,
  morecomment=[l]{--},
  moredelim=[s][\itshape]{`}{`},
  showspaces=false,
  xleftmargin=2.5em,
  commentstyle=\itshape\color{black!60},
  basicstyle=\ttfamily\footnotesize,%\sffamily\small,%
  flexiblecolumns=true,
  columns=[l]flexible,
  columns=fullflexible,
  keepspaces=true,
  literate=%
  {<}{\textcolor{effectColor}{<}}1
  {>}{\textcolor{effectColor}{>}}1
  {[}{\textcolor{coeffectColor}{[}}1
  {]}{\textcolor{coeffectColor}{]}}1
  {[r' : R']}{[{\textcolor{coeffectColor}{r' : R'}}]}1
  {[([}{\textcolor{coeffectColor}{[[}}2 
  {])]}{\textcolor{coeffectColor}{]]}}2
  {!a}{\textcolor{coeffectColor}{!a}}1
  {!b}{\textcolor{coeffectColor}{!b}}1
  {!Colour}{\textcolor{coeffectColor}{!Colour}}1
  {!Int}{\textcolor{coeffectColor}{!Int}}1
  {BpColour}{\textcolor{borrowColor}{\& p Colour}}1
  {Bp(MaybeColour)}{\textcolor{borrowColor}{\& p (Maybe Colour)}}1
  {Bp(FloatArrayid)}{\textcolor{borrowColor}{\& p (FloatArray id)}}1
  {B1Int}{\textcolor{borrowColor}{\& 1 Int}}1
  {B1(Refida)}{\textcolor{borrowColor}{\& 1 (Ref id a)}}1
  {*Coffee}{\textcolor{uniqueColor}{*Coffee}}1
  {*Colour}{\textcolor{uniqueColor}{*Colour}}1
  {*a}{\textcolor{uniqueColor}{*a}}1
  {*FloatArray}{\textcolor{uniqueColor}{*FloatArray}}1
  {*(FloatArrayid)}{\textcolor{uniqueColor}{*(FloatArray id)}}1
  {*(Refid'Float)}{\textcolor{uniqueColor}{*(Ref id' Float)}}1
  {*(Refid'Float,FloatArrayid)}{\textcolor{uniqueColor}{*(Ref id' Float, FloatArray id)}}1
  {forall}{$\forall$}1
  {exists}{$\exists$}1
  {Inf}{$\infty$}1
  {->}{$\rightarrow$}1
  {-o}{$\multimap$}1
  {=>}{$\Rightarrow$}1
  {<-}{\textcolor{effectColor}{$\leftarrow$}}1
  {/\\}{$\sqcap$}1
  {\\/}{$\sqcup$}1
  {<=}{$\leqslant$}1
  {>=}{$\geqslant$}1
  {\\}{$\lambda$}1
  {_1}{$\mathtt{_1}$}1
  {_2}{$\mathtt{_2}$}1
  {_3}{$\mathtt{_3}$}1
  {_4}{$\mathtt{_4}$}1
  {_L}{$\mathtt{_{L}}$}1
  {_LH}{$\mathtt{_{LH}}$}1
  {_Gr}{$\mathtt{_{Gr}}$}1
  {_p}{$\mathtt{_{p}}$}1
  {_q}{$\mathtt{_{q}}$}1
  {-o}{$\multimap$}1
  {\\times}{$\times$}1
  {--BLANK}{}1,
  keywordstyle = \color{blue!40!black},%\color{WildStrawberry!90!Black},
  keywords = {data, type, let, in, case, of, if, then, else, where, import, Type, Semiring, Protocol, Permission, Name, Fraction, Droppable, Nat},
  keywordstyle = [4]\bfseries\color{effectColor},
  keywords     = [4]{pure},%{IO,
  numbers=left,
  numberstyle=\tiny\color{gray}
}

\lstnewenvironment{granule}
{\lstset{language=Granule}}{}
\lstnewenvironment{granuleInterface}
{\lstset{language=Granule,numbers=none}}{}

\newcommand{\granin}[1]{\lstinline[language=Granule]{#1}}

\definecolor{multiplicity}{rgb}{0,0.3,0.08}

\lstdefinelanguage{Haskell}{%
  mathescape=true,
  morecomment=[l]{--},
  comment=[l]{\{-},
  moredelim=[s][\itshape]{`}{`},
  showspaces=false,
  commentstyle=\itshape\color{black!60},
  basicstyle=\ttfamily\footnotesize,%\sffamily\small,%
  flexiblecolumns=true,
  columns=[l]flexible,
  columns=fullflexible,
  keepspaces=true,
  xleftmargin=2.5em,
  literate=%
   {\%r}{\textcolor{multiplicity}{\%r}}1
   {\%'Many}{\textcolor{multiplicity}{\%'Many}}1
   {\%1}{\textcolor{multiplicity}{\%1}}1,
  keywordstyle = \color{blue!40!black},%\color{WildStrawberry!90!Black},
  keywords = {data, type, let, in, case, of, if, then, else, where,
  import, class, instance},
  numbers=left,
  numberstyle=\tiny\color{gray}
}

\newcommand{\haskin}[1]{\lstinline[language=Haskell]{#1}}

\usepackage[most]{tcolorbox}

\makeatletter
\newtcblisting[auto counter]{granulep}[1][]{%
  enhanced jigsaw,
  colframe=black!30,
  boxrule=0.5pt,
  toprule=0.5pt,
  bottomrule=0.5pt,
  top=0em,
  left=0em,
  bottom=0em,
  boxsep=0em,
  after skip=1em,
  before skip=1em,
  listing options={language={granule}},
  listing only,
  colback=white,
  sharp corners,
  overlay={%\node[inner sep=4pt,xshift=-2cm,fill=white] (A) at (frame.south east) {Granule};
           \node[inner sep=2pt,xshift=-2cm,fill=white] (B) at (frame.south east) {Granule};
  },
  #1,
}
\makeatother

\makeatletter
\newtcblisting[auto counter]{granulepill}[1][]{%
  enhanced jigsaw,
  colframe=black!30,
  boxrule=0.5pt,
  toprule=0.5pt,
  bottomrule=0.5pt,
  top=0em,
  left=0em,
  bottom=0em,
  boxsep=0em,
  after skip=1em,
  before skip=1em,
  listing options={language={granule}},
  listing only,
  colback=white,
  sharp corners,
  overlay={%\node[inner sep=4pt,xshift=-2cm,fill=white] (A) at (frame.south east) {Granule};
           \node[inner sep=2pt,xshift=-2cm,fill=white] (B) at (frame.south east) {Granule \textcolor{uniqueColor}{\xmark}};
  },
  #1,
}
\makeatother

\definecolor{GrayCodeBlock}{RGB}{241,241,241}
\definecolor{BlackText}{RGB}{110,107,94}
\definecolor{RedTypename}{RGB}{182,86,17}
\definecolor{GreenString}{RGB}{96,172,57}
\definecolor{PurpleKeyword}{RGB}{184,84,212}
\definecolor{GrayComment}{RGB}{170,170,170}
\definecolor{GoldDocumentation}{RGB}{180,165,45}
\lstdefinelanguage{rust}
{
    mathescape=true,
    morecomment=[l]{--},
    comment=[l]{\{-},
    moredelim=[s][\itshape]{`}{`},
    showspaces=false,
    commentstyle=\itshape\color{black!60},
    basicstyle=\ttfamily\footnotesize,%\sffamily\small,%
    flexiblecolumns=true,
    columns=[l]flexible,
    columns=fullflexible,
    keepspaces=true,
    xleftmargin=2.5em,
    basicstyle=\ttfamily\footnotesize,%\sffamily\small,%
    keywords={
        true,false,
        unsafe,async,await,move,
        use,pub,crate,super,self,mod,
        struct,enum,fn,const,static,let,mut,ref,type,impl,dyn,trait,where,as,
        break,continue,if,else,while,for,loop,match,return,yield,in
    },
    keywordstyle=\color{PurpleKeyword},
    ndkeywords={
        bool,u8,u16,u32,u64,u128,i8,i16,i32,i64,i128,char,str,
        Self,Option,Some,None,Result,Ok,Err,String,Box,Vec,Rc,Arc,Cell,RefCell,HashMap,BTreeMap,
        macro_rules
    },
    ndkeywordstyle=\color{RedTypename},
    comment=[l][\color{GrayComment}\slshape]{//},
    morecomment=[s][\color{GrayComment}\slshape]{/*}{*/},
    morecomment=[l][\color{GoldDocumentation}\slshape]{///},
    morecomment=[s][\color{GoldDocumentation}\slshape]{/*!}{*/},
    morecomment=[l][\color{GoldDocumentation}\slshape]{//!},
    morecomment=[s][\color{RedTypename}]{\#![}{]},
    morecomment=[s][\color{RedTypename}]{\#[}{]},
    stringstyle=\color{GreenString},
    string=[b]",
    numbers=left,
    numberstyle=\tiny\color{gray}
}

\lstnewenvironment{rust}
{\lstset{language=Rust}}{}
\newcommand{\rustin}[1]{\lstinline[language=Rust]{#1}}

\newtcblisting[auto counter]{rustp}[1][]{%
  enhanced jigsaw,
  colframe=black!30,
  boxrule=0.5pt,
  toprule=0.5pt,
  top=0em,
  left=0em,
  bottom=0em,
  bottomrule=0.5pt,
  boxsep=0em,
  after skip=1em,
  before skip=1em,
  listing options={language={rust}},
  listing only,
  colback=white,
  sharp corners,
  overlay={\node[inner sep=4pt,xshift=-2cm,fill=white] (A) at (frame.south east) {Rust \textcolor{borrowColor}{\checkmark}};
  },
  #1,
}
\makeatother

\newtcblisting[auto counter]{rustpill}[1][]{%
  enhanced jigsaw,
  colframe=black!30,
  boxrule=0.5pt,
  toprule=0.5pt,
  top=0em,
  left=0em,
  bottom=0em,
  bottomrule=0.5pt,
  boxsep=0em,
  after skip=1em,
  before skip=1em,
  listing options={language={rust}},
  listing only,
  colback=white,
  sharp corners,
  overlay={\node[inner sep=4pt,xshift=-2cm,fill=white] (A) at (frame.south east) {Rust \textcolor{uniqueColor}{\xmark}};
  },
  #1,
}
\makeatother
\definecolor{coeffectColor}{HTML}{0750d0}
\definecolor{uniqueColor}{HTML}{c20232}
\definecolor{highlightPropColor}{rgb}{0,0.35,0}
\definecolor{darkGreen}{rgb}{0,0.5,0.35}

\title{Functional Ownership through Fractional Uniqueness}

\begin{abstract}
Ownership and borrowing systems, designed to enforce safe memory management
without the need for garbage collection, have been brought to the fore by the
Rust programming language. Rust also aims to bring some guarantees offered by
functional programming into the realm of performant systems code, but the type
system is largely separate from the ownership model, with type and borrow
checking happening in separate compilation phases. Recent models such as
RustBelt and Oxide aim to formalise Rust in depth, but there is less focus on
integrating the basic ideas into more traditional type systems. An approach
designed to expose an essential core for ownership and borrowing would open the
door for functional languages to borrow concepts found in Rust and other
ownership frameworks, so that more programmers can enjoy their benefits.

One strategy for managing memory in a functional setting is through
\emph{uniqueness types}, but these offer a coarse-grained view: either a value
has exactly one reference, and can be mutated safely, or it cannot, since other
references may exist. Recent work demonstrates that \emph{linear} and
\emph{uniqueness} types can be combined in a single system to offer restrictions
on program behaviour and guarantees about memory usage. We develop this
connection further, showing that just as \emph{graded} type systems like those
of Granule and Idris generalise linearity, Rust's \emph{ownership} model arises
as a graded generalisation of uniqueness. We combine fractional permissions with
grading to give the first account of ownership and borrowing that smoothly
integrates into a standard type system alongside linearity and graded types, and
extend Granule accordingly with these ideas.
\end{abstract}

\begin{document}
\maketitle

\section{Introduction}
\label{sec:introduction}
The Rust programming language has dramatically grown in popularity in recent
years, having been adopted as the second official language of the Linux
kernel,\footnote{\url{https://www.zdnet.com/article/rust-takes-a-major-step-forward-as-linuxs-second-official-language/}}
and deployed in production code by companies including AWS, Huawei, Google,
Microsoft and Mozilla, all of whom are founding members of the Rust
Foundation.\footnote{\url{https://foundation.rust-lang.org/news/2021-02-08-hello-world/}}
This popularity is in large part due to its focus on memory safety; Rust finds a happy
medium between systems programming languages like C which offer precise control
but little in the way of safety guarantees and the contrasting approach of
higher-level languages like Java or Go where memory is managed automatically
through garbage collection.

The intricate ownership system which characterises Rust's approach to memory
management is inspired in part by the literature on using \emph{linear types}~\cite{wadler1990linear} for tracking
resource usage, based on \emph{linear
logic} \cite{girard1987linear}.
Linear types have also been brought into the realm of practical programming in
recent years, for example in Haskell via the GHC compiler's linear types extension~\cite{bernardy2017linear}. The precise theoretical relationship between
 linear types and the properties that can be enforced by Rust's borrow
checker remains unclear, however, particularly since Rust lies outside the
traditional functional paradigm to which linearity is most suited, opting for a
more mixed imperative and functional approach.

Linear types require that every value is used exactly once, with
the $!$ modality classifying non-linearity (arbitrary use).
Modern resourceful type systems go beyond this coarse
restriction of linearity, enabling usage to be classified more finely.
Quantitative types, inspired by bounded linear logic~\cite{girard1992bounded},
capture an upper bound on the amount of (re)use of
a value by indexing the $!$ modality with natural numbers or polynomial terms. This quantitative analysis is generalised further by \emph{graded modal types}, as exemplified
by the Granule programming language~\cite{orchard2019quantitative}, which allows
various properties of data use and data flow to be tracked smoothly in a single type system via an indexed modality $!_r$ with some algebraic structure on $r$~\cite{gaboardi2016combining,orchard2019quantitative}. In this work, we seek a shared understanding between the precise
resource reasoning of graded type systems and concepts of uniqueness, borrowing, and ownership as typified by Rust but also appearing in various other languages such as C++.

Many models of ownership from a type-theoretic perspective already exist. For
example, $\lambda_{\text{Rust}}$ developed as part of the RustBelt framework
provides a low-level model of Rust's ownership and borrowing systems suitable
for formally verifying properties of Rust programs~\cite{jung2017rustbelt}.
Oxide~\cite{weiss2019oxide} and FR~\cite{pearce2021lightweight} provide
more high-level theoretical models. Our focus is \emph{not} on
replicating these efforts; we do not aim to directly
model all the details of Rust's particular approach to ownership.
Rather, we offer a
different and complementary perspective: a single system that relates
general ownership and borrowing ideas to linear
and graded types. This integration enables
 type, resource, and memory safety to be enforced via a single type system.

A natural place to start is \emph{uniqueness types}, which
provide the concept that a uniquely typed value has exactly one reference, and
therefore is safe to mutate~\cite{clean, barendsen1996uniqueness,
simplified}. This gives a simplistic model of \emph{ownership} without
borrowing---a resource with a unique owner may be modified by that owner, but a
broken guarantee of uniqueness can never be recovered. Recent work
developed a type system that captures the relationship between linear and
unique types: linear values are restricted from being copied or discarded in the
future, whilst unique values are guaranteed to have never been copied in the
past~\cite{marshall2022linearity}. We build upon this work as our foundation here
but leverage the generalisations of grading.
In the ownership framework of \citet{mycroft}, while uniqueness allows
for memory-safe \emph{temporal} aliasing, borrowing extends uniqueness
by introducing the potential for safe patterns of \emph{spatial}
aliasing, dramatically increasing flexibility for the programmer.
Just as the non-linearity modality $!$ can be generalised to a graded
modality $!_r$ for fine-grained reasoning about resource usage via $r$
and its algebraic structure, we show that uniqueness types,
represented modally, can be `graded' to capture this idea that
borrowing is a controlled relaxation of the uniqueness guarantee.

Our approach allows many ideas from Rust, such as immutable and mutable borrows,
partial borrows and reborrowing to be explained explicitly in a functional setting, all
through the application of an elegant form of grading based on Boyland's
fractional permissions~\cite{boyland2003checking}. Our work thus integrates and
relates within a unified framework the substructural type systems of linearity,
uniqueness, grading, ownership, and borrowing. This offers a pathway to
expressing memory safety properties naturally in functional
languages such as Haskell or Idris.

The outline of the paper is as follows. First, in Section~\ref{sec:key} we recap
the key concepts of uniqueness and borrowing using Rust as a convenient
exemplar. We discuss how various programs making use of these ideas will later
be rewritten as Granule programs through our unified type system, in order to
motivate the rest of our work. We then recap Granule's pre-existing core
calculus in Section~\ref{sec:core}, before moving on to the primary
contributions of this paper:
\begin{itemize}

  \item In Section~\ref{sec:uniqueness}, we connect Granule's unique and graded modal
  types with ideas from Rust, discussing how unique ownership allows for safe
  mutation. We generalise the connection between uniqueness and linearity,
  demonstrating that uniqueness and precise grading can coexist within a single
  type system. We leverage existential types over `identifiers' for situations where
  multiple references pointing to the same value need tracking.
  \item In Section~\ref{sec:fractional}, we extend this idea to allow for
  multiple immutably borrowed references at a time, by carefully tracking
  references that exist, similarly to Boyland's fractional permissions. We show
  that this allows for a more fine-grained approach to tracking uniqueness in
  much the same way that grading increases expressivity over pure linearity.
  \begin{itemize}
  \item In Section~\ref{sec:equational}, we develop an equational theory for our
  extension to Granule's type system, showing that the modality for unique
  ownership induces a \emph{relative functor} over the new modality for mutable
  and immutable borrowing.
  \item In Section~\ref{sec:distrib}, we discuss how using distributive laws can
  allow for borrowing only part of a larger data structure while leaving the
  rest uniquely owned, enabling a much wider variety of practical programming
  patterns.
  \end{itemize}
  \item In Section~\ref{sec:semantics}, we detail the semantics for the calculus
   developed thus far, and prove various key properties---both standard
  notions like progress and preservation but also borrow safety and uniqueness.
  The semantics is call-by-value, representing a practical system,
  departing from most previous operational models of grading in the literature.
\end{itemize}

Sections~\ref{sec:related} and \ref{sec:conclusion} discuss the many and varied
areas of similar research that have inspired this paper, and look at some
possible avenues for future work. All source code discussed will be made
available in the artefact, along with the implementation of our type system atop
Granule.\footnote{If the reader wishes to experiment with the Granule language,
the latest releases are available from
\url{https://github.com/granule-project/granule/releases}.}

\section{Key Concepts in Ownership and Borrowing}
\label{sec:key}
In order to develop a type system which integrates ownership and borrowing concepts with the linear and graded types already present in Granule, we will
first need to understand the ideas in question. This section presents six key
patterns which we capture in our system in the rest of this paper,
along with simple Rust code examples to demonstrate the patterns in action. Each
of these examples relies on a single base value of type \rustin{Colour(u32, u32,
u32)} - a struct containing three unsigned integers, representing a colour with
red, green and blue components.

\paragraph{\textbf{Ownership}}

The first crucial concept is \emph{owned} values, where a value is `owned' by a
particular \emph{identifier}. Below, the value \rustin{Colour(220, 20, 60)} is
owned by the identifier \rustin{scarlet}. Each value can only be owned by a
single identifier at any given time. Rust enforces this via \emph{move
semantics}: on the second line, ownership of the value is \emph{moved} to the
identifier \rustin{x}. Now the identifier \rustin{scarlet} no longer owns the
value, so attempting to use it again on the third line gives an error.
\begin{rustpill}
let scarlet = Colour(220, 20, 60);
let x = scarlet;
let y = scarlet; // error: use of moved value: `scarlet`
\end{rustpill}
It is fairly clear that the idea here in some way relates to linearity, since
linear values can only be used once which restricts them to being passed around
sequentially; indeed, much of the literature on Rust makes mention of linear (or
affine) types. We demonstrate through our unified type system in
Chapter~\ref{sec:uniqueness} that in fact ownership can be better understood as
an extension of \emph{uniqueness} types, since each value having a single owner
implies that the owner holds the unique reference to said value. These are
similar to linear types in some ways but with important differences.

\paragraph{\textbf{Immutable borrowing}}

Borrowing generalises the concept of ownership, permitting multiple
references to point to a single value simultaneously. Rust's \emph{borrow
checker} manages all borrows that exist at a given time, ensuring that
memory safety properties are maintained until
values are eventually returned to their owners. \emph{Immutable} borrows are one
flavour of reference; any number of these can exist at a given time, but
mutation of a value is disallowed through an immutable borrow,
since this could result in data races. Below, two immutable
borrows (\rustin{x} and \rustin{y}) both reference the original
\rustin{persimmon} value.
\begin{rustp}
let persimmon = Colour(252, 118, 5);
let x = &persimmon; // & denotes a borrow of `persimmon`
let y = &persimmon; // ok!
\end{rustp}

\paragraph{\textbf{Mutable borrowing}}

It is also possible to borrow values and retain the capacity to mutate them, but
this comes at a price: as mentioned above, allowing mutation through multiple
references simultaneously is harmful to memory safety, so in order to prevent
this only a single mutable borrow is permitted at a time. Much like owned
values, mutably borrowing a value guarantees the sole capability for destructive
access to the underlying data. This is demonstrated below, where attempting to
create two \emph{mutable} borrows following the pattern of the above example is
disallowed.
\begin{rustpill}
let mut viridian = Colour(52, 161, 128);
let x = &mut viridian; // error: cannot borrow `viridian` as mutable
let y = &mut viridian; //                   more than once at a time
\end{rustpill}
Note, however, that the above example copied verbatim into a Rust file will in
fact compile successfully unless additional code is introduced that makes use of
the variables \rustin{x} and \rustin{y}; this is due to a feature called
\emph{non-lexical lifetimes}, through which the Rust compiler is able to
\emph{infer} that allowing two mutable borrows is safe as long as one of them is
never used. We do not linger on this, since we will not capture
this behaviour in our type system; our goal is to embed some essential ownership
and borrowing patterns \emph{explicitly}. However, we will
mention the possibility for extending our system with more advanced ideas in
Section~\ref{sec:conclusion}.

\paragraph{\textbf{Mixing mutable and immutable borrows}}

Rust also disallows immutable borrows of values that are already
borrowed mutably, since this invites similar problems: e.g., a value could be updated
through the mutable borrow whilst being read through the immutable borrow. Hence, the following example wherein the second reference is
borrowed immutably rather than mutably is also forbidden by the borrow checker.
In Section~\ref{sec:fractional}, we demonstrate that both notions
can be represented through a graded generalisation of uniqueness typing
to represent a borrower's level of access.

\begin{rustpill}
let mut cerulean = Colour(3, 109, 230);
let x = &mut cerulean; // error: cannot borrow `cerulean` as immutable
let y = &cerulean;     //       because it is also borrowed as mutable
\end{rustpill}

\paragraph{\textbf{Partial borrowing}}

A useful pattern for working with larger data structures is to
borrow only \emph{part} of the structure while the original owner retains access
to what remains. This is valuable for allowing one part of a program to work
with a particular piece of data without restricting access to the entire
structure, enabling tasks to be carried out in parallel more easily. In the below
example, the red and green components of the \rustin{indigo} value
are simultaneously borrowed mutably. The Rust compiler allows this since the
references point to disjoint parts of the original struct:
\begin{rustp}
let mut indigo = Colour(32, 36, 209);
let r = &mut indigo.0;
let g = &mut indigo.1; // ok!
\end{rustp}
In Section~\ref{sec:distrib}, we take advantage of Granule's
functional nature to present a cohesive way of managing this kind of partial
borrow at the type level, including the possibility for working with disjoint
components of the original structure concurrently.

\paragraph{\textbf{Reborrowing}}

The final pattern we aim to capture in our system, also essential for practical
programming, is the notion of \emph{reborrowing}, through which it is possible
to create a borrow of an identifier that itself references another value. We
will show in Section~\ref{sec:fractional} that this behaviour falls out of our
generalisation of uniqueness naturally, without a need for additional
constructs. To illustrate the general idea, consider the following example;
here, the red component of the \rustin{amethyst} value is borrowed mutably as
\rustin{r}, then another immutable borrow \rustin{x} is created pointing to
\rustin{r}. The value can now only be read through \rustin{x}---both \rustin{r}
and \rustin{amethyst} are inaccessible until the borrow is complete.

\begin{rustp}
let mut amethyst = Colour(98, 1, 170);
let r = &mut amethyst.0;
let x = & *r; // ok!
\end{rustp}

\section{Core Calculus}
\label{sec:core}

\newcommand{\coeff}[1]{\textcolor{coeffectColor}{#1}}

We recap Granule's core calculus which incorporates graded modal types into a
linear type theory~\cite{orchard2019quantitative}; we later extend this calculus
with our functional model of uniqueness and borrowing. This section covers the
syntax and static semantics (type system), whilst Section~\ref{sec:semantics}
gives an operational semantics incorporating the various extensions discussed
throughout this work.

The calculus extends the linear $\lambda$-calculus with multiplicative products
and unit, and a \emph{semiring-graded necessity modality} $\Box_{\coeff{r}} A$ where $r$ is an element of a pre-ordered semiring $(\coeff{\mathcal{R}},
\coeff{*}, \coeff{1}, \coeff{+}, \coeff{0}, \coeff{\sqsubseteq})$ which includes
a requirement that $\coeff{*}$ and $\coeff{+}$ must be monotonic with respect to
the ordering $\coeff{\sqsubseteq}$. This calculus gives a simplified monomorphic
subset of Granule~\cite{orchard2019quantitative}, and closely resembles other
graded systems from the literature~\cite{brunel2014core, petricek2014coeffects,
gaboardi2016combining, mcbride2016got, atkey2018syntax,
DBLP:journals/pacmpl/AbelB20, choudhury2021graded,  moon2021graded,
10.1145/3622843}. Our extensions for ownership and borrowing could be made
compatible with such systems, as they do not rely on particular details of the
core calculus in question.

Beyond existing work, we add existential types and type variables (restricted to
a particular kind) to track identifiers associated with resources (whose
dataflow paths may fork and join, due to the borrowing patterns we discuss
later). These identifiers explicitly name and identify a resource at the type
level such that references to different resources cannot be interchanged or
joined as they will have different types, distinguished by the identifier. This
will be illustrated in depth in Section~\ref{sec:fractional} when we discuss
particular examples of resources that can be managed by ownership.

\subsection{Syntax}

Our syntax consists of the linear $\lambda$-calculus with multiplicative products
and unit (first line of syntax below), graded modal terms (second line) and existentially quantified
identifiers (third line):
\begin{align*}
  \SYSTEMnt{t} ::= \; & \SYSTEMmv{x} \mid \lambda  \SYSTEMmv{x}  .  \SYSTEMnt{t} \mid \SYSTEMnt{t_{{\mathrm{1}}}} \, \SYSTEMnt{t_{{\mathrm{2}}}} \mid \SYSTEMsym{(}  \SYSTEMnt{t_{{\mathrm{1}}}}  \SYSTEMsym{,}  \SYSTEMnt{t_{{\mathrm{2}}}}  \SYSTEMsym{)} \mid \textbf{let} \, \SYSTEMsym{(x,}  \SYSTEMsym{y)}  \SYSTEMsym{=}  \SYSTEMnt{t_{{\mathrm{1}}}} \, \textbf{in} \, \SYSTEMnt{t_{{\mathrm{2}}}} \mid () \mid \textbf{let}\ () =  \SYSTEMnt{t_{{\mathrm{1}}}} \ \textbf{in}\  \SYSTEMnt{t_{{\mathrm{2}}}} \\
  \mid \; & \SYSTEMsym{[}  \SYSTEMnt{t}  \SYSTEMsym{]} \mid \textbf{let} \, \SYSTEMsym{[}  \SYSTEMmv{x}  \SYSTEMsym{]}  \SYSTEMsym{=}  \SYSTEMnt{t_{{\mathrm{1}}}} \, \textbf{in} \, \SYSTEMnt{t_{{\mathrm{2}}}} \\
  \mid \; & \textbf{pack}\ \langle{  \SYSTEMnt{id}  ,  \SYSTEMnt{t}  }\rangle \mid \textbf{unpack}\ \langle{  \SYSTEMnt{id}  ,  \SYSTEMmv{x}  }\rangle
     =  \SYSTEMnt{t_{{\mathrm{1}}}} \ \textbf{in}\  \SYSTEMnt{t_{{\mathrm{2}}}}
\tag{terms}
\end{align*}
Following the syntax of variables, we group terms above into pairs of
introduction and elimination forms, for functions, products, units, the
graded modality, and existential types respectively. The meaning of these terms is explained in the
next subsection with reference to their typing.

\subsection{Type system}

Typing judgments have the form $\Gamma  \vdash  \SYSTEMnt{t}  :  \SYSTEMnt{A}$, assigning type $A$ to term
$t$ under context $\Gamma$. Types are:
\begin{align*}
  A, B ::= \; & \SYSTEMnt{A}  \multimap  \SYSTEMnt{B} \mid \SYSTEMnt{A}  \otimes  \SYSTEMnt{B} \mid \mathsf{unit} \mid {\textcolor{coeffectColor}{\Box_{ r } } }  \SYSTEMnt{A}
  \mid \exists  \SYSTEMnt{id}  .  \SYSTEMnt{A}
\tag{types}
\end{align*}
Hence, our type syntax comprises linear function types $\SYSTEMnt{A}  \multimap  \SYSTEMnt{B}$, linear
multiplicative products $\SYSTEMnt{A}  \otimes  \SYSTEMnt{B}$, a linear multiplicative unit $(\mathsf{unit})$,
the graded modality ${\textcolor{coeffectColor}{\Box_{ r } } }  \SYSTEMnt{A}$ where $\coeff{r} \in \coeff{\mathcal{R}}$, and existentially quantified types where
$\SYSTEMnt{id} : \mathsf{Name}$ for an abstract kind of names $\mathsf{Name}$.

Contexts $\Gamma$ contain both linear assumptions $x : A$, graded assumptions
$\textcolor{coeffectColor}{ \SYSTEMmv{x}  : [  \SYSTEMnt{A}  ]_{ r } }$ which have originated from inside a graded modality, and type
variables which we write as $\SYSTEMnt{id}$ due to their restricted purpose
here, omitting the kind which is the abstract type $\mathsf{Name}$:
\begin{align*}
  \Gamma ::= \; \emptyset \mid \Gamma ,   \SYSTEMmv{x}  :  \SYSTEMnt{A} \mid \Gamma ,   \textcolor{coeffectColor}{ \SYSTEMmv{x}  : [  \SYSTEMnt{A}  ]_{ r } } \mid \Gamma ,   \SYSTEMnt{id}
\tag{contexts}
\end{align*}
Typing of the $\lambda$-calculus fragment is then by the following rules:
\begin{align*}
\begin{array}{c}
  \SYSTEMdruleTyvar{}
  \;\;\;\;
  \SYSTEMdruleTyabs{}
  \;\;\;\;
  \SYSTEMdruleTyapp{}
\end{array}
\end{align*}
The \textsc{var}, \textsc{abs}, and \textsc{app} rules are the standard rules of
the linear $\lambda$-calculus, augmented with a notion of \emph{contraction}
captured by the $+$ operation on contexts coming from multiple sub-terms, which is only defined when contexts
are disjoint with respect to linear assumptions, and on overlapping graded
assumptions we add their grades, e.g. $(\Gamma_1, \textcolor{coeffectColor}{ \SYSTEMmv{x}  : [  \SYSTEMnt{A}  ]_{ r } }) + (\Gamma_2, \textcolor{coeffectColor}{ \SYSTEMmv{x}  : [  \SYSTEMnt{A}  ]_{ s } })
= (\Gamma_1 + \Gamma_2), \textcolor{coeffectColor}{ \SYSTEMmv{x}  : [  \SYSTEMnt{A}  ]_{   r  +  s   } }$. More explicitly, context addition is
declaratively specified as follows:
\begin{align*}
  \setlength{\arraycolsep}{0.1em}
  \begin{array}{cc}
  \begin{array}{rl}
  \SYSTEMsym{(}   \Gamma ,   \SYSTEMmv{x}  :  \SYSTEMnt{A}    \SYSTEMsym{)}  +  \Gamma' & = \SYSTEMsym{(}   \Gamma  +  \Gamma'   \SYSTEMsym{)} ,   \SYSTEMmv{x}  :  \SYSTEMnt{A} \quad \text{iff} \,\; x
  \not\in | \Gamma' | \\
  \Gamma  +  \SYSTEMsym{(}   \Gamma' ,   \SYSTEMmv{x}  :  \SYSTEMnt{A}    \SYSTEMsym{)} & = \SYSTEMsym{(}   \Gamma  +  \Gamma'   \SYSTEMsym{)} ,   \SYSTEMmv{x}  :  \SYSTEMnt{A} \quad \text{iff} \,\; x \not\in | \Gamma | \\
  \SYSTEMsym{(}   \Gamma ,   \textcolor{coeffectColor}{ \SYSTEMmv{x}  : [  \SYSTEMnt{A}  ]_{ r } }    \SYSTEMsym{)}  +  \SYSTEMsym{(}   \Gamma' ,   \textcolor{coeffectColor}{ \SYSTEMmv{x}  : [  \SYSTEMnt{A}  ]_{ s } }    \SYSTEMsym{)} & = \SYSTEMsym{(}   \Gamma  +  \Gamma'   \SYSTEMsym{)} ,   \textcolor{coeffectColor}{ \SYSTEMmv{x}  : [  \SYSTEMnt{A}  ]_{  (   r  +  s   )  } }
  \end{array}
    \! & \!
  \begin{array}{rl}
    \emptyset   +  \Gamma & = \Gamma \quad \text{(context addition)} \\
    \Gamma  +   \emptyset & = \Gamma \\
    \SYSTEMsym{(}   \Gamma ,   \SYSTEMnt{id}    \SYSTEMsym{)}  +  \SYSTEMsym{(}   \Gamma' ,   \SYSTEMnt{id}    \SYSTEMsym{)} & = \SYSTEMsym{(}   \Gamma  +  \Gamma'   \SYSTEMsym{)} ,   \SYSTEMnt{id} \\
  \end{array}
  \end{array}
\end{align*}
In the first two cases, $x$ may be $id$ with $A = \mathsf{Name}$ implicitly. This
is a declarative rather than algorithmic specification of $+$ as graded
assumptions of the same variable may appear in different positions within the
two contexts---for example, when typechecking the program \granin{((x, y), (y,
x))}. The \textsc{var} rule also embeds the notion of \emph{weakening}, allowing
a context of variables graded by $0$, using the partial operation of
\emph{scalar multiplication of a context}:
\begin{align*}
  r  \cdot   \emptyset = \emptyset
  \quad\;\,
      r  \cdot  \SYSTEMsym{(}   \Gamma ,   \textcolor{coeffectColor}{ \SYSTEMmv{x}  : [  \SYSTEMnt{A}  ]_{ s } }    \SYSTEMsym{)} = \SYSTEMsym{(}   r  \cdot  \Gamma   \SYSTEMsym{)} ,   \textcolor{coeffectColor}{ \SYSTEMmv{x}  : [  \SYSTEMnt{A}  ]_{   r  *  s   } }
      \quad\;
      r  \cdot  \SYSTEMsym{(}   \Gamma ,   \SYSTEMnt{id}    \SYSTEMsym{)} = \SYSTEMsym{(}   r  \cdot  \Gamma   \SYSTEMsym{)} ,   \SYSTEMnt{id}
\tag{context multiplication}
\end{align*}
scaling graded assumptions by $r$, preserving
$\SYSTEMnt{id}$s, but undefined if $\Gamma$ contains linear
assumptions.

The rules involving graded modalities are then:
\begin{align*}
\begin{array}{c}
  \SYSTEMdruleTypr{}
  \quad
  \SYSTEMdruleTyder{}
  \\[1.5em]
  \SYSTEMdruleTyelim{}
  \quad
  \SYSTEMdruleTyapprox{}
\end{array}
\end{align*}
The \textsc{pr} rule (\emph{promotion}) introduces a graded modality with grade $r$,
implying that the result of $t$ can be used in an `$r$-like' way and thus
all of the dependencies of $\SYSTEMnt{t}$ must be scaled by $r$ to propagate usage to
the dependencies, none of which are allowed to be linear. Promotion also
contains an explicit restriction that the value to be promoted must not be a
\emph{resource allocator}, designed to avoid problems that otherwise arise when
promoting values with certain behaviours in a call-by-value setting; we will
discuss this in detail when we describe our interface for mutable arrays in
Section~\ref{sec:fractional}.

The \textsc{elim} rule eliminates a graded modality, capturing the idea that a
requirement for $x$ to be used in an `$r$-like' way in $t_2$ can be matched with
the capability of $t_1$ described by its graded modal type. In Granule, this
construct is folded into pattern matching: we can `unbox' (eliminate) a graded
modality to provide a graded variable in the body of the function (the analogue
to $t_2$ in this rule).

The \textsc{der} rule (\emph{dereliction}) connects linear typing to graded typing, stating that a
requirement for a linear assumption is satisfied by an assumption graded by $1$.
The \textsc{approx} rule converts a grade $r$ to $s$
if $s$ \textit{approximates} $r$ according to
the semiring's pre-order $\sqsubseteq$.

One possible choice of semiring is that of
natural numbers $(\mathbb{N}, *, +, 0, 1, \equiv)$
with \emph{discrete ordering}
$\equiv$ such that there is no approximation and we
track the exact number of times a term has been used. We could instead use
the standard $\leq$ ordering on natural numbers which would permit
approximation, allowing for an \emph{upper bound} on a term's usage. Another
useful semiring is on \emph{intervals} of natural numbers. Here, we can use a
grade such as \granin{0..1} to represent a value which can be used \emph{either}
zero or one times; this captures the notion of an \emph{affine} value.
Other interesting semirings include
lattices for security levels~\cite{gaboardi2016combining,DBLP:journals/pacmpl/AbelB20},
hardware schedules~\cite{ghica2014coeffects}, and sets for abstract property tracking.

Finally, we have the rules for introducing and eliminating tensor products and
the multiplicative unit, which are standard, though it is important to remember
that products are \emph{linear} and so it is not possible to freely discard
either side: both elements must be used when consuming a product.
\begin{align*}
\begin{array}{c}
  \SYSTEMdruleTypairIntro{}
  \;\;\;\;
  \SYSTEMdruleTypairElim{}
  \\[1.25em]
  \SYSTEMdruleTyunitIntro{}
  \;\;\;\;
  \SYSTEMdruleTyunitElim{}
\end{array}
\end{align*}
\begin{example}
  The following gives an example derivation assuming
  the natural number semiring:
  \begin{gather*}
    \vspace{-1em}
    \inferrule*[right=\SYSTEMdruleTyappName]
               {
                 \inferrule*[right=\SYSTEMdruleTyabsName{}]
                            {
                              \inferrule*[right=\SYSTEMdruleTypairIntroName]
                                         {\inferrule*[right=\SYSTEMdruleTyvarName]
                                           {\;}{\SYSTEMmv{x}  :   \SYSTEMnt{A}  \otimes  \SYSTEMnt{A}    \vdash  \SYSTEMmv{x}  :   \SYSTEMnt{A}  \otimes  \SYSTEMnt{A}}
                                           \inferrule*[right=\SYSTEMdruleTyvarName']
                                           {\;}{\textcolor{coeffectColor}{ \SYSTEMmv{y}  : [  \SYSTEMnt{A}  ]_{ \SYSTEMsym{1} } }   \vdash  \SYSTEMmv{y}  :  \SYSTEMnt{A}}}
                                {\textcolor{coeffectColor}{ \SYSTEMmv{y}  : [  \SYSTEMnt{A}  ]_{ \SYSTEMsym{1} } }  ,   \SYSTEMmv{x}  :   \SYSTEMnt{A}  \otimes  \SYSTEMnt{A}     \vdash  \SYSTEMsym{(}  \SYSTEMmv{x}  \SYSTEMsym{,}  \SYSTEMmv{y}  \SYSTEMsym{)}  :     (   \SYSTEMnt{A}  \otimes  \SYSTEMnt{A}   )   \otimes  \SYSTEMnt{A}}
                            }
                            {\textcolor{coeffectColor}{ \SYSTEMmv{y}  : [  \SYSTEMnt{A}  ]_{ \SYSTEMsym{1} } }   \vdash   \lambda  \SYSTEMmv{x}  .  \SYSTEMsym{(}  \SYSTEMmv{x}  \SYSTEMsym{,}  \SYSTEMmv{y}  \SYSTEMsym{)}   :    \SYSTEMnt{A}  \otimes  \SYSTEMnt{A}    \multimap     (   \SYSTEMnt{A}  \otimes  \SYSTEMnt{A}   )   \otimes  \SYSTEMnt{A}}
                 \;
                 \inferrule*[right=\SYSTEMdruleTypairIntroName{}]
                            {\emptyset   \vdash  \SYSTEMnt{v}  :  \SYSTEMnt{A} \quad
                             \inferrule*[right=\SYSTEMdruleTyvarName']
                                           {\;}{\textcolor{coeffectColor}{ \SYSTEMmv{y}  : [  \SYSTEMnt{A}  ]_{ \SYSTEMsym{1} } }   \vdash  \SYSTEMmv{y}  :  \SYSTEMnt{A}}}
                            {\textcolor{coeffectColor}{ \SYSTEMmv{y}  : [  \SYSTEMnt{A}  ]_{ \SYSTEMsym{1} } }   \vdash  \SYSTEMsym{(}  \SYSTEMnt{v}  \SYSTEMsym{,}  \SYSTEMmv{y}  \SYSTEMsym{)}  :   \SYSTEMnt{A}  \otimes  \SYSTEMnt{A}}
               }
               {\textcolor{coeffectColor}{ \SYSTEMmv{y}  : [  \SYSTEMnt{A}  ]_{ \SYSTEMsym{2} } }   \vdash  \SYSTEMsym{(}   \lambda  \SYSTEMmv{x}  .  \SYSTEMsym{(}  \SYSTEMmv{x}  \SYSTEMsym{,}  \SYSTEMmv{y}  \SYSTEMsym{)}   \SYSTEMsym{)} \, \SYSTEMsym{(}  \SYSTEMnt{v}  \SYSTEMsym{,}  \SYSTEMmv{y}  \SYSTEMsym{)}  :    (   \SYSTEMnt{A}  \otimes  \SYSTEMnt{A}   )   \otimes  \SYSTEMnt{A} }
  \end{gather*}
  where (\SYSTEMdruleTyvarName{}') is a synonym for the (\SYSTEMdruleTyvarName{}) rule followed
  by dereliction (\SYSTEMdruleTyderName{}).

  As a first taste of Granule syntax (which resembles Haskell, apart from the presence of
  the graded modal type constructs), the following captures the same idea as this example:
  \begin{granulep}
exampleNat : forall {a : Type} . a -> a [2] -> ((a, a), a)
exampleNat v yb = let [y] = yb in (\x -> (x, y)) (v, y)
  \end{granulep}
  The type ${\textcolor{coeffectColor}{\Box_{ r } } }  \SYSTEMnt{A}$ is instead written postfix as \granin{A [r]}, and
  Granule uses \granin{->} for its linear function types.
\end{example}

Existential types have standard introduction and elimination
typing forms, but restricted only to type variables of kind $\mathsf{Name}$
(whose kind is omitted here for simplicity due to this restriction):
\begin{align*}
  \inferrule*[right=\SYSTEMdruleTypackName{}]
             {\Gamma  \vdash  \SYSTEMnt{t}  :  \SYSTEMnt{A} \quad\;
               \SYSTEMnt{id}  \not\in \mathsf{dom}( \Gamma )}
             {\Gamma  \vdash   \textbf{pack}\ \langle{  \SYSTEMnt{id'}  ,  \SYSTEMnt{t}  }\rangle   :   \exists  \SYSTEMnt{id}  .    \SYSTEMnt{A}  [  \SYSTEMnt{id}  /  \SYSTEMnt{id'}  ]}
  \;\;\;
  \inferrule*[right=\SYSTEMdruleTyunpackName{}]
             {\Gamma_{{\mathrm{1}}}  \vdash  \SYSTEMnt{t_{{\mathrm{1}}}}  :   \exists  \SYSTEMnt{id}  .  \SYSTEMnt{A}
               \quad\;
                 \Gamma_{{\mathrm{2}}} ,   \SYSTEMnt{id}    ,   \SYSTEMmv{x}  :  \SYSTEMnt{A}    \vdash  \SYSTEMnt{t_{{\mathrm{2}}}}  :  \SYSTEMnt{B}
                 \quad\;
                   \SYSTEMnt{id}  \not\in \mathsf{fv}(  \SYSTEMnt{B}  )
             }
             {\Gamma_{{\mathrm{1}}}  +  \Gamma_{{\mathrm{2}}}   \vdash   \textbf{unpack}\ \langle{  \SYSTEMnt{id}  ,  \SYSTEMmv{x}  }\rangle
     =  \SYSTEMnt{t_{{\mathrm{1}}}} \ \textbf{in}\  \SYSTEMnt{t_{{\mathrm{2}}}}   :  \SYSTEMnt{B}}
\end{align*}

\paragraph{Note on Linear Haskell}

As mentioned at the start of this section, we aim for the extensions we develop
throughout this work to be compatible with other graded type systems after some
adaptation. One particular system of interest is the calculus underlying the
recent extension which introduces linear types to
Haskell~\cite{bernardy2017linear}. This extension uses a graded type system
below the surface in order to implement linearity; all function types are given
a `multiplicity' annotation $r$ written  \haskin{(a \%r -> b)} akin to a type
$\coeff{\Box_r} A \multimap B$ here, where the annotation can be either
\haskin{'One} or \haskin{'Many} representing either linear or unrestricted usage
respectively. Work on formalising the connection between systems like this where
all function types come with a grade (sometimes called `graded base') and
systems such as Granule's where values are linear by default and graded values
are wrapped inside a modality (called `linear base') is
ongoing~\cite{vollmer2024mixed}.

\section{One of a Kind: Uniqueness and Sharing}
\label{sec:uniqueness}
The first extension we make to the core calculus is
to represent \emph{ownership}: values that have
a unique owner can mutate the value freely because they are
in possession of the only reference that exists. It turns out this matches
closely with the idea of \emph{uniqueness} types; if it is possible to guarantee
that a reference to a value is unique (i.e., it is the only
reference that exists), then whichever part of the program (e.g., a given thread or process) holds that reference must be the owner.

Uniqueness types were introduced into Granule's core type system in previous
work~\cite{marshall2022linearity}, but we will extend this in two ways:
first, we generalise the rules of \citet{marshall2022linearity} to allow for a graded necessity
modality parameterised by an arbitrary semiring rather than the simple
non-linear $\oc$, and second, we incorporate the identifiers described in
Section~\ref{sec:core} into the typing, which will become important when
multiple references need to be tracked.

As with uniqueness typing, the crucial insight for integrating owned values with linear and graded values is to consider any linear or graded value to have no ownership information
attached; these values have their memory managed in other ways, such as explicitly by the programmer or implicitly by a garbage collector. We then introduce a modality,
written $\textcolor{uniqueColor}{\ast}$ (akin to Marshall et al.'s uniqueness modality) to type values that are \emph{uniquely owned}, meaning that the type system must ensure only one reference exists at any given time. We extend the
syntax of terms and types:
\begin{align*}
\tag{terms \& types, extended}
\begin{array}{c}
  \SYSTEMnt{t} ::= \ldots \mid \SYSTEMkw{share} \, \SYSTEMnt{t} \mid \SYSTEMkw{clone} \, \SYSTEMnt{t_{{\mathrm{1}}}} \, \textbf{as} \, \SYSTEMmv{x} \, \textbf{in} \, \SYSTEMnt{t_{{\mathrm{2}}}}
   \qquad\quad
  \SYSTEMnt{A}, \SYSTEMnt{B} ::= \ldots \mid {\textcolor{uniqueColor}{\ast} }{ \SYSTEMnt{A} }
\end{array}
\end{align*}
The two constructs for this uniqueness modality $\textcolor{uniqueColor}{\ast}$
provide sharing and cloning, with types:
\begin{align*}
  \begin{array}{c}
    \SYSTEMdruleTyreturnGen{}
    \;\;\;\;
    \SYSTEMdruleTybindGenFresh{}
  \end{array}
\end{align*}
The \textsc{Share} construct allows a guarantee of unique ownership to be discarded,
though this means that memory must now once again be managed automatically just
as with any other Granule values where ownership is not tracked. Any grade
$r$ can be selected here, though the shared value must eventually be either
discarded (requiring $\SYSTEMsym{0}  \sqsubseteq  r$) or fully consumed via pattern matching.

The \textsc{Clone} construct makes a \emph{deep copy} of the value $\SYSTEMnt{t_{{\mathrm{1}}}}$,
where we take a value that does not have an owner and make a unique copy of it;
we take ownership of the copy, binding it the scope of $\SYSTEMnt{t_{{\mathrm{2}}}}$. We can
guarantee uniqueness since this is now the only reference that exists to the
copied value. We must update any identifiers along the way, however, to ensure
that this is understood as a separate value to the original, in case other
references still exist. This is mediated by existential types, where all
identifiers $\overline{ \SYSTEMnt{id} }$ used in typing $\SYSTEMnt{t_{{\mathrm{1}}}}$ are bound under an
existential quantifier.

Note, however, that cloning requires some additional conditions for soundness.
The predicate $\mathsf{cloneable} (  \SYSTEMnt{A}  )$ ensures the value is of a resource
type---such as a reference---or a product of resources. (The full definition of
$\mathsf{cloneable}$ is presented in Section~\ref{sec:fractional}, where we
discuss said resource types in more depth.) This is necessary to prevent values
which do not come with identifiers from being managed via ownership, since once
we introduce borrowing in Section~\ref{sec:fractional} this would lead to
borrows of different values of the same type (such as two floating point
numbers) being indistinguishable at the type level and thus interchangeable. The
side-condition $\SYSTEMsym{1}  \sqsubseteq  r$ explains that we must be able to accommodate a
usage of the input $\SYSTEMnt{t_{{\mathrm{1}}}}$ since $\textbf{clone}$ consumes the $\SYSTEMnt{t_{{\mathrm{1}}}}$
value once in copying it.

Our implementation of these ideas in Granule follows the above typing. We recap
the initial ownership example from Section~\ref{sec:key} here in our new
extension of Granule using a data type \granin{Colour} which acts as an alias
for a triple of type \granin{(Int, (Int, Int))}.

The following code illustrates that linear types
already require us to obey the laws of \emph{move semantics} for
owned values, where we move ownership of \granin{scarlet} to \granin{x}
and then to \granin{y}:
\begin{granulep}
exampleMove : *Colour -> *Colour
exampleMove scarlet = let x = scarlet in -- ownership "moved" to x
               -- let y = scarlet in ... -- would cause an error: non-linear use of `scarlet`
                  let y = x in y         -- but "moving" ownership from x to y is allowed
\end{granulep}
We can have multiple variables which all point to the
initial value of type \granin{Colour} via the
\textbf{share} operation; this explicitly exempts the value from being
managed by the ownership system. Here, we share at the interval grade \granin{0..2}, capturing the
idea that the value must be used somewhere between zero and two times (where
here it happens to be used twice).

\begin{granulep}
exampleShare : *Colour -> (Colour, Colour)
exampleShare scarlet = let [s] : (Colour [0..2]) = share scarlet in
                       let x = s in
                       let y = s in (x, y)
\end{granulep}

The $\textbf{share}$ and $\textbf{clone}$ operations have the following
equations showing their interaction:
\begin{align}
  \SYSTEMkw{clone} \, \SYSTEMsym{(}  \SYSTEMkw{share} \, \SYSTEMnt{v}  \SYSTEMsym{)} \, \textbf{as} \, \SYSTEMmv{x} \, \textbf{in} \, \SYSTEMnt{t} & \equiv \SYSTEMnt{t}  [   \textbf{pack}\ \langle{ \overline{ \SYSTEMnt{id} } ,  \SYSTEMnt{v}  }\rangle   /  \SYSTEMmv{x}  ] \tag{$\beta_\ast$} \\
  \SYSTEMkw{clone} \, \SYSTEMnt{t_{{\mathrm{1}}}} \, \textbf{as} \, \SYSTEMmv{x} \, \textbf{in} \, \SYSTEMsym{(}  \SYSTEMkw{clone} \, \SYSTEMnt{t_{{\mathrm{2}}}} \, \textbf{as} \, \SYSTEMmv{y} \, \textbf{in} \, \SYSTEMnt{t_{{\mathrm{3}}}}  \SYSTEMsym{)}
   & \equiv \SYSTEMkw{clone} \, \SYSTEMsym{(}  \SYSTEMkw{clone} \, \SYSTEMnt{t_{{\mathrm{1}}}} \, \textbf{as} \, \SYSTEMmv{x} \, \textbf{in} \, \SYSTEMnt{t_{{\mathrm{2}}}}  \SYSTEMsym{)} \, \textbf{as} \, \SYSTEMmv{y} \, \textbf{in} \, \SYSTEMnt{t_{{\mathrm{3}}}}  \quad(x \not\in \mathsf{FV}(t_3)) \tag{$\ast$assoc}
\end{align}
The ($\beta_\ast$) axiom states that sharing a value $\SYSTEMnt{v}$ (term with no
further reductions---see Section~\ref{sec:semantics}) and cloning it to create a
new owned $x$ in the scope of $t$ is equivalent to substituting the original $\SYSTEMnt{v}$  for $x$ in $t$ (with its identifiers packed in an existential). The
($\ast$assoc) axiom is associativity of cloning.

The presence of identifiers necessitates the existential typing. The need for
identifiers is shown in the next section once we move to a generalisation of
uniqueness with the ability to separate \emph{immutable borrows} from
\emph{mutable borrows}, where existential types will allow for tighter control
of the \emph{lifetime} of borrowed values with identifiers.

\section{Immutably Borrowed is Fractionally Unique}
\label{sec:fractional}
We now generalise from the above system which incorporates uniqueness into
the core calculus to a system that allows for the
uniqueness guarantees to be temporarily broken in controlled ways, such that we
can continue to ensure memory safety and more closely approximate Rust-like
ownership and borrowing rules. We take inspiration from the
fractional permissions of \citet{boyland2003checking} (which themselves served as
partial inspiration for Rust's notion of ownership) and from recent literature on integrating linear types with fractional
permissions~\cite{makwana2019numlin}.

Just as grading non-linearity gives a more precise account
of resource usage, we introduce borrowing as a graded form of non-\emph{uniqueness} denoted ${\textcolor{borrowColor}{\&_{ p } } }  \SYSTEMnt{A}$, allowing precise control of references:
\begin{align*}
\tag{types, extended}
  \SYSTEMnt{A}, \SYSTEMnt{B} & ::= \ldots \mid {\textcolor{borrowColor}{\&_{ p } } }  \SYSTEMnt{A}
\\
\tag{permissions}
  p, q & ::= \ast \mid f \qquad\qquad (\textit{where $f \in \mathbb{Q}, 0 < f \leq 1$})
\end{align*}
where $p$ is a new form of grade for
tracking borrowing, called a \emph{permission}. Permissions are either rational
numbers $f$ between $0$ and $1$ or a special permission $\ast$ representing
unique ownership (described below).
A \emph{mutable borrow} is represented by ${\textcolor{borrowColor}{\&_{ \SYSTEMsym{1} } } }  \SYSTEMnt{A}$ which allows temporary mutable access to
a value while preserving the guarantee that it will be eventually returned to
its original owner. Note that $0$ is excluded: the typing rules we provide can never
produce a value with permission $0$.

For working with the borrowing graded modality, terms
are extended as follows:
\begin{align*}
\tag{terms, extended}
  \SYSTEMnt{t} ::= \ldots \mid \textbf{withBorrow}\  \SYSTEMnt{t_{{\mathrm{1}}}} \  \SYSTEMnt{t_{{\mathrm{2}}}} \mid \textbf{split}\  \SYSTEMnt{t} \mid \textbf{join}\  \SYSTEMnt{t_{{\mathrm{1}}}} \  \SYSTEMnt{t_{{\mathrm{2}}}}
\end{align*}
The creation of mutable borrows is via $\textbf{withBorrow}$ with
the following typing:
\begin{align*}
  \SYSTEMdruleTywithBorrow{}
\end{align*}
Here, we allow for a uniquely owned ${\textcolor{uniqueColor}{\ast} }{ \SYSTEMnt{A} }$ value to be borrowed and
manipulated in a mutable way by some function $\SYSTEMnt{t_{{\mathrm{2}}}}$ which expects a ${\textcolor{borrowColor}{\&_{ \SYSTEMsym{1} } } }  \SYSTEMnt{A}$ as
input, so long as it returns the mutable borrow as a ${\textcolor{borrowColor}{\&_{ \SYSTEMsym{1} } } }  \SYSTEMnt{B}$ in the output
so that the original owner can reclaim this as a unique reference (${\textcolor{uniqueColor}{\ast} }{ \SYSTEMnt{B} }$).
By encapsulating the mutable borrow behaviour inside a continuation, we ensure
that it is impossible to construct a closed term with a borrowed type, which
means borrowed references to values must always eventually return full access to
the owner of said value.

\paragraph{Uniqueness as uniquely borrowed}
We absorb the development of Section~\ref{sec:uniqueness} by the
inclusion of $* \in p$ and a type identity ${\textcolor{uniqueColor}{\ast} }{ \SYSTEMnt{A} } \equiv {\textcolor{borrowColor}{\&_{ \ast } } }  \SYSTEMnt{A}$.
Thus, the borrowing graded modality
captures both (uniquely) owned and borrowed values and hence functions polymorphic
in their permission can range over both kinds of ownership. Furthermore, this enables us to later define operations that can work over uniquely owned or mutably borrowed values without needing multiple primitives.

\paragraph{Immutable borrows}
We generalise further to \emph{immutable
borrows}, wherein multiple references to a value can exist at any given time as
long as they are unable to be mutated (as otherwise this would violate memory safety
and lead to, e.g., data races).
Mutable borrows are already graded with permission $1$, representing full
unfettered access to a value, so both reads and writes are permissible (shown later for resources). The $\textbf{split}$ and $\textbf{join}$ constructs
generalise mutable borrows, with typing:
\begin{align*}
  \begin{array}{c}
    \SYSTEMdruleTysplit{}
    \;\;\;\;
    \SYSTEMdruleTyjoin{}
  \end{array}
\end{align*}
where $\textbf{split}$ takes a borrowed reference and splits it into two borrowed references to the same value, each graded by
half of the original permission.\footnote{\SYSTEMdruleTysplitName{}
could be generalised to produce a vector with any number $n$ of ${\textcolor{borrowColor}{\&_{  { p }/{ \SYSTEMmv{n} }  } } }  \SYSTEMnt{A}$,
or even to arbitrary permissions $q$ and $q'$ such that $q +
q' = p$, but here we restrict to halving for simplicity.} The dual
\textbf{join} (re)combines two borrows into one
with their permissions added. Due to the interfaces for manipulating resources (shown later), the type $A$ above will always contain a resource
identifier, so it is never possible to combine references to two different values.
The addition and halving operations are defined only for
fractions $f$, not $*$.

These rules do not encompass every detail of Rust's
intricate ownership system; they instead aim to represent the \emph{core}
concepts of ownership and borrowing in such a way that they can integrate into
an existing graded type system, leaving room for potential extensions. In
particular, as was mentioned in Section~\ref{sec:key}, we only support
\emph{lexical} lifetimes, meaning that there is a class of programs which are
accepted by Rust's compiler but cannot be represented in this calculus. We focus
on being able to represent notions of ownership \emph{explicitly} in the types,
rather than using a complex static analysis like Rust's borrow checker to
\emph{infer} which programs are safe.

The following code demonstrates how all of these new primitives can be put
together in a simple Granule program. More practical examples will be presented
when we introduce interfaces for resources. This particular example, which is
purely illustrative, makes use of a function called \granin{observe} which
operates over borrowed values; the precise behaviour of this function is elided,
but it does not perform any mutation and so it is polymorphic in the permission:
\begin{granulep}
observe : forall {p : Fraction} . BpColour -> BpColour

exampleBorrow : *Colour -> *Colour
exampleBorrow persimmon = withBorrow (\b -> let (x, y) = split b   in
                                            let     x' = observe x in
                                            let      b = join (x', y) in b) persimmon
\end{granulep}
\noindent
In \granin{exampleBorrow}, we \emph{mutably} borrow the value \granin{persimmon}
as \granin{b}, split this into two \emph{immutable} borrows \granin{x} and
\granin{y}, and apply the \granin{observe} function to \granin{x} before
rejoining the borrows and returning the value.

The unsafe patterns demonstrated by the \granin{viridian} and \granin{cerulean}
examples in Section~\ref{sec:key} are also unrepresentable in Granule. It is
impossible for two mutable borrows or for a mutable borrow and an immutable
borrow to coexist here, since splitting a borrow must reduce the permission in
its type by the nature of the \granin{split} primitive. We can, however,
represent \emph{reborrowing}; this simply involves continuing to split immutable
borrows further, where recovering the original value now requires collecting all
of the borrows once more. The order in which the borrows are rejoined is
immaterial. The following example illustrates this.
\begin{granulep}
exampleReborrow : *Colour -> *Colour
exampleReborrow amethyst = withBorrow (\b -> let (x, y) = split b in
                                             let (l, r) = split x in
                                             let     x' = join (r, l) in
                                             let     b  = join (y, x') in b) amethyst
\end{granulep}

The benefits of uniqueness typing for memory management are well-trodden in
prior work~\cite{marshall2022linearity}; the idea is that the guarantee of a
unique reference both allows safe in-place update but also obviates the need for
garbage collection, since memory can be reclaimed as soon as a reference is
deleted. The system described here augments this by allowing increased
flexibility while preserving the same guarantees--multiple references can be
created via \granin{split} but they are tracked precisely through fractional
grades so that uniqueness is still guaranteed once they are re\granin{join}ed,
with in-place update only allowed for references with unique access.
\paragraph{Resources}
Enforcing which behaviours should be allowed for different permissions on a
given resource is mediated by said resource's interface. We will give two
in-depth examples in this paper to demonstrate the flexibility of our ownership
and borrowing interface. The first is mutable arrays of floats that can be
created, read from, written to and deleted (each at differing levels of access).
This allows us to illustrate one of the key practical benefits of ownership,
which is that unique access to an owned value allows for safe mutation (the
original pun behind ``Linear Types can Change the
World''~\cite{wadler1990linear}). The second example will be polymorphic
references which store a pointer to a value of any type. First, we extend our
calculus with a primitive type of resources, along with some additional base
types used to parametrise said resources:
\begin{align*}
\tag{types, extended}
\SYSTEMnt{A} ::= \ldots \mid \SYSTEMnt{Res} _{ \SYSTEMnt{id} } \  \SYSTEMnt{A} \mid \mathbb{N} \mid \mathbb{F} \qquad
\SYSTEMnt{Res} ::= \mathsf{Array} \mid \mathsf{Ref}
\end{align*}
where $\mathbb{N}$ are natural numbers used for sizes and indices and $\mathbb{F}$ are floating-point numbers (which we treat as inherently non-linear, much as
the \rustin{f32} and \rustin{f64} types implement the \rustin{Copy} trait in
Rust), and resources $\SYSTEMnt{Res}$ (ranging over arrays or references for our
purposes) are indexed by an identifier. Our interface for mutable arrays
provides the following primitives (with built-in weakening):
\begin{align*}
  \begin{array}{lll}
    \SYSTEMsym{0}  \cdot  \Gamma \vdash & \textbf{newArray} & :  \mathbb{N}   \multimap   \exists  \SYSTEMnt{id}  .   {\textcolor{uniqueColor}{\ast} }{  (    \mathsf{Array}  _{ \SYSTEMnt{id} } \   \mathbb{F}    )  } \\
    \SYSTEMsym{0}  \cdot  \Gamma \vdash & \textbf{readArray} & : {\textcolor{borrowColor}{\&_{ p } } }   (    \mathsf{Array}  _{ \SYSTEMnt{id} } \   \mathbb{F}    )    \multimap   \mathbb{N}   \multimap   \mathbb{F}   \otimes    {\textcolor{borrowColor}{\&_{ p } } }   (    \mathsf{Array}  _{ \SYSTEMnt{id} } \   \mathbb{F}    ) \\
    \SYSTEMsym{0}  \cdot  \Gamma \vdash & \textbf{writeArray} & : {\textcolor{borrowColor}{\&_{ p } } }   (    \mathsf{Array}  _{ \SYSTEMnt{id} } \   \mathbb{F}    )    \multimap   \mathbb{N}   \multimap   \mathbb{F}   \multimap   {\textcolor{borrowColor}{\&_{ p } } }   (    \mathsf{Array}  _{ \SYSTEMnt{id} } \   \mathbb{F}    ) \quad (\textit{where}\ p  \equiv  \SYSTEMsym{1}  \ \vee\   p  \equiv  \ast) \\
    \SYSTEMsym{0}  \cdot  \Gamma \vdash & \textbf{deleteArray} & : {\textcolor{uniqueColor}{\ast} }{  (    \mathsf{Array}  _{ \SYSTEMnt{id} } \   \mathbb{F}    )  }   \multimap   \mathsf{unit}
\end{array}
\end{align*}
where, unless bound, $\SYSTEMnt{id}$ and $p$ are metavariables for identifier
types and permissions respectively. Thus, we treat these primitives as part of
the typing rules rather than functions that are in scope. Note that the
implementation allows for full universal quantification over identifiers and
permissions, enabling polymorphism; for example, the Granule type signature for
$\textbf{readArray}$ is as follows.

\begin{granulep}
readFloatArray : forall {p : Fraction, id : Name} 
               . Bp(FloatArrayid) -> Int -> (Float, Bp(FloatArrayid))
\end{granulep}

This interface resembles Granule's existing interface for array mutation
using modal uniqueness types, but with some crucial differences owing to the
introduction of borrowing. First, array types are now indexed by an
identifier pointing to a particular array object in the heap, generated
by the existential type in $\textbf{newArray}$ and within the scope
of the \SYSTEMdruleTybindGenFreshName{} rule. This allows us to keep track of
which array is being referenced now that multiple borrows pointing to the same
array are permitted to exist.
The other key difference is that reading and writing no longer require sole
ownership. Reading can be carried out at any permission, since this is safe no
matter how many references exist, but writing is restricted to either a mutable-borrowed or uniquely-owned array.

The following example illustrates the application of these various primitives in
a Granule program, and also demonstrates the usage of existentially typed
identifiers from the perspective of the programmer.

\begin{granulep}
exampleArray : Float
exampleArray = unpack <id , a> = newFloatArray 3 in let
                      a'       = writeFloatArray a 1 4.2;
                      (f, a'') = readFloatArray a' 1;
                      ()       = deleteFloatArray a'' in f
\end{granulep}

As per the typing rules of existentials presented in Section~\ref{sec:core},
once we unpack the existential in the second line the type variable \granin{id}
must not occur in the type of the body. In this example, the array is deleted
and never returned in the body’s result, thus satisfying the type restrictions
of \granin{unpack}. We can also apply the \granin{clone} function to make a deep
copy of a unique array as described in Section~\ref{sec:uniqueness}, in the
following way.

\begin{granulep}
exampleClone : ()
exampleClone = unpack <id , a> = newFloatArray 3 in
                 clone (share a) as x in
                   unpack <id' , a'> = x in (deleteFloatArray a')
\end{granulep}

The interface for polymorphic references is similar, with primitives:
\begin{align*}
  \begin{array}{lll}
    \SYSTEMsym{0}  \cdot  \Gamma \vdash & \textbf{newRef} & :  \SYSTEMnt{A}  \multimap   \exists  \SYSTEMnt{id}  .   {\textcolor{uniqueColor}{\ast} }{  (    \mathsf{Ref}  _{ \SYSTEMnt{id} } \  \SYSTEMnt{A}   )  } \\
    \SYSTEMsym{0}  \cdot  \Gamma \vdash & \textbf{readRef} & : {\textcolor{borrowColor}{\&_{ p } } }   (    \mathsf{Ref}  _{ \SYSTEMnt{id} } \   (   {\textcolor{coeffectColor}{\Box_{  r  +  \SYSTEMsym{1}  } } }  \SYSTEMnt{A}   )    )    \multimap  \SYSTEMnt{A}  \otimes    {\textcolor{borrowColor}{\&_{ p } } }   (    \mathsf{Ref}  _{ \SYSTEMnt{id} } \   (   {\textcolor{coeffectColor}{\Box_{ r } } }  \SYSTEMnt{A}   )    ) \\
\SYSTEMsym{0}  \cdot  \Gamma \vdash & \textbf{swapRef} & : {\textcolor{borrowColor}{\&_{ p } } }   (    \mathsf{Ref}  _{ \SYSTEMnt{id} } \  \SYSTEMnt{A}   )    \multimap  \SYSTEMnt{A}  \multimap  \SYSTEMnt{A}  \otimes    {\textcolor{borrowColor}{\&_{ p } } }   (    \mathsf{Ref}  _{ \SYSTEMnt{id} } \  \SYSTEMnt{A}   ) \quad (\textit{where}\ p  \equiv  \SYSTEMsym{1}  \ \vee\   p  \equiv  \ast) \\
    \SYSTEMsym{0}  \cdot  \Gamma \vdash & \textbf{freezeRef} & : {\textcolor{uniqueColor}{\ast} }{  (    \mathsf{Ref}  _{ \SYSTEMnt{id} } \  \SYSTEMnt{A}   )  }   \multimap  \SYSTEMnt{A}
\end{array}
\end{align*}
The crucial differences follow from the value encapsulated in the reference
being of any type, rather than only floats which are inherently duplicable and
discardable. Thus, $\textbf{swapRef}$ must enforce linear usage for both the
existing and new values. Similarly $\textbf{freezeRef}$ must return the value in
the reference to obey linearity. The next example illustrates this interface, by
creating a reference to a float, updating the value and then deleting the
reference. 

\begin{granulep}
exampleReference : (Float, Float)
exampleReference = unpack <id , ref> = newRef 0.0 in let
                           (x, ref') = swapRef ref 42.0;
                                   y = freezeRef ref' in (x, y)
\end{granulep}

In order to allow for non-linear behaviours, we also provide a $\textbf{readRef}$
primitive which requires the value to be of a graded type which permits the
additional usage accrued with the appropriate type-level accounting; this can be
applied as follows.

\begin{granulep}
test : Float [6]
test = [42.0]

exampleGraded : (Float [4], Float [2])
exampleGraded =                  unpack <id , ref> = newRef test in let
  ([x], ref') : (Float [4], *(Ref id (Float [2]))) = readRef ref;
                                   [y] : Float [2] = freezeRef ref' in ([x], [y])
\end{granulep}

One important benefit of introducing polymorphic references is that these allow
standard Granule values such as integers or floating point numbers to be treated
uniquely and managed by the ownership system, since otherwise this would be
prevented by the $\mathsf{cloneable}$ predicate described in
Section~\ref{sec:uniqueness}. This allows for the use of general imperative
programming patterns as might be applied in languages like Rust (where ownership
is the default) to be transferred to the setting of Granule (where functional
patterns are more typical).

We are now able to give the full inductive definition of the
$\mathsf{cloneable}$ predicate, displayed below. As described in
Section~\ref{sec:uniqueness}, this permits only resources and products
containing resources to be cloned, ensuring that all cloneable values come with
an identifier. Note that in order to preserve both linearity and this condition,
for a polymorphic reference to be cloneable it must itself point either to a
cloneable value or to a value that is freely copyable, such as a floating
point number.

\begin{definition}[Copyable predicate]
  Predicate definition:
  \begin{gather*}
        \begin{array}{c}
        \dfrac{}{\mathsf{copyable} (   \mathsf{unit}   )}
        \quad
        \dfrac{}{\mathsf{copyable} (   \mathbb{F}   )}
        \quad
        \dfrac{\mathsf{copyable} (  \SYSTEMnt{A}  ) \quad \mathsf{copyable} (  \SYSTEMnt{B}  )}{\mathsf{copyable} (   \SYSTEMnt{A}  \otimes  \SYSTEMnt{B}   )}
        \end{array}
  \end{gather*}
\end{definition}
\begin{definition}[Cloneable predicate]
  Predicate definition:
  \begin{gather*}
        \begin{array}{c}
        \dfrac{}{\mathsf{cloneable} (    \mathsf{Array}  _{ \SYSTEMnt{id} } \   \mathbb{F}    )}
        \quad
        \dfrac{\mathsf{cloneable} (  \SYSTEMnt{A}  )  \ \vee\   \mathsf{copyable} (  \SYSTEMnt{A}  )}{\mathsf{cloneable} (    \mathsf{Ref}  _{ \SYSTEMnt{id} } \  \SYSTEMnt{A}   )}
        \quad
        \dfrac{\mathsf{cloneable} (  \SYSTEMnt{A}  ) \quad \mathsf{cloneable} (  \SYSTEMnt{B}  )}{\mathsf{cloneable} (   \SYSTEMnt{A}  \otimes  \SYSTEMnt{B}   )}
        \end{array}
  \end{gather*}
\end{definition}

\paragraph{Resource allocators} Here, the restriction introduced on the
promotion rule in Section~\ref{sec:core} meaning that `resource allocators' are
forbidden from being promoted becomes crucial for ensuring soundness in the
call-by-value setting of this work. Consider the following pseudocode (eliding
packing and unpacking of existentials), which would be allowed in Granule given
unrestricted promotion:

\begin{granulepill}
let [x] : ((*(Array id Float)) [2]) = [newArray 1] in
let                              () = deleteArray x in writeArray x 0 1.0
\end{granulepill}

On the first line, this program creates a reference \granin{x} to a new array of
size $1$, but under a promotion, with the type explaining that we want to use
the resulting value twice (given by the explicit type signature here). This
promotion then allows two uses of the array on the second line.

Under a call-by-name semantics, as in previous work on
embedding uniqueness types in Granule~\cite{marshall2022linearity} (and
accessible in Granule via the extension \granin{language CBN}), this program
executes successfully and produces an array which contains the value written on line 2. The key is that call-by-name reduction substitutes the call to
\granin{newArray} into the two uses of the variable \granin{x}, and so these
point to two entirely separate arrays. However, under the call-by-value semantics
of this paper (and Granule's default), the first line is
fully evaluated, and so both uses of \granin{x} point to the same
array. Thus the second line attempts to write to an array
after it has been deleted, which would cause a runtime error. In a
call-by-value setting this program must not be permitted.

The solution we apply here (also used in recent work on graded session
types~\cite{marshall2022replicate}) is to syntactically restrict promotion to
terms which do not allocate resources, i.e., that do not use \granin{newArray}
or \granin{newRef} in a reduction position. The predicate $\mathsf{resourceAllocator}(  \SYSTEMnt{t}  )$ classifies precisely these terms, as a kind of specialised ``value
restriction''; note in particular that $\neg \mathsf{resourceAllocator}(   \lambda  \SYSTEMmv{x}  .    \textbf{newArray}   \, \SYSTEMnt{t}   )$ since reduction does not happen underneath an abstraction
(Section~\ref{sec:semantics} defines the reduction semantics). The appendix
includes the full inductive definition.

Other work resolves this same problem relating to
promotion of resource allocators through different techniques. Originally, the
linear types extension to Haskell got around this difficulty by only ever
allocating resources inside a specialised continuation, which is passed around
at every step until deallocation to prevent the allocator itself from ever being
used non-linearly. More recently, this strategy has been generalised by
introducing a ``linear constraint''~\cite{10.1145/3547626} called
\textcolor{coeffectColor}{\granin{Linearly}}. This constraint, which must itself
be used in a linear fashion, is assumed whenever a new resource is allocated. A
continuation is still necessary for the initial assumption of
\textcolor{coeffectColor}{\granin{Linearly}}, but the same qualification may now
be used generically for varying resource types.

\subsection{Equational theory}
\label{sec:equational}

The \SYSTEMdruleTywithBorrowName{} typing suggests a close
relationship between ${\textcolor{borrowColor}{\&_{ \ast } } }  \SYSTEMnt{A}$ and ${\textcolor{borrowColor}{\&_{ \SYSTEMsym{1} } } }  \SYSTEMnt{A}$. By analogy to the notion of a
relative monad used in prior work~\cite{marshall2022linearity},
${\textcolor{borrowColor}{\&_{ \ast } } }  \SYSTEMnt{A}$ acts as a `relative
functor' with regard to ${\textcolor{borrowColor}{\&_{ \SYSTEMsym{1} } } }  \SYSTEMnt{A}$, where $\textbf{withBorrow}$ equates to mapping a function involving mutable
borrows onto a function between uniquely owned values. The following axioms for functors then hold:
\begin{align*}
  \textbf{withBorrow}\  \SYSTEMsym{(}   \lambda  \SYSTEMmv{x}  .  \SYSTEMmv{x}   \SYSTEMsym{)} \  \SYSTEMnt{t}
  &\equiv
  \SYSTEMnt{t} \tag{$\with$unit}\\
  \textbf{withBorrow}\  \SYSTEMsym{(}   \lambda  \SYSTEMmv{x}  .   f   \, \SYSTEMsym{(}  \SYSTEMnt{g} \, \SYSTEMmv{x}  \SYSTEMsym{)}  \SYSTEMsym{)} \  \SYSTEMnt{t}
  &\equiv
  \textbf{withBorrow}\   f  \  \SYSTEMsym{(}   \textbf{withBorrow}\  \SYSTEMnt{g} \  \SYSTEMnt{t}   \SYSTEMsym{)} \tag{$\with$assoc} %\\
\end{align*}

The first axiom states that borrowing a reference to an owned value and simply
returning it via the identity function without making use of it is equivalent to
doing nothing at all, as one might expect. The second axiom is an associativity
axiom, giving us the result that borrowing a value to apply one function and
then borrowing the value again to apply a second function is equivalent to
simply borrowing the value once and applying the functions in sequence.

Lastly, \textbf{split} and \textbf{join} have additional properties which form an isomorphism:
\begin{align*}
  (\textbf{let} \, \SYSTEMsym{(x,}  \SYSTEMsym{y)}  \SYSTEMsym{=}  \SYSTEMsym{(}   \textbf{split}\  \SYSTEMnt{t}   \SYSTEMsym{)} \, \textbf{in} \, \SYSTEMsym{(}   \textbf{join}\  \SYSTEMmv{x} \  \SYSTEMmv{y}   \SYSTEMsym{)}) & \equiv \SYSTEMnt{t} \tag{$\with$rejoin}\\
  \textbf{split}\  \SYSTEMsym{(}   \textbf{join}\  \SYSTEMnt{t_{{\mathrm{1}}}} \  \SYSTEMnt{t_{{\mathrm{2}}}}   \SYSTEMsym{)} & \equiv \SYSTEMsym{(}  \SYSTEMnt{t_{{\mathrm{1}}}}  \SYSTEMsym{,}  \SYSTEMnt{t_{{\mathrm{2}}}}  \SYSTEMsym{)} \tag{$\with$resplit}
\end{align*}

\subsection{Divide and conquer: partial views via distributive laws}
\label{sec:distrib}

One particularly useful borrowing pattern when writing practical programs is the
ability to take a composite data structure and borrow only \emph{part} of it,
such that the original owner retains access to the remaining structure. This
introduces the possibility of multiple threads working with parts of a data
structure in parallel, where it is no longer necessary to take ownership of the
full data structure in order to mutate the only part for which access is
required.

Our core calculus provides some capacity for structured data in the form of
product types. We introduce two additional constructs for distributing the
borrowing graded modality into and out of products, enabling the pattern of
partial borrowing:\footnote{One might wonder whether the $\Box_r$ modality also
distributes in this way; the answer is not in general, though some choices of
semiring permit this behaviour. This question has been explored in depth in
prior work~\cite{hughes2021linear}.}
\begin{align*}
    \begin{array}{c}
     \hspace{-3em}
      \SYSTEMdruleTypush{}
      \;\;\;\;
      \SYSTEMdruleTypull{}
    \end{array}
\end{align*}
We can derive an operation akin to \textbf{withBorrow} for borrowing one
component of a product: use \textbf{push} to move ownership onto the values
inside the product, use \textbf{withBorrow} to borrow the required component,
and then once we are done, use \textbf{pull} to recover the original pair. We
demonstrate this concept in action with the following example. We assume a
function \granin{alter : & 1 Int -> & 1 Int} which operates over borrowed
components of \granin{Colour} and requires a whole permission, in order to
illustrate the idea. We then employ this strategy for partial borrowing:
\begin{granulep}
examplePartial : *Colour -> *Colour
examplePartial indigo = let (r, p) = push indigo in
                        let     r' = withBorrow alter r in pull (r', p)
\end{granulep}
This notion extends gracefully to structures containing more than two
components. For example, to borrow a single element from a
unique tuple of three values, leaving the remaining two values available
to their original owner, we can use nested products, e.g. ${\textcolor{uniqueColor}{\ast} }{  (   \SYSTEMnt{A}  \otimes   (   \SYSTEMnt{B}  \otimes  \SYSTEMnt{C}   )    )  }$,
and \textbf{push} and \textbf{pull} to
then access just one part, similarly to in \granin{indigo}. We may need to \textbf{push} twice in order to distribute the modality over the
entire product depending on which component we are borrowing, but in the same way as above we can use \textbf{withBorrow} to extract the desired data (for
example, ${\textcolor{borrowColor}{\&_{ \SYSTEMsym{1} } } }  \SYSTEMnt{A}$) while applying \textbf{pull} to recover the
remaining structure (in this case, of type ${\textcolor{uniqueColor}{\ast} }{  (   \SYSTEMnt{B}  \otimes  \SYSTEMnt{C}   )  }$).

Partial borrowing allows safe mutation
of disjoint parts of a
single data structure \emph{concurrently} without risking a data race,
e.g., using \granin{par} for concurrent
threads~\cite{marshall2022replicate}:
\begin{granulep}
par : forall {a b : Type} . (() -> a) -> (() -> b) -> (a, b) -- Granule's parallel composition

exampleConcurrent : *Colour -> *Colour
exampleConcurrent indigo = let (r, p) = push indigo in
                           let (g, b) = push p in
                         let (r', b') = par (\() -> withBorrow alter r) 
                                            (\() -> withBorrow alter b) in
                           let     p' = pull (g, b') in pull (r', p')
\end{granulep}
Building on this core concept in the more expressive setting of Granule's full
type system could involve notions such as borrowing from any algebraic data
type, or borrowing a `slice' of multiple values. One low-hanging fruit is to
extend the \granin{push} and \granin{pull} primitives in order to derive generic
operations for arbitrary data structures, following patterns described in prior
work~\cite{hughes2021deriving}. We do not formalise this again here, but the
approach allows for writing programs like:
\begin{granulep}
exampleDeriving : forall {p : Fraction} . Bp(MaybeColour) -> Either () (BpColour)
exampleDeriving octarine = case (push@Maybe octarine) of Nothing -> Left ();
                                                         Just x  -> Right x
\end{granulep}
Here, the borrow modality is distributed into the \granin{Maybe} data type
using a derived \granin{push} operation. This general pattern for partial
borrowing relates closely to McBride's notion of computing the \emph{derivative}
of a data type by finding its type of one-hole
contexts~\cite{mcbride2001derivative}, where the derivative describes the
structure that remains after borrowing one element. Under this interpretation,
borrowing a slice of multiple values would be akin to taking the derivative of a
data type \emph{with respect to} another type---an idea described elsewhere in recent
work~\cite{marshall2022take}.

\paragraph{\textbf{Parallel sum example}} To conclude this section, we present one last
example which combines all of the elements we have introduced here in a more
practical setting. The result will be a function which splits a unique mutable
array of floats into two immutably borrowed halves and computes their sums
concurrently, returning the total sum inside a polymorphic reference.

First, in order to implement this function elegantly we will need two additional
basic functions for working with arrays and references. Within Granule it is
possible to derive a \granin{writeRef} operation which destructively mutates the
value inside a reference, as long as the type of said value allows for it to be
discarded freely; this makes use of Granule's existing mechanism for deriving a
\granin{drop} operation for this kind of value~\cite{hughes2021deriving}.
Granule also provides an additional primitive for taking the length of an array,
which is useful for iteration but was elided from our formal calculus for
brevity. The types of both of these operations are given below.

\begin{granulep}
lengthFloatArray : forall {id : Name, p : Fraction} . Bp(FloatArrayid) 
                                                -> (!Int, Bp(FloatArrayid))
writeRef         : forall {id : Name, a : Type}     . {Droppable a} => a -> B1(Refida) 
                                                -> B1(Refida)
\end{granulep}

We begin by defining a simple recursive auxiliary function for iterating through
an array of floats and summing all values between two given indices.

\begin{granulep}
sumFromTo : forall {id : Name, p : Fraction} . Bp(FloatArrayid) ->  !Int -> !Int 
                                         -> (Float, Bp(FloatArrayid))
sumFromTo array [i] [n] =
  if i == n then (0.0, array)
    else
      let (x, a) = readFloatArray array i;
          (y, arr) = sumFromTo a [i+1] [n]
      in  (x + y, arr)
\end{granulep}

Finally, we define the core \granin{parSum} function as described above. Note in
particular the usage of \granin{withBorrow} to borrow an initial reference to
the mutable array, \granin{split} and \granin{join} to manage the forking
dataflow when computing the sums for the two separate halves in parallel, and
\granin{push} and \granin{pull} to distribute modalities into and out of
products. A more comprehensive version of this example is presented in the
appendices, including full definitions for all auxiliary functions and an
illustration of how this function could be applied in practice to a freshly
created pair of an array (which is generated from a length-indexed vector) and a
polymorphic reference.

\begin{granulep}
parSum : forall {id id' : Name} . *(FloatArrayid) -> *(Refid'Float) 
                            -> *(Refid'Float,FloatArrayid)
parSum array ref = let
      ([n], array) : (!Int, *(FloatArrayid))     = lengthFloatArray array;
      compIn                                      = pull (ref, array)
  in withBorrow (\compIn ->
                let (ref, array)      = push compIn;
                    (array1, array2)  = split array; -- immutable borrow happens here

            -- Compute in parallel
                    ((x, array1), (y, array2)) =
                                par (\() -> sumFromTo array1 [0] [div n 2])
                                    (\() -> sumFromTo array2 [div n 2] [n]);

            -- Update the reference
                    ref'        = writeRef ((x : Float) + y) ref;
                    compOut     = pull (ref', join (array1, array2)) -- end immutable borrow

                  in compOut) compIn
\end{granulep}

\section{Semantics and Meta-theory}
\label{sec:semantics}
We define an operational semantics here for the calculus which serves to further
explain the details of arrays, mutation, copying, and borrowing.  We adapt the
approach of \citet{choudhury2021graded} and \citet{marshall2022linearity} for
giving an operational semantics to a graded system, with some degree of
accounting for grades and references in order to relate the
dynamic semantics back to the static semantics of typing
(Section~\ref{sec:theorems}). With the exception of Granule
and Multi-Graded Featherweight Java~\cite{DBLP:conf/ecoop/BianchiniDGZ23},
much of the preceding work on operational models for graded systems is based on
call-by-name. Here we opted for call-by-value for the purpose of describing
real-world practical functional languages; call-by-name is prohibitively
expensive with unpredictable performance and poor interaction with side effects.

\subsection{Preliminary definitions}

\paragraph{Values}

We first define the subset of terms that are \emph{values} in the semantics,
i.e., normal forms (terms that have no further reduction), via the grammar:
\begin{align}
  \SYSTEMnt{v} ::= \SYSTEMsym{(}  \SYSTEMnt{v_{{\mathrm{1}}}}  \SYSTEMsym{,}  \SYSTEMnt{v_{{\mathrm{2}}}}  \SYSTEMsym{)} \mid () \mid \SYSTEMsym{[}  \SYSTEMnt{v}  \SYSTEMsym{]} \mid
  \lambda  \SYSTEMmv{x}  .  \SYSTEMnt{t} \mid \SYSTEMmv{n} \mid p \mid \textbf{pack}\ \langle{  \SYSTEMnt{id}  ,  \SYSTEMnt{v}  }\rangle
  \tag{value terms sub-grammar}
\end{align}
including pairs of values, the unit value, boxed values, abstractions, natural
numbers $n$, or primitives $p$ which may also be partially applied to other
values: $\textbf{newArray}$, $\textbf{readArray}$, $\textbf{readArray}  \, \SYSTEMnt{v}$, etc.

\paragraph{Runtime terms and typing}

We extend the syntax of terms (and values) with several runtime representations
which appear only in the semantics, i.e., they cannot be written
by users in programs:
\begin{align*}
  \SYSTEMnt{t} & ::= \ldots \mid \ast  \SYSTEMnt{t} \mid \textbf{unborrow}\  \SYSTEMnt{t} \mid \SYSTEMmv{ref} \mid [ { \SYSTEMnt{t} } ]_{\color{coeffectColor}{ r } } \tag{runtime terms} \\
  \SYSTEMnt{v} & ::= \ldots \mid \ast  \SYSTEMnt{v} \mid \textbf{unborrow}\  \SYSTEMnt{v} \mid \SYSTEMmv{ref}
  \tag{runtime values}
\end{align*}
where $\ast  \SYSTEMnt{t}$ represents unique and borrowed terms, $\textbf{unborrow}\  \SYSTEMnt{t}$ is the inverse to
borrowed terms (used to implement $\textbf{withBorrow}$),
and $\SYSTEMmv{ref}$ are references to resources
bound in the heap. Furthermore, the syntax for promotion is augmented with an annotation
$[ { \SYSTEMnt{t} } ]_{\color{coeffectColor}{ r } }$ of the grade at which the term $t$ is promoted,
i.e., Church-style with respect to grades, and thus typing
this grade-annotated version produces $r  \cdot  \Gamma   \vdash   [ { \SYSTEMnt{t} } ]_{\color{coeffectColor}{ r } }   :   {\textcolor{coeffectColor}{\Box_{ r } } }  \SYSTEMnt{A}$.
This annotation is only needed
to prove that type preservation respects resourcing (as per~\citet{choudhury2021graded})
and can be ignored when
 actually computing the reduction behaviour of a term using the operational model.

In order to type runtime terms with references,
the syntax of contexts is extended to include
assumptions $\SYSTEMmv{ref}  :   \SYSTEMnt{Res} _{ \SYSTEMnt{id} } \  \SYSTEMnt{A}$ which are treated
as a different syntactic category of variables.
A \emph{runtime context} $\gamma$ is a context containing only references, i.e.:
$ \gamma ::= \emptyset \mid \gamma ,   \SYSTEMmv{ref}  :   \SYSTEMnt{Res} _{ \SYSTEMnt{id} } \  \SYSTEMnt{A} $.

Context scalar multiplication and addition extend as follows (eliding
a symmetric case for brevity):
\begin{align*}
\tag{runtime context addition}
  \SYSTEMsym{(}   \Gamma ,   \SYSTEMmv{ref}  :   \SYSTEMnt{Res} _{ \SYSTEMnt{id} } \  \SYSTEMnt{A}     \SYSTEMsym{)}  +  \SYSTEMsym{(}   \Gamma' ,   \SYSTEMmv{ref}  :   \SYSTEMnt{Res} _{ \SYSTEMnt{id} } \  \SYSTEMnt{A}     \SYSTEMsym{)}
  & = \SYSTEMsym{(}   \Gamma  +  \Gamma'   \SYSTEMsym{)} ,   \SYSTEMmv{ref}  :   \SYSTEMnt{Res} _{ \SYSTEMnt{id} } \  \SYSTEMnt{A} \\
  \SYSTEMsym{(}   \Gamma ,   \SYSTEMmv{ref}  :   \SYSTEMnt{Res} _{ \SYSTEMnt{id} } \  \SYSTEMnt{A}     \SYSTEMsym{)}  +  \Gamma'
  & = \SYSTEMsym{(}   \Gamma  +  \Gamma'   \SYSTEMsym{)} ,   \SYSTEMmv{ref}  :   \SYSTEMnt{Res} _{ \SYSTEMnt{id} } \  \SYSTEMnt{A} \; (\textit{where}\ \SYSTEMmv{ref} \not\in \Gamma') \\
r  \cdot  \SYSTEMsym{(}   \Gamma ,   \SYSTEMmv{ref}  :   \SYSTEMnt{Res} _{ \SYSTEMnt{id} } \  \SYSTEMnt{A}     \SYSTEMsym{)} & = \SYSTEMsym{(}   r  \cdot  \Gamma   \SYSTEMsym{)} ,   \SYSTEMmv{ref}  :   \SYSTEMnt{Res} _{ \SYSTEMnt{id} } \  \SYSTEMnt{A}
\tag{runtime context multiplication}
\end{align*}
Importantly, references are not treated linearly here since they may be shared
in the runtime context (such as via the \textbf{split} operation). Runtime terms
are typed by the following rules:
\begin{align*}
  \begin{array}{c}
  \SYSTEMdruleRTynec{}
  \;
  \SYSTEMdruleRTyunborrow{}
  \;
  \SYSTEMdruleRTyres{}
  %\\
  %\SYSTEMdruleRTyarrInit{} \quad
  %\SYSTEMdruleRTyarrAt{} \quad
  %\SYSTEMdruleRTyrefStore{} %\quad
%  \SYSTEMdruleRTypr{}
  \end{array}
\end{align*}
\paragraph{Heaps and configurations}

Heaps map program variables to values (our semantics does not
use syntactic substitution), and also map references to identifiers $\SYSTEMnt{id}$ and
identifiers to values:
\begin{align}
  H & ::= \emptyset \mid H ,   \SYSTEMmv{x}  \textcolor{coeffectColor}{\mapsto_{ r } }  \SYSTEMnt{v} \mid H ,   \SYSTEMmv{ref}  \textcolor{uniqueColor}{\mapsto_{ p } }  \SYSTEMnt{id}
  \mid H ,   \SYSTEMnt{id}  \mapsto  v_r
  \tag{heaps} \\
  v_r & ::= \textbf{arr} \mid \mathbf{ref} (  \SYSTEMnt{v}  )
  \qquad \textbf{arr} ::= \mathsf{init} \mid \textbf{arr}  [  \SYSTEMmv{n}  ] =  \SYSTEMnt{v}
\tag{heap resource terms}
\end{align}
Thus, a heap can be extended in three ways:
(1) with an assignment of a program variable $x$ to a value $\SYSTEMnt{v}$,
storing the grade of the variable ${\textcolor{coeffectColor}{r}}$
(which comes from the typing);
(2) with an assignment of a reference $\SYSTEMmv{ref}$ to an identifier
$id$ with permission ${\textcolor{uniqueColor}{p}}$;
(3) with an assignment of an identifier $id$ to a
resource term. Identifiers and references are separated since we may
need multiple references to point to the same identifier, e.g., in the case of immutable borrows
(one can think of identifiers like abstract memory addresses in the semantics).

Array terms $\textbf{arr}$ in the heap are either
an empty array $\mathsf{init}$ or an array storing value $\SYSTEMnt{v}$ at
index $\SYSTEMmv{n}$. Reference values $\mathbf{ref} (  \SYSTEMnt{v}  )$ in the heap store a value $\SYSTEMnt{v}$.
These runtime heap terms are typed according to straightforward rules presented
in Appendix A.

A \emph{configuration} comprises a pair of a \emph{heap} $H$
and a term $\SYSTEMnt{t}$, written as $H  \vdash  \SYSTEMnt{t}$, where
$\mathsf{fv}(v) \subseteq \mathsf{dom}(H)$, and $\mathsf{refs}  (  \SYSTEMnt{v}  ) \subseteq \mathsf{dom}(H)$ (where $\mathsf{refs}$ is the set of all resource references $\SYSTEMmv{ref}$ in the
term).

\subsection{Single-step reduction}

Single-step reductions map source
configurations to target configurations, with the judgment:
\begin{align*}
  H_{{\mathrm{1}}}  \vdash  \SYSTEMnt{t_{{\mathrm{1}}}}  \,\leadsto_{ s } \,  H_{{\mathrm{2}}}  \vdash  \SYSTEMnt{t_{{\mathrm{2}}}}
\end{align*}
where $H_{{\mathrm{1}}}$ and $H_{{\mathrm{2}}}$ are input and output
heaps respectively, $\SYSTEMnt{t_{{\mathrm{1}}}}$ is the source term
and $\SYSTEMnt{t_{{\mathrm{2}}}}$ the target. The grade $s$ denotes
the \emph{usage context} of this rule.
We explain the reduction rules for our operational
semantics in detail. Throughout, $x \# \overline{t}$ means that
$x$ is a fresh name with respect to some terms $\overline{t}$.

\paragraph{Lambda calculus}

The $\lambda$-calculus core of the operational semantics has rules:
\begin{align*}
  \begin{array}{c}
  \SYSTEMdruleHeapvar{}
  \quad
  \SYSTEMdruleHeapbeta{}
  \\[1.5em]
  \SYSTEMdruleHeapappL{}
  \quad
  \SYSTEMdruleHeapappR{}
  \end{array}
\end{align*}
The rules for function application are fairly standard. In the
\SYSTEMdruleHeapvarName{} rule, a variable $x$ is reduced to a value $v$ which
was assigned to $x$ in the heap. The annotation $r$ is preserved in the output
heap, with the side condition in the premise ensuring that the grade $r$ will be
enough to capture the usage $s$ required by the reduction. In the
\SYSTEMdruleHeapbetaName{} rule, rather than using a substitution to enact
$\beta$-reduction, the resulting term is the body of the function $t$ with the heap
extended with a $y$ assigned to the argument value $v$. The grade $s$ (parameterising the reduction) annotates $x$ in the heap.

\paragraph{Existential types and names}

The semantics of existentials is standard, with a beta
rule:
\begin{align*}
  \SYSTEMdruleHeapexistentialBeta{}
\end{align*}
and two standard congruence rules for $\mathsf{pack}$ and $\mathsf{unpack}$
(elided for brevity).

\paragraph{Tensors and units} Tensor products have the following rules
for their introduction and elimination forms, with three congruence
rules and one $\beta$-rule:
 \begin{gather*}
  \begin{array}{c}
    \SYSTEMdruleHeapcongPairL{}
    \;
    \SYSTEMdruleHeapcongPairElim{}
    \\[1.5em]
    \SYSTEMdruleHeapcongPairR{}
    \;
    \SYSTEMdruleHeappairBeta{}
  \end{array}
\end{gather*}
In the case of (\SYSTEMdruleHeappairBetaName{}), we extend
the heap with assignments for (fresh names) $x'$ and $y'$ to $\SYSTEMnt{v_{{\mathrm{1}}}}$
and $\SYSTEMnt{v_{{\mathrm{2}}}}$ respectively, continuing on with the body term
$\SYSTEMnt{t}$. Similarly to the $\beta$-rule for functions, the
freshening avoids variable capture.
We elide the rules for the unit type as they are similar.

\begin{example}
  The following
  gives a reduction sequence for the term
  $\SYSTEMsym{(}   \lambda  \SYSTEMmv{x}  .  \SYSTEMsym{(}  \SYSTEMmv{x}  \SYSTEMsym{,}  \SYSTEMmv{y}  \SYSTEMsym{)}   \SYSTEMsym{)} \, \SYSTEMsym{(}  \SYSTEMnt{v}  \SYSTEMsym{,}  \SYSTEMmv{y}  \SYSTEMsym{)}$
  under a heap $H = \SYSTEMmv{y}  \textcolor{coeffectColor}{\mapsto_{ \SYSTEMsym{2} } }  \SYSTEMnt{v}$,
  until a normal form is reached:

  {\scalebox{0.95}{
  \hspace{-1em}\begin{minipage}{1\linewidth}
  \begin{align*}
    \setlength{\arraycolsep}{0.1em}
    \begin{array}{rll}
             & \SYSTEMmv{y}  \textcolor{coeffectColor}{\mapsto_{ \SYSTEMsym{2} } }  \SYSTEMnt{v}    \vdash  \SYSTEMsym{(}   \lambda  \SYSTEMmv{x}  .  \SYSTEMsym{(}  \SYSTEMmv{x}  \SYSTEMsym{,}  \SYSTEMmv{y}  \SYSTEMsym{)}   \SYSTEMsym{)} \, \SYSTEMsym{(}  \SYSTEMnt{v}  \SYSTEMsym{,}  \SYSTEMmv{y}  \SYSTEMsym{)} & \\
    (\textit{\SYSTEMdruleHeapappLName{},\SYSTEMdruleHeapcongPairRName{},\SYSTEMdruleHeapvarName{}})
    \leadsto_1 & \SYSTEMmv{y}  \textcolor{coeffectColor}{\mapsto_{ \SYSTEMsym{2} } }  \SYSTEMnt{v}    \vdash  \SYSTEMsym{(}   \lambda  \SYSTEMmv{x}  .  \SYSTEMsym{(}  \SYSTEMmv{x}  \SYSTEMsym{,}  \SYSTEMmv{y}  \SYSTEMsym{)}   \SYSTEMsym{)} \, \SYSTEMsym{(}  \SYSTEMnt{v}  \SYSTEMsym{,}  \SYSTEMnt{v}  \SYSTEMsym{)} & \\
    (\textsc{\SYSTEMdruleHeapbetaName{}})
    \leadsto_1 & \SYSTEMmv{y}  \textcolor{coeffectColor}{\mapsto_{ \SYSTEMsym{2} } }  \SYSTEMnt{v}   ,   \SYSTEMmv{x}  \textcolor{coeffectColor}{\mapsto_{ \SYSTEMsym{1} } }  \SYSTEMsym{(}  \SYSTEMnt{v}  \SYSTEMsym{,}  \SYSTEMnt{v}  \SYSTEMsym{)}    \vdash  \SYSTEMsym{(}  \SYSTEMmv{x}  \SYSTEMsym{,}  \SYSTEMmv{y}  \SYSTEMsym{)} & \\
    (\textsc{\SYSTEMdruleHeapappLName{},\SYSTEMdruleHeapcongPairLName{},\SYSTEMdruleHeapvarName{}})
    \leadsto_1 & \SYSTEMmv{y}  \textcolor{coeffectColor}{\mapsto_{ \SYSTEMsym{2} } }  \SYSTEMnt{v}   ,   \SYSTEMmv{x}  \textcolor{coeffectColor}{\mapsto_{ \SYSTEMsym{1} } }  \SYSTEMsym{(}  \SYSTEMnt{v}  \SYSTEMsym{,}  \SYSTEMnt{v}  \SYSTEMsym{)}    \vdash  \SYSTEMsym{(}  \SYSTEMsym{(}  \SYSTEMnt{v}  \SYSTEMsym{,}  \SYSTEMnt{v}  \SYSTEMsym{)}  \SYSTEMsym{,}  \SYSTEMmv{y}  \SYSTEMsym{)} &
    \end{array}
  \end{align*}
  \end{minipage}}}
\end{example}

\paragraph{Graded modalities} The rules for graded modalities are structured similarly to the standard lambda
calculus rules above, but we need
to do some additional management of grades in the heap.
\begin{gather*}
  \begin{array}{c}
    \SYSTEMdruleHeapcongPromotion{}
    \quad
    \SYSTEMdruleHeapcongBoxElim{}
    \\[1.25em]
    \SYSTEMdruleHeapbetaBox{}
  \end{array}
\end{gather*}
In the \SYSTEMdruleHeapcongPromotionName{} rule, to construct a reduction with the
required grade $s$ then we need to be able to reduce inside the box at grade
$s  *  r$, to account for the additional usage required by the modality's $r$
grade. In the \SYSTEMdruleHeapbetaBoxName{} rule, the $x$ in the heap is
annotated not only with $s$ as in the regular \SYSTEMdruleHeapbetaName{} rule
but with $s  *  r$, again to account for the additional usage the modality
requires.

\paragraph{Arrays}

In the body of the paper we present only the reduction rules for arrays, for
brevity; the corresponding (and very similar) rules for polymorphic references
are included in the appendix.
Note that we do not explicitly track the sizes of arrays in our semantics, for
simplicity. New arrays are created by the $\textbf{newArray}$
primitive, with reduction rule:
\begin{align*}
\SYSTEMdruleHeapnewArray{}
\end{align*}
where $\SYSTEMmv{ref}$ and $id$ are fresh for the heap
$H$. The resulting array is initialised to the empty
array ($\mathsf{init}$), and the result is a unique array reference
$\ast  \SYSTEMmv{ref}$. In the heap, $\SYSTEMmv{ref}$ is marked with the whole
permission $1$. Arrays are read and written via:
\begin{gather*}
  \SYSTEMdruleHeapreadArray{} \\
  \SYSTEMdruleHeapwriteArray{}
\end{gather*}
For $\textbf{writeArray}$, $p$ should be $\SYSTEMsym{1}$ or $\ast$, but this is mediated by the
type system rather than being enforced in the semantics; the following rule is
similar in this regard for deleting unique arrays:
\begin{align*}
  \SYSTEMdruleHeapdeleteArray{}
\end{align*}
\paragraph{Sharing and cloning}
Sharing reduces permissions to $0$ and cloning involves copying heap terms:
\begin{align*}
  \begin{array}{c}
    \SYSTEMdruleHeapcongShare{}
    \quad
    \SYSTEMdruleHeapshare{}
    \\[1.25em]
    \SYSTEMdruleHeapcopyBeta{}
  \end{array}
\end{align*}
In the \SYSTEMdruleHeapshareName{} rule, the incoming heap is split into two
parts, where $H$ is such that it provides the allocations for all resource
references in $\SYSTEMnt{v}$ (enforced by the premise). The unique value $\ast  \SYSTEMnt{v}$ is
wrapped in the graded box modality in the result as $\SYSTEMsym{[}  \SYSTEMnt{v}  \SYSTEMsym{]}$, and thus all
its references are now annotated with $\SYSTEMsym{0}$ in the heap via $\SYSTEMsym{(}   { {[  H  ]}_{\textcolor{uniqueColor}{ \SYSTEMsym{0} } } }   \SYSTEMsym{)}$, e.g.:
\begin{align*}
  H' ,   \SYSTEMnt{id}  \mapsto  v_r   ,   \SYSTEMmv{ref}  \textcolor{uniqueColor}{\mapsto_{ \SYSTEMsym{1} } }  \SYSTEMnt{id}    \vdash  \SYSTEMkw{share} \, \SYSTEMsym{(}  \ast  \SYSTEMmv{ref}  \SYSTEMsym{)}  \leadsto    H' ,   \SYSTEMnt{id}  \mapsto  v_r   ,   \SYSTEMmv{ref}  \textcolor{uniqueColor}{\mapsto_{ \SYSTEMsym{0} } }  \SYSTEMnt{id}    \vdash  \SYSTEMsym{[}  \SYSTEMmv{ref}  \SYSTEMsym{]}  \mid   \emptyset
\end{align*}
The \SYSTEMdruleHeapcopyBetaName{} rule enacts a `deep copy', where $\mathsf{dom}  (  H'  )   \equiv    \mathsf{refs}  (  \SYSTEMnt{v}  )$ marks the part of the heap with resource references coming from $\SYSTEMnt{v}$. Then
$\mathsf{copy}(  H'  )$ copies the resources in this part of the heap, creating a heap
fragment $H''$ and a renaming operator $\theta$ which maps from old
references to new copied references. This renaming is applied to $\SYSTEMnt{v}$ in the
freshly bound unique variable $x$, such that the value $\ast  \SYSTEMsym{(}   \theta( \SYSTEMnt{v} )   \SYSTEMsym{)}$
refers to any newly copied resources. Lastly, we pack the renamed unique value
with new identifiers $\overline{ \SYSTEMnt{id} }$ generated by $\mathsf{copy}$.

We elide the straightforward congruence rule for
$\textbf{clone}$.

\paragraph{Borrowing}

We elide the congruence rules for $\textbf{withBorrow}$ which
ensure that we reduce the two argument terms left to right
until they are values, after which we can reduce as follows:
\begin{align*}
  \begin{array}{c}
    \SYSTEMdruleHeapwithBorrow{}
  \end{array}
\end{align*}
Here, the $\beta$-reduction that comes from applying the function $\lambda  \SYSTEMmv{x}  .  \SYSTEMnt{t}$ to
the value $\SYSTEMnt{v}$ is enacted as a substitution, with the resulting value being
wrapped inside the uniqueness wrapper as a representation of the fact that within
the context of $\textbf{withBorrow}$ it is now a borrowed term.

For a term to escape $\textbf{withBorrow}$ it must eventually be `unborrowed';
the runtime term $\textbf{unborrow}$ encapsulating the resulting term represents
this idea, and obeys the following rules:
\begin{align*}
  \begin{array}{c}
    \SYSTEMdruleHeapcongUnborrow{} \quad
    \SYSTEMdruleHeapunborrowBorrow{}
  \end{array}
\end{align*}
Aside from the congruence, which is standard, the
\SYSTEMdruleHeapunborrowBorrowName{} rule simply unwraps the value from the
$\with$ modality, allowing the semantics to treat said value as a unique term
once more.

\paragraph{Push and pull}
The reduction rules for $\textbf{push}$ and $\textbf{pull}$ are fairly simple;
they simply distribute the modality into or out of the product term in a way
that matches the typing of the given rule.
\begin{align*}
  \begin{array}{c}
    \SYSTEMdruleHeappushUnique{}
    \quad
    \SYSTEMdruleHeappullUnique{}
  \end{array}
\end{align*}
The heap is left unchanged. The above rules are for unique terms; there are two
equivalent rules for $\textbf{push}$ and $\textbf{pull}$ on borrowed terms, but
these are identical aside from the modality on the term.

Push and pull also have congruence rules which are straightforward
and elided.

\paragraph{Split and join}
Join and split also have congruence rules which are straightforward
and elided. There are two primary reduction rules for each of $\textbf{split}$ and
$\textbf{join}$: one for the case where the terms are references and one
for the case where the terms are pairs of values:
\renewcommand{\SYSTEMdrule}[4][]{{\displaystyle\inferrule*[right=\,\SYSTEMdrulename{#4}]{\begin{array}{l}#2\end{array}}{#3}}}
\begin{gather*}
  \begin{array}{c}
    \SYSTEMdruleHeapsplitRef{}
    \\[1em]
    \SYSTEMdruleHeapjoinRef{}
    \\[1em]
    \SYSTEMdruleHeapsplitPair{}
    \SYSTEMdruleHeapjoinPair{}
  \end{array}
\end{gather*}
\renewcommand{\SYSTEMdrule}[4][]{{\displaystyle\frac{\begin{array}{l}#2\end{array}}{#3}\quad\SYSTEMdrulename{#4}}}
In the reference case, $\textbf{split}$ removes the initial reference from the
heap and generates two fresh references pointing to the same identifier. These
are each annotated with half of the permission belonging to the original
reference, to match the typing. As one might expect, $\textbf{join}$ for
references behaves dually; it deletes two existing references from the heap, and
generates one fresh reference, with its permission being the sum of the
constituent parts.

For pairs, the reduction rules are defined inductively:
as long as we can split or join on the two components of the pair, we are
allowed to reduce on the overall pair itself. In this way, we construct a
reduction which ensures that all references contained within the pair are split
or joined as required, no matter how deeply nested the pair may be.

\paragraph{Multi-step reductions}

Lastly, we define a relation that composes
 single-step reductions into a sequence of reductions, called a multi-reduction:
\begin{align*}
\SYSTEMdruleHeapMultirefl{}
\quad
\SYSTEMdruleHeapMultiext{}
\end{align*}

\subsection{Theorems}
\label{sec:theorems}
We now consider the relationship between typing and the operational
semantics. First, we must build a foundation via
checking some preliminary results and noting some crucial definitions.

Two useful results, which extend those found elsewhere in
the literature, are that substitution is admissible for our
calculus, coming in both linear and graded variants:

\begin{lemma}[Linear substitution is admissible, extending~\cite{orchard2019quantitative}]
\label{lemma:linear-subst}
  
  If $\Gamma_{{\mathrm{1}}}  \vdash  \SYSTEMnt{t_{{\mathrm{1}}}}  :  \SYSTEMnt{A}$ and $\Gamma_{{\mathrm{2}}} ,   \SYSTEMmv{x}  :  \SYSTEMnt{A}    \vdash  \SYSTEMnt{t_{{\mathrm{2}}}}  :  \SYSTEMnt{B}$ then
  $\Gamma_{{\mathrm{2}}}  +  \Gamma_{{\mathrm{1}}}   \vdash   \SYSTEMnt{t_{{\mathrm{2}}}}  [  \SYSTEMnt{t_{{\mathrm{1}}}}  /  \SYSTEMmv{x}  ]   :  \SYSTEMnt{B}$.
\end{lemma}

\begin{lemma}[Graded substitution is admissible, extending~\cite{orchard2019quantitative}]

  If $\textcolor{coeffectColor}{[  \Gamma_{{\mathrm{1}}}  ]}   \vdash  \SYSTEMnt{t_{{\mathrm{1}}}}  :  \SYSTEMnt{A}$ and $\Gamma_{{\mathrm{2}}} ,   \textcolor{coeffectColor}{ \SYSTEMmv{x}  : [  \SYSTEMnt{A}  ]_{ r } }    \vdash  \SYSTEMnt{t_{{\mathrm{2}}}}  :  \SYSTEMnt{B}$ (where $\textcolor{coeffectColor}{[  \Gamma_{{\mathrm{1}}}  ]}$ represents a context $\Gamma_{{\mathrm{1}}}$ containing only graded assumptions) and
  $\neg \mathsf{resourceAllocator}(  \SYSTEMnt{t_{{\mathrm{1}}}}  )$ then $\Gamma_{{\mathrm{2}}}  +   r  \cdot  \Gamma_{{\mathrm{1}}}    \vdash   \SYSTEMnt{t_{{\mathrm{2}}}}  [  \SYSTEMnt{t_{{\mathrm{1}}}}  /  \SYSTEMmv{x}  ]   :  \SYSTEMnt{B}$.
\end{lemma}

\paragraph{Type safety}

Key to ensuring type safety is the notion of \emph{heap compatibility with a
typing context}.

\begin{definition}[Heap compatibility]
A heap $H$ is compatible with free variable context $\Gamma$,
 denoted $H  \bowtie  \Gamma$, if the grades in the heap match those of the
 context, and any values stored in the heap have their resources
 accounted for in the rest of the heap. The relation is defined
 inductively over the syntax of heaps and contexts:
\begin{align*}
\begin{array}{c}
\SYSTEMdruleHeapCtxtCompatbase{}
\;\;
\SYSTEMdruleHeapCtxtCompatextRes{}
\;\;
\SYSTEMdruleHeapCtxtCompatgcArr{}
\end{array}
\end{align*}
\begin{align*}
\begin{array}{c}
\SYSTEMdruleHeapCtxtCompatextGr{}
\\[1.25em]
\SYSTEMdruleHeapCtxtCompatextLin{}
\end{array}
\end{align*}
Thus, a context extended with a runtime type of
a resource reference (\SYSTEMdruleHeapCtxtCompatextResName{})
is compatible with a heap which contains that resource reference, pointing to
some resource term with the corresponding identifier that the resource
points to. In the premise of (\SYSTEMdruleHeapCtxtCompatextResName{}), the resource term with its identifier
is preserved in the heap since there may be other references
pointing to it (e.g., generated from a \textsf{split}).
The (\SYSTEMdruleHeapCtxtCompatgcArrName{}) rule then allows
heap compatibility to `garbage collect' any remaining resources.

The (\SYSTEMdruleHeapCtxtCompatextGrName{}) rule says
that a context with graded assumption $\textcolor{coeffectColor}{ \SYSTEMmv{x}  : [  \SYSTEMnt{A}  ]_{ s } }$
is compatible with a heap as long as the heap contains
a binding for $x$ to some value $v$ and
heap grade $r$ that can accommodate the usage
of $s$ (via the constraint $\exists  r'  .\    s  +  r'   \equiv  r$)
and as long as the free variables $\Gamma'$ of $\SYSTEMnt{v}$
are also compatible with the heap (scaled by $s$ to reflect
the usage of $x$). The (\SYSTEMdruleHeapCtxtCompatextLinName{}) rule
is similar to (\SYSTEMdruleHeapCtxtCompatextGrName{}), but \emph{effectively} where
$s = 1$; the variable $x$ is used in a linear fashion.
\end{definition}

\begin{example}
As an illustrative example, the context $\textcolor{coeffectColor}{ \SYSTEMmv{x}  : [  \SYSTEMnt{A}  ]_{ \SYSTEMsym{1} } }  ,   \textcolor{coeffectColor}{ \SYSTEMmv{y}  : [  \SYSTEMnt{B}  ]_{ \SYSTEMsym{2} } }$
is compatible with the heap $\SYSTEMmv{x}  \textcolor{coeffectColor}{\mapsto_{ \SYSTEMsym{7} } }  \SYSTEMnt{v_{{\mathrm{1}}}}   ,   \SYSTEMmv{y}  \textcolor{coeffectColor}{\mapsto_{ \SYSTEMsym{2} } }  \SYSTEMnt{v_{{\mathrm{2}}}}$
assuming the typing $\textcolor{coeffectColor}{ \SYSTEMmv{x}  : [  \SYSTEMnt{A}  ]_{ \SYSTEMsym{3} } }   \vdash  \SYSTEMnt{v_{{\mathrm{2}}}}  :  \SYSTEMnt{B}$, with heap compatibility derivation:
\begin{align*}
\vspace{-1em}
\inferrule*[right=\SYSTEMdruleHeapCtxtCompatextGrName{}]
 {
    \inferrule*[right=\SYSTEMdruleHeapCtxtCompatextGrName{}]
      {
         \inferrule*[right=\SYSTEMdruleHeapCtxtCompatbaseName{}]
          {\quad}{\emptyset   \bowtie   \emptyset}
       \quad
     }{\emptyset  ,   \SYSTEMmv{x}  \textcolor{coeffectColor}{\mapsto_{ \SYSTEMsym{7} } }  \SYSTEMnt{v}    \bowtie   \textcolor{coeffectColor}{ \SYSTEMmv{x}  : [  \SYSTEMnt{A}  ]_{ \SYSTEMsym{7} } }}
     \quad
    \textcolor{coeffectColor}{ \SYSTEMmv{x}  : [  \SYSTEMnt{A}  ]_{ \SYSTEMsym{3} } }   \vdash  \SYSTEMnt{v_{{\mathrm{2}}}}  :  \SYSTEMnt{B}
 }
 {\SYSTEMsym{(}     \emptyset  ,   \SYSTEMmv{x}  \textcolor{coeffectColor}{\mapsto_{ \SYSTEMsym{7} } }  \SYSTEMnt{v_{{\mathrm{1}}}}   ,   \SYSTEMmv{y}  \textcolor{coeffectColor}{\mapsto_{ \SYSTEMsym{2} } }  \SYSTEMnt{v_{{\mathrm{2}}}}    \SYSTEMsym{)}  \bowtie  \SYSTEMsym{(}     \textcolor{coeffectColor}{ \SYSTEMmv{x}  : [  \SYSTEMnt{A}  ]_{ \SYSTEMsym{1} } }  ,   \textcolor{coeffectColor}{ \SYSTEMmv{y}  : [  \SYSTEMnt{B}  ]_{ \SYSTEMsym{2} } }     \SYSTEMsym{)}}
\end{align*}
\end{example}
Heap compatibility allows us to establish key properties for our calculus in
conjunction with the operational semantics. We begin with syntactic type safety,
by verifying progress and preservation. These are largely standard, though
preservation links with heap compatibility in its second conjunct.

\begin{theorem}[Progress]
    \label{lemma:progress}
    Given $\Gamma  \vdash  \SYSTEMnt{t}  :  \SYSTEMnt{A}$, then $\SYSTEMnt{t}$ is either a value, or for all grades $s$ and contexts $\Gamma_{{\mathrm{0}}}$
    then if $H  \bowtie   \Gamma_{{\mathrm{0}}}  +   s  \cdot  \Gamma$ there exists a heap $H'$ and term $\SYSTEMnt{t'}$
    such that $H  \vdash  \SYSTEMnt{t}  \,\leadsto_{ s } \,  H'  \vdash  \SYSTEMnt{t'}$.
\end{theorem}

\begin{theorem}[Type Preservation]
\label{thm:type-preservation}
For a well-typed term
% TYPING
$\Gamma  \vdash  \SYSTEMnt{t}  :  \SYSTEMnt{A}$,
under a restriction that reference resources are restricted
to non-function types,
and all $s$, $\Gamma_{{\mathrm{0}}}$, and $H$ such that
% HEAP COMPAT
$H  \bowtie  \SYSTEMsym{(}   \Gamma_{{\mathrm{0}}}  +   s  \cdot  \Gamma    \SYSTEMsym{)}$
% REDUCTION
and a reduction
$H  \vdash  \SYSTEMnt{t}  \,\leadsto_{ s } \,  H'  \vdash  \SYSTEMnt{t'}$
we have:
\begin{equation*}
  \exists \Gamma', H' . \;\;
% TYPING
          \Gamma'  \vdash  \SYSTEMnt{t'}  :  \SYSTEMnt{A}
          \;\ \wedge\;\
          H'  \bowtie  \SYSTEMsym{(}   \Gamma_{{\mathrm{0}}}  +   s  \cdot  \Gamma'    \SYSTEMsym{)}
\end{equation*}
\end{theorem}
Note the caveat to preservation: references $\mathsf{Ref}  _{ \SYSTEMnt{id} } \  \SYSTEMnt{A}$ are restricted such
that $\SYSTEMnt{A}$ cannot be of function type, or some other composite type
involving functions. The restriction is needed for preservation since it works
at the granularity of a single reduction, and so cannot rule out the possibility
that a reference is storing a $\lambda$ term with free variables. This
considerably complicates reasoning about resources and heaps, so we rule it out
for this theorem. Importantly, this is not a restriction that needs to be made
on the calculus and its implementation as a whole: for deterministic CBV
reduction starting from a closed term (i.e., a complete program) then all beta
redexes are on closed values and hence this problem does not exist in the
context of an overall reduction sequence. However, to make preservation work for
a single reduction, on potentially open terms, this minor restriction is needed
locally. This does not affect any of the examples discussed in the paper.

In addition to preservation, \citet{marshall2022linearity} also proved a
\emph{conservation} property, ensuring that resource usage is respected (in
particular, that resource usage accrued in a given reduction plus remaining
resources in the resulting heap are approximated by the resources in the
original heap plus the specified resource usage from any variable bindings
encountered along the way). Their semantics was call-by-name, which is more
natural for conservation; checking that a modified form of conservation still
holds in our setting is future work, though various other graded type systems
with similar properties are also call-by-value~\cite{orchard2019quantitative,
10.1145/3563319}.

We now move on to establishing perhaps the most interesting properties here,
which are related to the safety of ownership and borrowing and how these notions
relate to the references present in the heap. First, we ensure that taking a
single step preserves the total of fractional permision annotations on references (unless
we have stopped tracking ownership information for the given reference, in which
case it should be annotated with $0$):
\begin{lemma}[Borrow safety]
    \label{thm:single-step-uniqueness}
    For a well-typed term $\Gamma  \vdash  \SYSTEMnt{t}  :  \SYSTEMnt{A}$ and all $\Gamma_{{\mathrm{0}}}$ and heaps $H$ such that $H  \bowtie  \SYSTEMsym{(}   \Gamma_{{\mathrm{0}}}  +   s  \cdot  \Gamma    \SYSTEMsym{)}$, and given a single-step reduction $H  \vdash  \SYSTEMnt{t}  \,\leadsto_{ s } \,  H'  \vdash  \SYSTEMnt{t'}$ then for all $\SYSTEMnt{id} \in \mathsf{dom}  (  H  )$:
    \begin{align*}
       \sum_{\footnotesize{\begin{array}{c}\forall \SYSTEMmv{ref} \in \mathsf{refs}  (  \SYSTEMnt{t}  ) . \\
                                  \SYSTEMmv{ref}  \textcolor{uniqueColor}{\mapsto_{ p } }  \SYSTEMnt{id} \in H\end{array}}}\hspace{-2em}  p \;\;= 1
      \quad\implies
       \sum_{\footnotesize{\begin{array}{c}\forall \SYSTEMmv{ref} \in \mathsf{refs}  (  \SYSTEMnt{t'}  ) . \\ \SYSTEMmv{ref}  \textcolor{uniqueColor}{\mapsto_{ p' } }  \SYSTEMnt{id} \in H'\end{array}}} \hspace{-2em} p' \;\; \in \{0, 1\}
    \end{align*}
    i.e., for all resources with identifier $\SYSTEMnt{id}$ in the incoming
    heap and all references in the term pointing to this resource, if the
    sum of all permissions pointing to this resource are $1$
    in the incoming heap then
    either this is preserved in the outgoing heap \emph{or}
    the total permissions in the output heap is $0$, i.e., this resource
    has now been fully shared and has no ownership tracking now.

    Furthermore, any resources in the outgoing heap
    that did not appear in the initial heap
    with references in the final term should have
    permissions summing to $1$. That is, for all
    $\SYSTEMnt{id'} \in \mathsf{dom}  (  H'  ) \wedge \SYSTEMnt{id'} \not\in \mathsf{dom}( H )$:
    \begin{align*}
              \sum_{\footnotesize{\begin{array}{c}
                    \forall \SYSTEMmv{ref'} \in \mathsf{refs}  (  \SYSTEMnt{t'}  ) . \\
                           \SYSTEMmv{ref'}  \textcolor{uniqueColor}{\mapsto_{ q } }  \SYSTEMnt{id'} \in H'
                \end{array}}} \hspace{-2em} q \;\; = 1
    \end{align*}
\end{lemma}

When we extend this result to multi-step reductions, we are able to restrict the
result further, and prove that if the resulting term is of unique type, then not
only is the sum of fractions preserved throughout the overall reduction (which
we verify inductively using the above result) but also that in the final term the
reference to the resource we are considering must itself be unique.
\begin{theorem}[Multi-reduction borrow safety]
    \label{thm:single-step-uniqueness-multi}
    For a well-typed term $\Gamma  \vdash  \SYSTEMnt{t}  :  \SYSTEMnt{A}$ and all $\Gamma_{{\mathrm{0}}}$
    and $H$ such that $H  \bowtie  \SYSTEMsym{(}   \Gamma_{{\mathrm{0}}}  +   s  \cdot  \Gamma    \SYSTEMsym{)}$, and multi-step
    reduction $H  \vdash  \SYSTEMnt{t} \ \Rightarrow_{ s } \  H'  \vdash  \SYSTEMnt{v}$, then for all
    $\SYSTEMnt{id} \in \mathsf{dom}  (  H  )$:
    \begin{align*}
      \sum_{\footnotesize{\begin{array}{c}
          \forall
              \SYSTEMmv{ref} \in \mathsf{refs}  (  \SYSTEMnt{t}  ) . \\
              \SYSTEMmv{ref}  \textcolor{uniqueColor}{\mapsto_{ p } }  \SYSTEMnt{id} \in H
           \end{array}}} \hspace{-2em} p \;\; = 1
      \implies
      \exists! \SYSTEMmv{ref'} . \SYSTEMmv{ref'}  \textcolor{uniqueColor}{\mapsto_{ \SYSTEMsym{1} } }  \SYSTEMnt{id} \in H'
    \end{align*}
    i.e., for all resources with identifier $\SYSTEMnt{id}$ in the incoming
    heap and all references in the term pointing to this resource, if the
    sum of all permissions pointing to this resource are $1$
    in the incoming heap then
    their total permission of $1$ is preserved from the incoming heap to the
    resulting term, with this permission now contained in a single reference
    $\SYSTEMmv{ref'}$.

    Furthermore, any new references
    in the final term should uniquely point to an identifier,
    and thus have permission $1$. That is, for all $\SYSTEMnt{id'} \in \mathsf{dom}  (  H'  )
    \wedge \SYSTEMnt{id'} \not\in \mathsf{dom}  (  H  )$ then:
    \begin{align*}
      \forall \SYSTEMmv{ref} \in \mathsf{refs}  (  \SYSTEMnt{v}  ) . \;
      \exists! \SYSTEMmv{ref'} . \;\, \SYSTEMmv{ref'}  \textcolor{uniqueColor}{\mapsto_{ \SYSTEMsym{1} } }  \SYSTEMnt{id'} \in H'
    \end{align*}
\end{theorem}

The uniqueness theorem presented by \citet{marshall2022linearity} now follows as
a direct corollary of the multi-step borrow safety theorem we described above.
\begin{corollary}[Uniqueness]
    \label{thm:uniqueness}
    For a well-typed term  $\Gamma  \vdash  \SYSTEMnt{t}  :   {\textcolor{uniqueColor}{\ast} }{ \SYSTEMnt{A} }$ and all $\Gamma_{{\mathrm{0}}}$ and $H$ such that $H  \bowtie  \SYSTEMsym{(}   \Gamma_{{\mathrm{0}}}  +   s  \cdot  \Gamma    \SYSTEMsym{)}$ and multi-reduction to a value
    $H  \vdash  \SYSTEMnt{t} \ \Rightarrow_{ s } \  H'  \vdash  \ast  \SYSTEMnt{v}$, for all $\SYSTEMnt{id} \in \mathsf{dom}  (  H  )$
    then:
    \begin{align*}
      & \forall \SYSTEMmv{ref} \in \mathsf{refs}  (  \SYSTEMnt{t}  ) . (\SYSTEMmv{ref}  \textcolor{uniqueColor}{\mapsto_{ \SYSTEMsym{1} } }  \SYSTEMnt{id} \in H \implies
      \SYSTEMmv{ref}  \textcolor{uniqueColor}{\mapsto_{ \SYSTEMsym{1} } }  \SYSTEMnt{id} \in H') \\
      \wedge & \quad \forall \SYSTEMnt{id'} \in \mathsf{dom}  (  H'  ) \wedge \SYSTEMnt{id'} \not\in \mathsf{dom}  (  H'  ) . \;
             \forall \SYSTEMmv{ref} \in \mathsf{refs}  (  \SYSTEMnt{v}  ) . \; \exists ! \SYSTEMmv{ref'}
    . \SYSTEMmv{ref}  \textcolor{uniqueColor}{\mapsto_{ \SYSTEMsym{1} } }  \SYSTEMnt{id'} \in H'
    \end{align*}
    i.e., any references contributing to the final term that are unique in the
    incoming heap stay unique in the resulting term, and any new references
    contributing to the final term are also unique.
\end{corollary}

Finally, we see that the operational semantics we have defined (extended to full
$\beta$-reduction) supports the equational theory we have gradually developed in
this work. Note that the proof includes only those equations defined by the
$\equiv$ relation presented throughout Sections~\ref{sec:uniqueness}
and~\ref{sec:fractional}; the remaining equations, presented in the appendix,
are as standard for a graded linear calculus.
\begin{theorem}[Soundness with respect to the equational theory]
  For all $\SYSTEMnt{t_{{\mathrm{1}}}}, \SYSTEMnt{t_{{\mathrm{2}}}}$ such that
  $\Gamma  \vdash  \SYSTEMnt{t_{{\mathrm{1}}}}  :  \SYSTEMnt{A}$ and $\Gamma  \vdash  \SYSTEMnt{t_{{\mathrm{2}}}}  :  \SYSTEMnt{A}$
  and $\SYSTEMnt{t_{{\mathrm{1}}}} \equiv \SYSTEMnt{t_{{\mathrm{2}}}}$ and
  given $H$ such that $H  \bowtie  \Gamma$, there exist multi-reductions
  to values that are equal
  under full $\beta$-reduction and
  evaluating any references to the value they point to in
  the resulting heaps:
  \begin{align*}
  H  \vdash  \SYSTEMnt{t_{{\mathrm{1}}}} \ \Rightarrow_{ \SYSTEMsym{1} } \  H'  \vdash  \SYSTEMnt{v_{{\mathrm{1}}}}  \quad \wedge
  \quad H  \vdash  \SYSTEMnt{t_{{\mathrm{2}}}} \ \Rightarrow_{ \SYSTEMsym{1} } \  H''  \vdash  \SYSTEMnt{v_{{\mathrm{2}}}} \quad \wedge \quad
  H'( \SYSTEMnt{v_{{\mathrm{1}}}} ) \equiv H''( \SYSTEMnt{v_{{\mathrm{2}}}} )
  \end{align*}
  %
  %where full $\beta$-reduction includes all congruences to evaluate
  %inside $\lambda$s, etc.
\end{theorem}

Full proofs for the above theorems may be found in the appendices for this
paper, provided as supplementary material, along with collected typing and
reduction rules.

\section{Related Work}
\label{sec:related}
\subsection{Linear and graded types}

The notion of linearity originated with Girard's linear
logic~\cite{girard1987linear}, which treats information in a resourceful way by
restricting the structural rules of intuitionistic logic. This was soon adopted
by programming language researchers who were interested in the resourceful
behaviour of data, rapidly developing into the concept of \emph{linear
types}~\cite{wadler1990linear, DBLP:conf/mfps/Wadler93}, but it took substantial
time for them to emerge into the realm of practical programming. Lately, linear
types have seen a renaissance due to their adoption as an extension to the
Haskell language~\cite{bernardy2017linear}; other languages incorporating pure
linearity (or affine types, which are similar but allow weakening) include
ATS~\cite{ats}, Alms~\cite{10.1145/1925844.1926436} and Mezzo~\cite{mezzo}.

This binary view was later refined by the notion of \emph{bounded} linear
logic~\cite{girard1992bounded}, which introduces a family of operators indexed
by a polynomial giving an upper bound on the usage of a resource. Further
generalisations of this idea to be able to track a broader range of increasingly
fine-grained properties are what led to the introduction of \emph{graded} types,
which allow for annotating values with precise information about how they
interact with their environment.

Granule's particular approach to graded modalities draws heavily from literature
on \emph{coeffects}, which describe how programs \emph{depend} on their
context~\cite{petricek2014coeffects}, though it also incorporates ideas from
effect systems which we focus on less here. This was developed concurrently with
other work which approached the same target from a different angle---through
attempting to find generalisations for bounded linear
logic~\cite{ghica2014coeffects}. Other instantiations of graded types include
Quantitative Type Theory (QTT)~\cite{atkey2018syntax} upon which the type system
for Idris 2 is built~\cite{idris2}, the core calculus underlying Linear
Haskell~\cite{bernardy2017linear, 10.1145/3547626}, and
others~\cite{DBLP:journals/pacmpl/AbelB20, gaboardi2016combining,
DBLP:conf/esop/WoodA22}.

\subsection{Uniqueness types}

Uniqueness types originated in the Clean language \cite{clean}, where they are
used in lieu of monadic computation and for efficiency gains offered by in-place
update. In Clean, computation is based on graph rewriting and reduction;
constants are graphs, and functions are graph rewriting formulas. This gives the
type system a different feel to those of more recent functional programming
languages. Recent work has attempted to capture benefits of uniqueness while
allowing a more modern programming style; examples include Cogent~\cite{cogent},
Mercury~\cite{mercury} and the prior iteration of uniqueness in
Granule~\cite{marshall2022linearity}.

Theoretical work on understanding uniqueness began with Harrington's uniqueness
logic~\cite{harrington2006uniqueness}; this was followed by a substantial amount
of theoretical groundwork on Clean's particular strategy for
uniqueness~\cite{simplified}. This paper clarified the distinction between
Clean's type system and systems based on the $\lambda$-calculus. Further work
made headway on distinguishing uniqueness from other substructural systems,
allowing for applications such as polymorphic programming and
concurrency~\cite{uniquepoly, uniquemessage}. This laid a foundation for
Granule's orthogonal approach to uniqueness, allowing for uniqueness and
linearity to be integrated in a single system, which lead to the extensions
developed in the present work.

\subsection{Region-based memory management}

Regions~\cite{region-retrospective} were conceived to bring benefits of
traditional stack-based memory management into higher-order functional
languages. Regions divide values using lifetimes; as with modern ownership
systems, they eliminate the need for garbage collection, by using region type
information to allow safe allocation and deallocation. Historically, region type
systems have typically been used for effect
systems~\cite{algebraic-reconstruction-effects, polymorphic-effects}.

Later work on regions extends this stack-based foundation by making use of
uniqueness information~\cite{static-capabilities}; a unique reference ensures
that a region has no aliases, so it can be deallocated efficiently. Similarly to
lifetimes in ownership systems, regions act as equivalence classes for a ``may
alias'' relation; values which do not share a region may not alias with one
another, so if a value does not share a region with any other then it can be
safely mutated.

Work on Cyclone (a `safe dialect' of C~\cite{cyclone-memory}) clarified the
relationship between references and regions, observing that ``unique pointers
are essentially lightweight, dynamic regions that hold exactly one
object~\cite{regions}.'' Rust's lifetimes were inspired by regions. One
extension of ML supports both linearity and borrowing using regions~\cite{affe}.

\subsection{Ownership and borrowing}

Ownership was developed as a framework for understanding aliasing in
object-oriented languages~\cite{mycroft, 10.1145/286936.286947}, with related
work in this area having introduced notions like
islands~\cite{10.1145/118014.117975}, balloon types~\cite{almeida1997balloon}
and external uniqueness~\cite{clarke2003external}. The intent of ownership is to
give a high-level structural view of objects and references, akin to type
systems which allow for a high-level structural view of data.

More recently, ownership is pervasively used in Rust in order to help ensure
memory safety. Multiple formalisations for Rust's ownership model have been
attempted; RustBelt~\cite{jung2017rustbelt} gives a lower-level encoding of Rust
intended for formal verification while Oxide~\cite{weiss2019oxide} is a
higher-level encoding designed for theoretical work, among
others~\cite{jung2019stacked, pearce2021lightweight}. Rust is not the only
modern language to make use of ownership; other languages like Swift incorporate
similar ideas, and work on introducing ownership into languages with manual
memory management is ongoing~\cite{sammler2021refinedc}. \citet{lorenzen2023fp}
describe how to use related ideas to determine which functional programs can be
executed in-place without allocation, and recent work on \emph{reachability
types}~\cite{10.1145/3485516, 10.1145/3632856} scales reasoning about lifetimes
and sharing into higher-order and polymorphic settings, taking inspiration from
separation logic.

Rust also provides an \emph{unsafe} mode (a superset of the safe portion of
Rust). While in Rust memory is managed through ownership and borrowing by
default with memory for unsafe code needing to be managed manually, in languages
like Granule managing memory automatically through a garbage collector is the
default. Using our extension to obviate the need for garbage collection is a
special case applying to some subset of a program. If the user does not wish for
ownership to be taken into consideration, they can use Granule's pre-existing
type system as before.

\section{Conclusions and Future Work}
\label{sec:conclusion}
\paragraph{Performance improvements.}
This paper has generalised uniqueness and linearity in order to develop a
unified framework for reasoning about ownership and grading. While we have an
implementation of this built upon the Granule compiler, the practical benefits
it can offer for resourceful programming have not been fully explored. We hope
to evaluate \textbf{performance improvements} that introducing precise ownership
tracking into a functional language can offer through obviating garbage
collection for a subset of a program, by collecting examples involving ownership
and borrowing and benchmarking them against equivalent functional code.

\paragraph{Guarantees.} Linearity is only one property that can be tracked via
\emph{coeffects}, the flavour of grading we consider; many others have been
described in prior work~\cite{petricek2014coeffects}. If uniqueness dualises
linearity, translating the same relationship onto other coeffects may result in
other interesting properties that can be tracked in a similar manner; for
example, it has been noted that in the realm of information flow tracking for
security, \emph{confidentiality} can be understood as a coeffect with
\emph{integrity} as its dual~\cite{marshall2022integrity}. It would be valuable
to find an analogy for the ownership generalisation developed here in the
context of security, and also to go further and develop a \emph{general}
algebraic theory for \textbf{guarantees} about global program behaviour.

\paragraph{Counting permissions.} Fractional permissions and the general notion
of dividing a single mutable borrow into many immutable borrows is not the only
model for representing borrowing that we considered. One interesting path for
future research would be to explore an alternative strategy for graded
uniqueness that is less symmetrical but instead privileges the original owner,
allowing for a more exact count of other extant references. This model, based on
\textbf{counting permissions}~\cite{10.1145/1040305.1040327}, involves having
many references with only read permissions, and a designated owner that keeps a
count of how many such references exist, so that uniqueness can still be
recovered.

\paragraph{Non-lexical lifetimes.} The model of ownership developed here was
\emph{not} intended to be a complete model of every aspect of Rust's ownership
system, instead aiming to extract essential features and develop a framework
where they coexist with more traditional linear and graded types. It would still
be of interest to pursue this further and capture Rust's more advanced features,
however---in particular, incorporating notions such as \textbf{non-lexical
lifetimes} would increase the power of the described system. This may eventually
allow for a potential encoding study between Granule's calculus and a more
faithful representation of Rust's type system such as
$\lambda_\text{Rust}$~\cite{jung2017rustbelt}. It would also be useful to
uncover general principles for extending our type system with further resource
interfaces; in this work we opted to introduce the minimal machinery required to
illustrate our approach with some key examples, but this would allow for a more
fully extensible system.

\paragraph{Categorical model.} Finally, alongside the operational heap model
presented in this paper it would be interesting to explore a \textbf{categorical
model} based on adjunctions. Benton's linear/non-linear (LNL)
logic~\cite{benton1995mixed} is well-known, and progress has been made on using
similar tools to understand more advanced substructural systems involving graded
types~\cite{eades2020grading, vollmer2024mixed}, much like Granule's core
calculus as extended in this paper. Despite the close theoretical relationship
between linearity and uniqueness, the categorical background of uniqueness has
been only briefly explored~\cite{harrington2006uniqueness}, and for more complex
ownership and borrowing systems even less so. This would be a fruitful pathway
for further research.

\paragraph{Conclusion}

Graded type systems and ownership with borrowing are both ways of carefully and
precisely managing the usage of data built upon the foundation of linear logic,
but these two approaches have developed through very different pathways on the
road to being incorporated in modern-day programming languages. In this work, we
have developed a core calculus that captures many of the key concepts for
ownership tracking in a graded setting, connecting these fine-grained
substructural notions with prior work on simpler systems such as linear and
uniqueness types which sit closer to the theory. This has allowed us to not only
better understand the relationship between these disparate approaches to
resourceful reasoning but also to examine how they interact.

By developing a framework through which both ownership properties and precise
grades for reasoning about data can be tracked within the setting of the Granule
language, we demonstrate that careful management of both resource and memory
usage are not only compatible but complementary in a functional context. This
paper represents a piece of two larger puzzles---one aiming to develop an
in-depth theoretical understanding of Rust's comprehensive approach to memory
management, and one aiming to expand the range of properties about programs that
can be represented explicitly through graded types---and we look forward to
seeing more ideas being shared (or, indeed, borrowed) across the boundary
between these two closely related worlds in the future.

\section*{Data-Availability Statement}
All of the examples presented throughout this work, as well as an installation
of Granule with the ownership and borrowing extensions that we describe, are
included in the artefact for this paper. The artefact will be available on
Zenodo following the conclusion of the artefact evaluation process.

\bibliography{references}


\begin{thebibliography}{65}

%%% the Web address.

\ifx \showCODEN    \undefined \def \showCODEN     #1{\unskip}     \fi
\ifx \showDOI      \undefined \def \showDOI       #1{#1}\fi
\ifx \showISBNx    \undefined \def \showISBNx     #1{\unskip}     \fi
\ifx \showISBNxiii \undefined \def \showISBNxiii  #1{\unskip}     \fi
\ifx \showISSN     \undefined \def \showISSN      #1{\unskip}     \fi
\ifx \showLCCN     \undefined \def \showLCCN      #1{\unskip}     \fi
\ifx \shownote     \undefined \def \shownote      #1{#1}          \fi
\ifx \showarticletitle \undefined \def \showarticletitle #1{#1}   \fi
\ifx \showURL      \undefined \def \showURL       {\relax}        \fi
% invisible to TeX
\providecommand\bibfield[2]{#2}
\providecommand\bibinfo[2]{#2}
\providecommand\natexlab[1]{#1}
\providecommand\showeprint[2][]{arXiv:#2}

\bibitem[Abel and Bernardy(2020)]%
        {DBLP:journals/pacmpl/AbelB20}
\bibfield{author}{\bibinfo{person}{Andreas Abel} {and}
  \bibinfo{person}{Jean{-}Philippe Bernardy}.} \bibinfo{year}{2020}\natexlab{}.
\newblock \showarticletitle{A Unified View of Modalities in Type Systems}.
\newblock \bibinfo{journal}{\emph{Proc. {ACM} Program. Lang.}}
  \bibinfo{volume}{4}, \bibinfo{number}{{ICFP}} (\bibinfo{year}{2020}),
  \bibinfo{pages}{90:1--90:28}.
\newblock
\urldef\tempurl%
\url{https://doi.org/10.1145/3408972}
\showDOI{\tempurl}


\bibitem[Almeida(1997)]%
        {almeida1997balloon}
\bibfield{author}{\bibinfo{person}{Paulo~S{\'e}rgio Almeida}.}
  \bibinfo{year}{1997}\natexlab{}.
\newblock \showarticletitle{Balloon Types: Controlling Sharing of State in Data
  Types}. In \bibinfo{booktitle}{\emph{ECOOP'97—Object-Oriented Programming:
  11th European Conference Jyv{\"a}skyl{\"a}, Finland, June 9--13, 1997
  Proceedings 11}}. Springer, \bibinfo{pages}{32--59}.
\newblock


\bibitem[Atkey(2018)]%
        {atkey2018syntax}
\bibfield{author}{\bibinfo{person}{Robert Atkey}.}
  \bibinfo{year}{2018}\natexlab{}.
\newblock \showarticletitle{Syntax and Semantics of Quantitative Type Theory}.
  In \bibinfo{booktitle}{\emph{Proceedings of the 33rd Annual ACM/IEEE
  Symposium on Logic in Computer Science}}. \bibinfo{pages}{56--65}.
\newblock


\bibitem[Balabonski et~al\mbox{.}(2016)]%
        {mezzo}
\bibfield{author}{\bibinfo{person}{Thibaut Balabonski},
  \bibinfo{person}{Fran\c{c}ois Pottier}, {and} \bibinfo{person}{Jonathan
  Protzenko}.} \bibinfo{year}{2016}\natexlab{}.
\newblock \showarticletitle{The Design and Formalization of Mezzo, a
  Permission-Based Programming Language}.
\newblock \bibinfo{journal}{\emph{ACM Trans. Program. Lang. Syst.}}
  \bibinfo{volume}{38}, \bibinfo{number}{4}, Article \bibinfo{articleno}{14}
  (\bibinfo{date}{aug} \bibinfo{year}{2016}), \bibinfo{numpages}{94}~pages.
\newblock
\showISSN{0164-0925}
\urldef\tempurl%
\url{https://doi.org/10.1145/2837022}
\showDOI{\tempurl}


\bibitem[Bao et~al\mbox{.}(2021)]%
        {10.1145/3485516}
\bibfield{author}{\bibinfo{person}{Yuyan Bao}, \bibinfo{person}{Guannan Wei},
  \bibinfo{person}{Oliver Bra\v{c}evac}, \bibinfo{person}{Yuxuan Jiang},
  \bibinfo{person}{Qiyang He}, {and} \bibinfo{person}{Tiark Rompf}.}
  \bibinfo{year}{2021}\natexlab{}.
\newblock \showarticletitle{Reachability Types: Tracking Aliasing and
  Separation in Higher-Order Functional Programs}.
\newblock \bibinfo{journal}{\emph{Proc. ACM Program. Lang.}}
  \bibinfo{volume}{5}, \bibinfo{number}{OOPSLA}, Article
  \bibinfo{articleno}{139} (\bibinfo{date}{oct} \bibinfo{year}{2021}),
  \bibinfo{numpages}{32}~pages.
\newblock
\urldef\tempurl%
\url{https://doi.org/10.1145/3485516}
\showDOI{\tempurl}


\bibitem[Barendsen and Smetsers(1996)]%
        {barendsen1996uniqueness}
\bibfield{author}{\bibinfo{person}{Erik Barendsen} {and} \bibinfo{person}{Sjaak
  Smetsers}.} \bibinfo{year}{1996}\natexlab{}.
\newblock \showarticletitle{Uniqueness Typing for Functional Languages with
  Graph Rewriting Semantics}.
\newblock \bibinfo{journal}{\emph{Mathematical Structures in Computer Science}}
  \bibinfo{volume}{6}, \bibinfo{number}{6} (\bibinfo{year}{1996}),
  \bibinfo{pages}{579--612}.
\newblock


\bibitem[Benton(1995)]%
        {benton1995mixed}
\bibfield{author}{\bibinfo{person}{P~Nick Benton}.}
  \bibinfo{year}{1995}\natexlab{}.
\newblock \showarticletitle{A Mixed Linear and Non-Linear Logic: Proofs, Terms
  and Models}. In \bibinfo{booktitle}{\emph{Computer Science Logic: 8th
  Workshop, CSL'94 Kazimierz, Poland, September 25--30, 1994 Selected Papers
  8}}. Springer, \bibinfo{pages}{121--135}.
\newblock


\bibitem[Bernardy et~al\mbox{.}(2017)]%
        {bernardy2017linear}
\bibfield{author}{\bibinfo{person}{Jean-Philippe Bernardy},
  \bibinfo{person}{Mathieu Boespflug}, \bibinfo{person}{Ryan~R Newton},
  \bibinfo{person}{Simon Peyton~Jones}, {and} \bibinfo{person}{Arnaud
  Spiwack}.} \bibinfo{year}{2017}\natexlab{}.
\newblock \showarticletitle{Linear Haskell: Practical Linearity in a
  Higher-Order Polymorphic Language}.
\newblock \bibinfo{journal}{\emph{Proceedings of the ACM on Programming
  Languages}} \bibinfo{volume}{2}, \bibinfo{number}{POPL}
  (\bibinfo{year}{2017}), \bibinfo{pages}{1--29}.
\newblock


\bibitem[Bianchini et~al\mbox{.}(2023a)]%
        {DBLP:conf/ecoop/BianchiniDGZ23}
\bibfield{author}{\bibinfo{person}{Riccardo Bianchini},
  \bibinfo{person}{Francesco Dagnino}, \bibinfo{person}{Paola Giannini}, {and}
  \bibinfo{person}{Elena Zucca}.} \bibinfo{year}{2023}\natexlab{a}.
\newblock \showarticletitle{Multi-Graded Featherweight Java}. In
  \bibinfo{booktitle}{\emph{37th European Conference on Object-Oriented
  Programming, {ECOOP} 2023, July 17-21, 2023, Seattle, Washington, United
  States}} \emph{(\bibinfo{series}{LIPIcs}, Vol.~\bibinfo{volume}{263})},
  \bibfield{editor}{\bibinfo{person}{Karim Ali} {and} \bibinfo{person}{Guido
  Salvaneschi}} (Eds.). \bibinfo{publisher}{Schloss Dagstuhl - Leibniz-Zentrum
  f{\"{u}}r Informatik}, \bibinfo{pages}{3:1--3:27}.
\newblock
\urldef\tempurl%
\url{https://doi.org/10.4230/LIPIcs.ECOOP.2023.3}
\showDOI{\tempurl}


\bibitem[Bianchini et~al\mbox{.}(2023b)]%
        {10.1145/3622843}
\bibfield{author}{\bibinfo{person}{Riccardo Bianchini},
  \bibinfo{person}{Francesco Dagnino}, \bibinfo{person}{Paola Giannini}, {and}
  \bibinfo{person}{Elena Zucca}.} \bibinfo{year}{2023}\natexlab{b}.
\newblock \showarticletitle{Resource-Aware Soundness for Big-Step Semantics}.
\newblock \bibinfo{journal}{\emph{Proc. ACM Program. Lang.}}
  \bibinfo{volume}{7}, \bibinfo{number}{OOPSLA2}, Article
  \bibinfo{articleno}{267} (\bibinfo{date}{oct} \bibinfo{year}{2023}),
  \bibinfo{numpages}{29}~pages.
\newblock
\urldef\tempurl%
\url{https://doi.org/10.1145/3622843}
\showDOI{\tempurl}


\bibitem[Bianchini et~al\mbox{.}(2022)]%
        {10.1145/3563319}
\bibfield{author}{\bibinfo{person}{Riccardo Bianchini},
  \bibinfo{person}{Francesco Dagnino}, \bibinfo{person}{Paola Giannini},
  \bibinfo{person}{Elena Zucca}, {and} \bibinfo{person}{Marco Servetto}.}
  \bibinfo{year}{2022}\natexlab{}.
\newblock \showarticletitle{Coeffects for Sharing and Mutation}.
\newblock \bibinfo{journal}{\emph{Proc. ACM Program. Lang.}}
  \bibinfo{volume}{6}, \bibinfo{number}{OOPSLA2}, Article
  \bibinfo{articleno}{156} (\bibinfo{date}{oct} \bibinfo{year}{2022}),
  \bibinfo{numpages}{29}~pages.
\newblock
\urldef\tempurl%
\url{https://doi.org/10.1145/3563319}
\showDOI{\tempurl}


\bibitem[Bornat et~al\mbox{.}(2005)]%
        {10.1145/1040305.1040327}
\bibfield{author}{\bibinfo{person}{Richard Bornat}, \bibinfo{person}{Cristiano
  Calcagno}, \bibinfo{person}{Peter O'Hearn}, {and} \bibinfo{person}{Matthew
  Parkinson}.} \bibinfo{year}{2005}\natexlab{}.
\newblock \showarticletitle{Permission Accounting in Separation Logic}. In
  \bibinfo{booktitle}{\emph{Proceedings of the 32nd ACM SIGPLAN-SIGACT
  Symposium on Principles of Programming Languages}} (Long Beach, California,
  USA) \emph{(\bibinfo{series}{POPL '05})}. \bibinfo{publisher}{Association for
  Computing Machinery}, \bibinfo{address}{New York, NY, USA},
  \bibinfo{pages}{259–270}.
\newblock
\showISBNx{158113830X}
\urldef\tempurl%
\url{https://doi.org/10.1145/1040305.1040327}
\showDOI{\tempurl}


\bibitem[Boyland(2003)]%
        {boyland2003checking}
\bibfield{author}{\bibinfo{person}{John Boyland}.}
  \bibinfo{year}{2003}\natexlab{}.
\newblock \showarticletitle{Checking Interference with Fractional Permissions}.
  In \bibinfo{booktitle}{\emph{Static Analysis: 10th International Symposium,
  SAS 2003 San Diego, CA, USA, June 11--13, 2003 Proceedings}}. Springer,
  \bibinfo{pages}{55--72}.
\newblock


\bibitem[Brady(2021)]%
        {idris2}
\bibfield{author}{\bibinfo{person}{Edwin~C. Brady}.}
  \bibinfo{year}{2021}\natexlab{}.
\newblock \bibinfo{title}{Idris 2: Quantitative Type Theory in Practice}.
\newblock , \bibinfo{numpages}{9:1--9:26}~pages.
\newblock
\urldef\tempurl%
\url{https://doi.org/10.4230/LIPIcs.ECOOP.2021.9}
\showDOI{\tempurl}


\bibitem[Brunel et~al\mbox{.}(2014)]%
        {brunel2014core}
\bibfield{author}{\bibinfo{person}{Alo{\"\i}s Brunel}, \bibinfo{person}{Marco
  Gaboardi}, \bibinfo{person}{Damiano Mazza}, {and} \bibinfo{person}{Steve
  Zdancewic}.} \bibinfo{year}{2014}\natexlab{}.
\newblock \showarticletitle{A Core Quantitative Coeffect Calculus}. In
  \bibinfo{booktitle}{\emph{ESOP}}, Vol.~\bibinfo{volume}{8410}. Springer,
  \bibinfo{pages}{351--370}.
\newblock


\bibitem[Choudhury et~al\mbox{.}(2021)]%
        {choudhury2021graded}
\bibfield{author}{\bibinfo{person}{Pritam Choudhury}, \bibinfo{person}{Harley
  Eades~III}, \bibinfo{person}{Richard~A Eisenberg}, {and}
  \bibinfo{person}{Stephanie Weirich}.} \bibinfo{year}{2021}\natexlab{}.
\newblock \showarticletitle{A Graded Dependent Type System with a Usage-Aware
  Semantics}.
\newblock \bibinfo{journal}{\emph{Proceedings of the ACM on Programming
  Languages}} \bibinfo{volume}{5}, \bibinfo{number}{POPL}
  (\bibinfo{year}{2021}), \bibinfo{pages}{1--32}.
\newblock


\bibitem[Clarke and Wrigstad(2003)]%
        {clarke2003external}
\bibfield{author}{\bibinfo{person}{Dave Clarke} {and} \bibinfo{person}{Tobias
  Wrigstad}.} \bibinfo{year}{2003}\natexlab{}.
\newblock \showarticletitle{External Uniqueness is Unique Enough}. In
  \bibinfo{booktitle}{\emph{ECOOP 2003--Object-Oriented Programming: 17th
  European Conference, Darmstadt, Germany, July 21-25, 2003. Proceedings 17}}.
  Springer, \bibinfo{pages}{176--200}.
\newblock


\bibitem[Clarke et~al\mbox{.}(1998)]%
        {10.1145/286936.286947}
\bibfield{author}{\bibinfo{person}{David~G. Clarke}, \bibinfo{person}{John~M.
  Potter}, {and} \bibinfo{person}{James Noble}.}
  \bibinfo{year}{1998}\natexlab{}.
\newblock \showarticletitle{Ownership Types for Flexible Alias Protection}. In
  \bibinfo{booktitle}{\emph{Proceedings of the 13th ACM SIGPLAN Conference on
  Object-Oriented Programming, Systems, Languages, and Applications}}
  (Vancouver, British Columbia, Canada) \emph{(\bibinfo{series}{OOPSLA '98})}.
  \bibinfo{publisher}{Association for Computing Machinery},
  \bibinfo{address}{New York, NY, USA}, \bibinfo{pages}{48–64}.
\newblock
\showISBNx{1581130058}
\urldef\tempurl%
\url{https://doi.org/10.1145/286936.286947}
\showDOI{\tempurl}


\bibitem[de~Vries(2013)]%
        {uniquepoly}
\bibfield{author}{\bibinfo{person}{Edsko de Vries}.}
  \bibinfo{year}{2013}\natexlab{}.
\newblock \showarticletitle{Modelling Unique and Affine Typing Using
  Polymorphism}. In \bibinfo{booktitle}{\emph{Essays Dedicated to Rinus
  Plasmeijer on the Occasion of His 61st Birthday on The Beauty of Functional
  Code - Volume 8106}}. \bibinfo{publisher}{Springer-Verlag},
  \bibinfo{address}{Berlin, Heidelberg}, \bibinfo{pages}{181–192}.
\newblock
\showISBNx{9783642403545}
\urldef\tempurl%
\url{https://doi.org/10.1007/978-3-642-40355-2{\_}13}
\showDOI{\tempurl}


\bibitem[de~Vries et~al\mbox{.}(2009)]%
        {uniquemessage}
\bibfield{author}{\bibinfo{person}{Edsko de Vries}, \bibinfo{person}{Adrian
  Francalanza}, {and} \bibinfo{person}{Matthew Hennessy}.}
  \bibinfo{year}{2009}\natexlab{}.
\newblock \showarticletitle{Uniqueness Typing for Resource Management in
  Message-Passing Concurrency}. In \bibinfo{booktitle}{\emph{Proceedings First
  International Workshop on Linearity, {LINEARITY} 2009, Coimbra, Portugal,
  12th September 2009}} \emph{(\bibinfo{series}{{EPTCS}},
  Vol.~\bibinfo{volume}{22})}, \bibfield{editor}{\bibinfo{person}{M{\'{a}}rio
  Florido} {and} \bibinfo{person}{Ian Mackie}} (Eds.). \bibinfo{pages}{26--37}.
\newblock
\urldef\tempurl%
\url{https://doi.org/10.4204/EPTCS.22.3}
\showDOI{\tempurl}


\bibitem[de~Vries et~al\mbox{.}(2008)]%
        {simplified}
\bibfield{author}{\bibinfo{person}{Edsko de Vries}, \bibinfo{person}{Rinus
  Plasmeijer}, {and} \bibinfo{person}{David~M. Abrahamson}.}
  \bibinfo{year}{2008}\natexlab{}.
\newblock \showarticletitle{Uniqueness Typing Simplified}. In
  \bibinfo{booktitle}{\emph{Implementation and Application of Functional
  Languages}}, \bibfield{editor}{\bibinfo{person}{Olaf Chitil},
  \bibinfo{person}{Zolt{\'a}n Horv{\'a}th}, {and} \bibinfo{person}{Vikt{\'o}ria
  Zs{\'o}k}} (Eds.). \bibinfo{publisher}{Springer Berlin Heidelberg},
  \bibinfo{address}{Berlin, Heidelberg}, \bibinfo{pages}{201--218}.
\newblock
\showISBNx{978-3-540-85373-2}


\bibitem[Eades~III and Orchard(2020)]%
        {eades2020grading}
\bibfield{author}{\bibinfo{person}{Harley Eades~III} {and}
  \bibinfo{person}{Dominic Orchard}.} \bibinfo{year}{2020}\natexlab{}.
\newblock \showarticletitle{Grading Adjoint Logic}.
\newblock \bibinfo{journal}{\emph{arXiv preprint arXiv:2006.08854}}
  (\bibinfo{year}{2020}).
\newblock


\bibitem[Fluet et~al\mbox{.}(2006)]%
        {regions}
\bibfield{author}{\bibinfo{person}{Matthew Fluet}, \bibinfo{person}{Greg
  Morrisett}, {and} \bibinfo{person}{Amal Ahmed}.}
  \bibinfo{year}{2006}\natexlab{}.
\newblock \showarticletitle{Linear Regions Are All You Need}. In
  \bibinfo{booktitle}{\emph{Programming Languages and Systems}},
  \bibfield{editor}{\bibinfo{person}{Peter Sestoft}} (Ed.).
  \bibinfo{publisher}{Springer Berlin Heidelberg}, \bibinfo{address}{Berlin,
  Heidelberg}, \bibinfo{pages}{7--21}.
\newblock
\showISBNx{978-3-540-33096-7}


\bibitem[Gaboardi et~al\mbox{.}(2016)]%
        {gaboardi2016combining}
\bibfield{author}{\bibinfo{person}{Marco Gaboardi}, \bibinfo{person}{Shin-ya
  Katsumata}, \bibinfo{person}{Dominic Orchard}, \bibinfo{person}{Flavien
  Breuvart}, {and} \bibinfo{person}{Tarmo Uustalu}.}
  \bibinfo{year}{2016}\natexlab{}.
\newblock \showarticletitle{Combining Effects and Coeffects via Grading}. In
  \bibinfo{booktitle}{\emph{Proceedings of the 21st ACM SIGPLAN International
  Conference on Functional Programming}} (Nara, Japan)
  \emph{(\bibinfo{series}{ICFP 2016})}. \bibinfo{publisher}{Association for
  Computing Machinery}, \bibinfo{address}{New York, NY, USA},
  \bibinfo{pages}{476–489}.
\newblock
\showISBNx{9781450342193}
\urldef\tempurl%
\url{https://doi.org/10.1145/2951913.2951939}
\showDOI{\tempurl}


\bibitem[Ghica and Smith(2014)]%
        {ghica2014coeffects}
\bibfield{author}{\bibinfo{person}{Dan~R. Ghica} {and} \bibinfo{person}{Alex~I.
  Smith}.} \bibinfo{year}{2014}\natexlab{}.
\newblock \showarticletitle{Bounded Linear Types in a Resource Semiring}. In
  \bibinfo{booktitle}{\emph{Proceedings of the 23rd European Symposium on
  Programming Languages and Systems - Volume 8410}}.
  \bibinfo{publisher}{Springer-Verlag}, \bibinfo{address}{Berlin, Heidelberg},
  \bibinfo{pages}{331–350}.
\newblock
\showISBNx{9783642548321}
\urldef\tempurl%
\url{https://doi.org/10.1007/978-3-642-54833-8_18}
\showDOI{\tempurl}


\bibitem[Girard(1987)]%
        {girard1987linear}
\bibfield{author}{\bibinfo{person}{Jean-Yves Girard}.}
  \bibinfo{year}{1987}\natexlab{}.
\newblock \showarticletitle{Linear Logic}.
\newblock \bibinfo{journal}{\emph{Theoretical Computer Science}}
  \bibinfo{volume}{50}, \bibinfo{number}{1} (\bibinfo{year}{1987}),
  \bibinfo{pages}{1--101}.
\newblock


\bibitem[Girard et~al\mbox{.}(1992)]%
        {girard1992bounded}
\bibfield{author}{\bibinfo{person}{Jean-Yves Girard}, \bibinfo{person}{Andre
  Scedrov}, {and} \bibinfo{person}{Philip~J Scott}.}
  \bibinfo{year}{1992}\natexlab{}.
\newblock \showarticletitle{Bounded Linear Logic: A Modular Approach to
  Polynomial-Time Computability}.
\newblock \bibinfo{journal}{\emph{Theoretical Computer Science}}
  \bibinfo{volume}{97}, \bibinfo{number}{1} (\bibinfo{year}{1992}),
  \bibinfo{pages}{1--66}.
\newblock


\bibitem[Harrington(2006)]%
        {harrington2006uniqueness}
\bibfield{author}{\bibinfo{person}{Dana Harrington}.}
  \bibinfo{year}{2006}\natexlab{}.
\newblock \showarticletitle{Uniqueness Logic}.
\newblock \bibinfo{journal}{\emph{Theoretical Computer Science}}
  \bibinfo{volume}{354}, \bibinfo{number}{1} (\bibinfo{year}{2006}),
  \bibinfo{pages}{24--41}.
\newblock


\bibitem[Hicks et~al\mbox{.}(2004)]%
        {cyclone-memory}
\bibfield{author}{\bibinfo{person}{Michael Hicks}, \bibinfo{person}{Greg
  Morrisett}, \bibinfo{person}{Dan Grossman}, {and} \bibinfo{person}{Trevor
  Jim}.} \bibinfo{year}{2004}\natexlab{}.
\newblock \showarticletitle{Experience with Safe Manual Memory-Management in
  {Cyclone}}. In \bibinfo{booktitle}{\emph{Proceedings of the 4th International
  Symposium on Memory Management}} (Vancouver, BC, Canada)
  \emph{(\bibinfo{series}{ISMM '04})}. \bibinfo{publisher}{Association for
  Computing Machinery}, \bibinfo{address}{New York, NY, USA},
  \bibinfo{pages}{73–84}.
\newblock
\showISBNx{1581139454}
\urldef\tempurl%
\url{https://doi.org/10.1145/1029873.1029883}
\showDOI{\tempurl}


\bibitem[Hogg(1991)]%
        {10.1145/118014.117975}
\bibfield{author}{\bibinfo{person}{John Hogg}.}
  \bibinfo{year}{1991}\natexlab{}.
\newblock \showarticletitle{Islands: Aliasing Protection in Object-Oriented
  Languages}.
\newblock \bibinfo{journal}{\emph{SIGPLAN Not.}} \bibinfo{volume}{26},
  \bibinfo{number}{11} (\bibinfo{date}{nov} \bibinfo{year}{1991}),
  \bibinfo{pages}{271–285}.
\newblock
\showISSN{0362-1340}
\urldef\tempurl%
\url{https://doi.org/10.1145/118014.117975}
\showDOI{\tempurl}


\bibitem[Hughes et~al\mbox{.}(2021a)]%
        {hughes2021linear}
\bibfield{author}{\bibinfo{person}{Jack Hughes}, \bibinfo{person}{Danielle
  Marshall}, \bibinfo{person}{James Wood}, {and} \bibinfo{person}{Dominic
  Orchard}.} \bibinfo{year}{2021}\natexlab{a}.
\newblock \showarticletitle{Linear Exponentials as Graded Modal Types}. In
  \bibinfo{booktitle}{\emph{5th International Workshop on Trends in Linear
  Logic and Applications (TLLA 2021)}}.
\newblock


\bibitem[Hughes et~al\mbox{.}(2021b)]%
        {hughes2021deriving}
\bibfield{author}{\bibinfo{person}{Jack Hughes}, \bibinfo{person}{Michael
  Vollmer}, {and} \bibinfo{person}{Dominic Orchard}.}
  \bibinfo{year}{2021}\natexlab{b}.
\newblock \showarticletitle{Deriving Distributive Laws for Graded Linear
  Types}.
\newblock \bibinfo{journal}{\emph{Electronic Proceedings in Theoretical
  Computer Science}}  \bibinfo{volume}{353} (\bibinfo{date}{dec}
  \bibinfo{year}{2021}), \bibinfo{pages}{109--131}.
\newblock
\urldef\tempurl%
\url{https://doi.org/10.4204/eptcs.353.6}
\showDOI{\tempurl}


\bibitem[Jouvelot and Gifford(1991)]%
        {algebraic-reconstruction-effects}
\bibfield{author}{\bibinfo{person}{Pierre Jouvelot} {and}
  \bibinfo{person}{David Gifford}.} \bibinfo{year}{1991}\natexlab{}.
\newblock \showarticletitle{Algebraic Reconstruction of Types and Effects}. In
  \bibinfo{booktitle}{\emph{Proceedings of the 18th ACM SIGPLAN-SIGACT
  Symposium on Principles of Programming Languages}} (Orlando, Florida, USA)
  \emph{(\bibinfo{series}{POPL '91})}. \bibinfo{publisher}{Association for
  Computing Machinery}, \bibinfo{address}{New York, NY, USA},
  \bibinfo{pages}{303–310}.
\newblock
\showISBNx{0897914198}
\urldef\tempurl%
\url{https://doi.org/10.1145/99583.99623}
\showDOI{\tempurl}


\bibitem[Jung et~al\mbox{.}(2019)]%
        {jung2019stacked}
\bibfield{author}{\bibinfo{person}{Ralf Jung}, \bibinfo{person}{Hoang-Hai
  Dang}, \bibinfo{person}{Jeehoon Kang}, {and} \bibinfo{person}{Derek Dreyer}.}
  \bibinfo{year}{2019}\natexlab{}.
\newblock \showarticletitle{Stacked Borrows: An Aliasing Model for Rust}.
\newblock \bibinfo{journal}{\emph{Proceedings of the ACM on Programming
  Languages}} \bibinfo{volume}{4}, \bibinfo{number}{POPL}
  (\bibinfo{year}{2019}), \bibinfo{pages}{1--32}.
\newblock


\bibitem[Jung et~al\mbox{.}(2017)]%
        {jung2017rustbelt}
\bibfield{author}{\bibinfo{person}{Ralf Jung}, \bibinfo{person}{Jacques-Henri
  Jourdan}, \bibinfo{person}{Robbert Krebbers}, {and} \bibinfo{person}{Derek
  Dreyer}.} \bibinfo{year}{2017}\natexlab{}.
\newblock \showarticletitle{RustBelt: Securing the Foundations of the Rust
  Programming Language}.
\newblock \bibinfo{journal}{\emph{Proceedings of the ACM on Programming
  Languages}} \bibinfo{volume}{2}, \bibinfo{number}{POPL}
  (\bibinfo{year}{2017}), \bibinfo{pages}{1--34}.
\newblock


\bibitem[Lorenzen et~al\mbox{.}(2023)]%
        {lorenzen2023fp}
\bibfield{author}{\bibinfo{person}{Anton Lorenzen}, \bibinfo{person}{Daan
  Leijen}, {and} \bibinfo{person}{Wouter Swierstra}.}
  \bibinfo{year}{2023}\natexlab{}.
\newblock \showarticletitle{FP$^2$: Fully in-Place Functional Programming}. In
  \bibinfo{booktitle}{\emph{ICFP'23}}. ACM SIGPLAN.
\newblock
\urldef\tempurl%
\url{https://www.microsoft.com/en-us/research/publication/fp2-fully-in-place-functional-programming-2/}
\showURL{%
\tempurl}
\newblock
\shownote{preprint}.


\bibitem[Lucassen and Gifford(1988)]%
        {polymorphic-effects}
\bibfield{author}{\bibinfo{person}{J.~M. Lucassen} {and} \bibinfo{person}{D.~K.
  Gifford}.} \bibinfo{year}{1988}\natexlab{}.
\newblock \showarticletitle{Polymorphic Effect Systems}. In
  \bibinfo{booktitle}{\emph{Proceedings of the 15th ACM SIGPLAN-SIGACT
  Symposium on Principles of Programming Languages}} (San Diego, California,
  USA) \emph{(\bibinfo{series}{POPL '88})}. \bibinfo{publisher}{Association for
  Computing Machinery}, \bibinfo{address}{New York, NY, USA},
  \bibinfo{pages}{47–57}.
\newblock
\showISBNx{0897912527}
\urldef\tempurl%
\url{https://doi.org/10.1145/73560.73564}
\showDOI{\tempurl}


\bibitem[Makwana and Krishnaswami(2019)]%
        {makwana2019numlin}
\bibfield{author}{\bibinfo{person}{Dhruv~C Makwana} {and} \bibinfo{person}{Neel
  Krishnaswami}.} \bibinfo{year}{2019}\natexlab{}.
\newblock \showarticletitle{NumLin: Linear Types for Linear Algebra}.
\newblock  (\bibinfo{year}{2019}).
\newblock


\bibitem[Marshall and Orchard(2022a)]%
        {marshall2022integrity}
\bibfield{author}{\bibinfo{person}{Daniellel Marshall} {and}
  \bibinfo{person}{Dominic Orchard}.} \bibinfo{year}{2022}\natexlab{a}.
\newblock \showarticletitle{Graded Modal Types for Integrity and
  Confidentiality}. In \bibinfo{booktitle}{\emph{17th Workshop on Programming
  Languages and Analysis for Security (PLAS 2022)}}.
\newblock
\showeprint[arxiv]{2309.04324}~[cs.PL]


\bibitem[Marshall and Orchard(2022b)]%
        {marshall2022take}
\bibfield{author}{\bibinfo{person}{Danielle Marshall} {and}
  \bibinfo{person}{Dominic Orchard}.} \bibinfo{year}{2022}\natexlab{b}.
\newblock \showarticletitle{{How to Take the Inverse of a Type}}. In
  \bibinfo{booktitle}{\emph{36th European Conference on Object-Oriented
  Programming (ECOOP 2022)}} \emph{(\bibinfo{series}{Leibniz International
  Proceedings in Informatics (LIPIcs)}, Vol.~\bibinfo{volume}{222})},
  \bibfield{editor}{\bibinfo{person}{Karim Ali} {and} \bibinfo{person}{Jan
  Vitek}} (Eds.). \bibinfo{publisher}{Schloss Dagstuhl -- Leibniz-Zentrum
  f{\"u}r Informatik}, \bibinfo{address}{Dagstuhl, Germany},
  \bibinfo{pages}{5:1--5:27}.
\newblock
\showISBNx{978-3-95977-225-9}
\showISSN{1868-8969}
\urldef\tempurl%
\url{https://doi.org/10.4230/LIPIcs.ECOOP.2022.5}
\showDOI{\tempurl}


\bibitem[Marshall and Orchard(2022c)]%
        {marshall2022replicate}
\bibfield{author}{\bibinfo{person}{Danielle Marshall} {and}
  \bibinfo{person}{Dominic Orchard}.} \bibinfo{year}{2022}\natexlab{c}.
\newblock \showarticletitle{Replicate, Reuse, Repeat: Capturing Non-Linear
  Communication via Session Types and Graded Modal Types}.
\newblock \bibinfo{journal}{\emph{Electronic Proceedings in Theoretical
  Computer Science}}  \bibinfo{volume}{356} (\bibinfo{date}{March}
  \bibinfo{year}{2022}), \bibinfo{pages}{1–11}.
\newblock
\showISSN{2075-2180}
\urldef\tempurl%
\url{https://doi.org/10.4204/eptcs.356.1}
\showDOI{\tempurl}


\bibitem[Marshall et~al\mbox{.}(2022)]%
        {marshall2022linearity}
\bibfield{author}{\bibinfo{person}{Danielle Marshall}, \bibinfo{person}{Michael
  Vollmer}, {and} \bibinfo{person}{Dominic Orchard}.}
  \bibinfo{year}{2022}\natexlab{}.
\newblock \showarticletitle{Linearity and Uniqueness: An Entente Cordiale}. In
  \bibinfo{booktitle}{\emph{Programming Languages and Systems: 31st European
  Symposium on Programming, ESOP 2022, Held as Part of the European Joint
  Conferences on Theory and Practice of Software, ETAPS 2022, Munich, Germany,
  April 2--7, 2022, Proceedings}}. Springer International Publishing Cham,
  \bibinfo{pages}{346--375}.
\newblock


\bibitem[McBride(2001)]%
        {mcbride2001derivative}
\bibfield{author}{\bibinfo{person}{Conor McBride}.}
  \bibinfo{year}{2001}\natexlab{}.
\newblock \showarticletitle{The Derivative of a Regular Type is its Type of
  One-Hole Contexts}.
\newblock \bibinfo{journal}{\emph{Unpublished manuscript}}
  (\bibinfo{year}{2001}), \bibinfo{pages}{74--88}.
\newblock


\bibitem[McBride(2016)]%
        {mcbride2016got}
\bibfield{author}{\bibinfo{person}{Conor McBride}.}
  \bibinfo{year}{2016}\natexlab{}.
\newblock \showarticletitle{I Got Plenty o' Nuttin'}.
\newblock \bibinfo{journal}{\emph{A List of Successes That Can Change the
  World: Essays Dedicated to Philip Wadler on the Occasion of His 60th
  Birthday}} (\bibinfo{year}{2016}), \bibinfo{pages}{207--233}.
\newblock


\bibitem[Moon et~al\mbox{.}(2021)]%
        {moon2021graded}
\bibfield{author}{\bibinfo{person}{Benjamin Moon}, \bibinfo{person}{Harley
  Eades~III}, {and} \bibinfo{person}{Dominic Orchard}.}
  \bibinfo{year}{2021}\natexlab{}.
\newblock \showarticletitle{Graded Modal Dependent Type Theory}.
\newblock \bibinfo{journal}{\emph{Programming Languages and Systems}}
  \bibinfo{volume}{12648} (\bibinfo{year}{2021}), \bibinfo{pages}{462}.
\newblock


\bibitem[Mycroft and Voigt(2013)]%
        {mycroft}
\bibfield{author}{\bibinfo{person}{Alan Mycroft} {and} \bibinfo{person}{Janina
  Voigt}.} \bibinfo{year}{2013}\natexlab{}.
\newblock \showarticletitle{Notions of Aliasing and Ownership}.
\newblock In \bibinfo{booktitle}{\emph{Aliasing in Object-Oriented Programming.
  Types, Analysis and Verification}}, \bibfield{editor}{\bibinfo{person}{Dave
  Clarke}, \bibinfo{person}{James Noble}, {and} \bibinfo{person}{Tobias
  Wrigstad}} (Eds.). \bibinfo{series}{Lecture Notes in Computer Science},
  Vol.~\bibinfo{volume}{7850}. \bibinfo{publisher}{Springer},
  \bibinfo{pages}{59--83}.
\newblock
\urldef\tempurl%
\url{https://doi.org/10.1007/978-3-642-36946-9\_4}
\showDOI{\tempurl}


\bibitem[O'Connor et~al\mbox{.}(2021)]%
        {cogent}
\bibfield{author}{\bibinfo{person}{Liam O'Connor}, \bibinfo{person}{Zilin
  Chen}, \bibinfo{person}{Christine Rizkallah}, \bibinfo{person}{Vincent
  Jackson}, \bibinfo{person}{Sidney Amani}, \bibinfo{person}{Gerwin Klein},
  \bibinfo{person}{Toby Murray}, \bibinfo{person}{Thomas Sewell}, {and}
  \bibinfo{person}{Gabriele Keller}.} \bibinfo{year}{2021}\natexlab{}.
\newblock \showarticletitle{Cogent: Uniqueness Types and Certified
  Compilation}.
\newblock \bibinfo{journal}{\emph{J. Funct. Program.}} (\bibinfo{year}{2021}).
\newblock


\bibitem[Orchard et~al\mbox{.}(2019)]%
        {orchard2019quantitative}
\bibfield{author}{\bibinfo{person}{Dominic Orchard},
  \bibinfo{person}{Vilem-Benjamin Liepelt}, {and} \bibinfo{person}{Harley
  Eades~III}.} \bibinfo{year}{2019}\natexlab{}.
\newblock \showarticletitle{Quantitative Program Reasoning with Graded Modal
  Types}.
\newblock \bibinfo{journal}{\emph{Proceedings of the ACM on Programming
  Languages}} \bibinfo{volume}{3}, \bibinfo{number}{ICFP}
  (\bibinfo{year}{2019}), \bibinfo{pages}{1--30}.
\newblock


\bibitem[Pearce(2021)]%
        {pearce2021lightweight}
\bibfield{author}{\bibinfo{person}{David~J Pearce}.}
  \bibinfo{year}{2021}\natexlab{}.
\newblock \showarticletitle{A Lightweight Formalism for Reference Lifetimes and
  Borrowing in Rust}.
\newblock \bibinfo{journal}{\emph{ACM Transactions on Programming Languages and
  Systems (TOPLAS)}} \bibinfo{volume}{43}, \bibinfo{number}{1}
  (\bibinfo{year}{2021}), \bibinfo{pages}{1--73}.
\newblock


\bibitem[Petricek et~al\mbox{.}(2014)]%
        {petricek2014coeffects}
\bibfield{author}{\bibinfo{person}{Tomas Petricek}, \bibinfo{person}{Dominic
  Orchard}, {and} \bibinfo{person}{Alan Mycroft}.}
  \bibinfo{year}{2014}\natexlab{}.
\newblock \showarticletitle{Coeffects: A Calculus of Context-Dependent
  Computation}. In \bibinfo{booktitle}{\emph{Proceedings of the 19th ACM
  SIGPLAN International Conference on Functional Programming}} (Gothenburg,
  Sweden) \emph{(\bibinfo{series}{ICFP '14})}. \bibinfo{publisher}{Association
  for Computing Machinery}, \bibinfo{address}{New York, NY, USA},
  \bibinfo{pages}{123–135}.
\newblock
\showISBNx{9781450328739}
\urldef\tempurl%
\url{https://doi.org/10.1145/2628136.2628160}
\showDOI{\tempurl}


\bibitem[Radanne et~al\mbox{.}(2020)]%
        {affe}
\bibfield{author}{\bibinfo{person}{Gabriel Radanne}, \bibinfo{person}{Hannes
  Saffrich}, {and} \bibinfo{person}{Peter Thiemann}.}
  \bibinfo{year}{2020}\natexlab{}.
\newblock \showarticletitle{Kindly Bent to Free Us}.
\newblock \bibinfo{journal}{\emph{Proc. ACM Program. Lang.}}
  \bibinfo{volume}{4}, \bibinfo{number}{ICFP}, Article \bibinfo{articleno}{103}
  (\bibinfo{date}{Aug.} \bibinfo{year}{2020}), \bibinfo{numpages}{29}~pages.
\newblock
\urldef\tempurl%
\url{https://doi.org/10.1145/3408985}
\showDOI{\tempurl}


\bibitem[Sammler et~al\mbox{.}(2021)]%
        {sammler2021refinedc}
\bibfield{author}{\bibinfo{person}{Michael Sammler}, \bibinfo{person}{Rodolphe
  Lepigre}, \bibinfo{person}{Robbert Krebbers}, \bibinfo{person}{Kayvan
  Memarian}, \bibinfo{person}{Derek Dreyer}, {and} \bibinfo{person}{Deepak
  Garg}.} \bibinfo{year}{2021}\natexlab{}.
\newblock \showarticletitle{RefinedC: Automating the Foundational Verification
  of C Code with Refined Ownership Types}. In
  \bibinfo{booktitle}{\emph{Proceedings of the 42nd ACM SIGPLAN International
  Conference on Programming Language Design and Implementation}}.
  \bibinfo{pages}{158--174}.
\newblock


\bibitem[Smetsers et~al\mbox{.}(1994)]%
        {clean}
\bibfield{author}{\bibinfo{person}{Sjaak Smetsers}, \bibinfo{person}{Erik
  Barendsen}, \bibinfo{person}{Marko van Eekelen}, {and} \bibinfo{person}{Rinus
  Plasmeijer}.} \bibinfo{year}{1994}\natexlab{}.
\newblock \showarticletitle{Guaranteeing Safe Destructive Updates Through a
  Type System with Uniqueness Information for Graphs}. In
  \bibinfo{booktitle}{\emph{Graph Transformations in Computer Science}},
  \bibfield{editor}{\bibinfo{person}{Hans~J{\"u}rgen Schneider} {and}
  \bibinfo{person}{Hartmut Ehrig}} (Eds.). \bibinfo{publisher}{Springer Berlin
  Heidelberg}, \bibinfo{address}{Berlin, Heidelberg},
  \bibinfo{pages}{358--379}.
\newblock
\showISBNx{978-3-540-48333-5}
\urldef\tempurl%
\url{https://doi.org/10.1007/3-540-57787-4\_23}
\showDOI{\tempurl}


\bibitem[Somogyi et~al\mbox{.}(1996)]%
        {mercury}
\bibfield{author}{\bibinfo{person}{Zoltan Somogyi}, \bibinfo{person}{Fergus
  Henderson}, {and} \bibinfo{person}{Thomas Conway}.}
  \bibinfo{year}{1996}\natexlab{}.
\newblock \showarticletitle{The Execution Algorithm of {Mercury}, an Efficient
  Purely Declarative Logic Programming Language}.
\newblock \bibinfo{journal}{\emph{The Journal of Logic Programming}}
  \bibinfo{volume}{29}, \bibinfo{number}{1} (\bibinfo{year}{1996}),
  \bibinfo{pages}{17--64}.
\newblock
\showISSN{0743-1066}
\urldef\tempurl%
\url{https://doi.org/10.1016/S0743-1066(96)00068-4}
\showDOI{\tempurl}
\newblock
\shownote{High-Performance Implementations of Logic Programming Systems}.


\bibitem[Spiwack et~al\mbox{.}(2022)]%
        {10.1145/3547626}
\bibfield{author}{\bibinfo{person}{Arnaud Spiwack}, \bibinfo{person}{Csongor
  Kiss}, \bibinfo{person}{Jean-Philippe Bernardy}, \bibinfo{person}{Nicolas
  Wu}, {and} \bibinfo{person}{Richard~A. Eisenberg}.}
  \bibinfo{year}{2022}\natexlab{}.
\newblock \showarticletitle{Linearly Qualified Types: Generic Inference for
  Capabilities and Uniqueness}.
\newblock \bibinfo{journal}{\emph{Proc. ACM Program. Lang.}}
  \bibinfo{volume}{6}, \bibinfo{number}{ICFP}, Article \bibinfo{articleno}{95}
  (\bibinfo{date}{aug} \bibinfo{year}{2022}), \bibinfo{numpages}{28}~pages.
\newblock
\urldef\tempurl%
\url{https://doi.org/10.1145/3547626}
\showDOI{\tempurl}


\bibitem[Tofte et~al\mbox{.}(2004)]%
        {region-retrospective}
\bibfield{author}{\bibinfo{person}{Mads Tofte}, \bibinfo{person}{Lars
  Birkedal}, \bibinfo{person}{Martin Elsman}, {and} \bibinfo{person}{Niels
  Hallenberg}.} \bibinfo{year}{2004}\natexlab{}.
\newblock \showarticletitle{A {Retrospective} on {Region}-{Based} {Memory}
  {Management}}.
\newblock \bibinfo{journal}{\emph{Higher-Order and Symbolic Computation}}
  \bibinfo{volume}{17}, \bibinfo{number}{3} (\bibinfo{date}{Sept.}
  \bibinfo{year}{2004}), \bibinfo{pages}{245--265}.
\newblock
\showISSN{1573-0557}
\urldef\tempurl%
\url{https://doi.org/10.1023/B:LISP.0000029446.78563.a4}
\showDOI{\tempurl}


\bibitem[Tov and Pucella(2011)]%
        {10.1145/1925844.1926436}
\bibfield{author}{\bibinfo{person}{Jesse~A. Tov} {and}
  \bibinfo{person}{Riccardo Pucella}.} \bibinfo{year}{2011}\natexlab{}.
\newblock \showarticletitle{Practical Affine Types}.
\newblock \bibinfo{journal}{\emph{SIGPLAN Not.}} \bibinfo{volume}{46},
  \bibinfo{number}{1} (\bibinfo{date}{Jan.} \bibinfo{year}{2011}),
  \bibinfo{pages}{447–458}.
\newblock
\showISSN{0362-1340}
\urldef\tempurl%
\url{https://doi.org/10.1145/1925844.1926436}
\showDOI{\tempurl}


\bibitem[Vollmer et~al\mbox{.}(2024)]%
        {vollmer2024mixed}
\bibfield{author}{\bibinfo{person}{Victoria Vollmer}, \bibinfo{person}{Danielle
  Marshall}, \bibinfo{person}{Harley~Eades III}, {and} \bibinfo{person}{Dominic
  Orchard}.} \bibinfo{year}{2024}\natexlab{}.
\newblock \bibinfo{title}{A Mixed Linear and Graded Logic: Proofs, Terms, and
  Models}.
\newblock
\newblock
\showeprint[arxiv]{2401.17199}~[cs.LO]


\bibitem[Wadler(1990)]%
        {wadler1990linear}
\bibfield{author}{\bibinfo{person}{Philip Wadler}.}
  \bibinfo{year}{1990}\natexlab{}.
\newblock \showarticletitle{Linear Types can Change the World!}. In
  \bibinfo{booktitle}{\emph{Programming Concepts and Methods}},
  Vol.~\bibinfo{volume}{3}. Citeseer, \bibinfo{pages}{5}.
\newblock


\bibitem[Wadler(1993)]%
        {DBLP:conf/mfps/Wadler93}
\bibfield{author}{\bibinfo{person}{Philip Wadler}.}
  \bibinfo{year}{1993}\natexlab{}.
\newblock \showarticletitle{A Syntax for Linear Logic}. In
  \bibinfo{booktitle}{\emph{Mathematical Foundations of Programming Semantics,
  9th International Conference, New Orleans, LA, USA, April 7-10, 1993,
  Proceedings}}. \bibinfo{pages}{513--529}.
\newblock
\urldef\tempurl%
\url{https://doi.org/10.1007/3-540-58027-1\_24}
\showDOI{\tempurl}


\bibitem[Walker et~al\mbox{.}(2000)]%
        {static-capabilities}
\bibfield{author}{\bibinfo{person}{David Walker}, \bibinfo{person}{Karl Crary},
  {and} \bibinfo{person}{Greg Morrisett}.} \bibinfo{year}{2000}\natexlab{}.
\newblock \showarticletitle{Typed Memory Management via Static Capabilities}.
\newblock \bibinfo{journal}{\emph{ACM Trans. Program. Lang. Syst.}}
  \bibinfo{volume}{22}, \bibinfo{number}{4} (\bibinfo{date}{jul}
  \bibinfo{year}{2000}), \bibinfo{pages}{701–771}.
\newblock
\showISSN{0164-0925}
\urldef\tempurl%
\url{https://doi.org/10.1145/363911.363923}
\showDOI{\tempurl}


\bibitem[Wei et~al\mbox{.}(2024)]%
        {10.1145/3632856}
\bibfield{author}{\bibinfo{person}{Guannan Wei}, \bibinfo{person}{Oliver
  Bra\v{c}evac}, \bibinfo{person}{Songlin Jia}, \bibinfo{person}{Yuyan Bao},
  {and} \bibinfo{person}{Tiark Rompf}.} \bibinfo{year}{2024}\natexlab{}.
\newblock \showarticletitle{Polymorphic Reachability Types: Tracking Freshness,
  Aliasing, and Separation in Higher-Order Generic Programs}.
\newblock \bibinfo{journal}{\emph{Proc. ACM Program. Lang.}}
  \bibinfo{volume}{8}, \bibinfo{number}{POPL}, Article \bibinfo{articleno}{14}
  (\bibinfo{date}{jan} \bibinfo{year}{2024}), \bibinfo{numpages}{32}~pages.
\newblock
\urldef\tempurl%
\url{https://doi.org/10.1145/3632856}
\showDOI{\tempurl}


\bibitem[Weiss et~al\mbox{.}(2019)]%
        {weiss2019oxide}
\bibfield{author}{\bibinfo{person}{Aaron Weiss}, \bibinfo{person}{Olek
  Gierczak}, \bibinfo{person}{Daniel Patterson}, {and} \bibinfo{person}{Amal
  Ahmed}.} \bibinfo{year}{2019}\natexlab{}.
\newblock \showarticletitle{Oxide: The Essence of Rust}.
\newblock \bibinfo{journal}{\emph{arXiv preprint arXiv:1903.00982}}
  (\bibinfo{year}{2019}).
\newblock


\bibitem[Wood and Atkey(2022)]%
        {DBLP:conf/esop/WoodA22}
\bibfield{author}{\bibinfo{person}{James Wood} {and} \bibinfo{person}{Robert
  Atkey}.} \bibinfo{year}{2022}\natexlab{}.
\newblock \showarticletitle{A Framework for Substructural Type Systems}. In
  \bibinfo{booktitle}{\emph{Programming Languages and Systems - 31st European
  Symposium on Programming, {ESOP} 2022, Held as Part of the European Joint
  Conferences on Theory and Practice of Software, {ETAPS} 2022, Munich,
  Germany, April 2-7, 2022, Proceedings}} \emph{(\bibinfo{series}{Lecture Notes
  in Computer Science}, Vol.~\bibinfo{volume}{13240})},
  \bibfield{editor}{\bibinfo{person}{Ilya Sergey}} (Ed.).
  \bibinfo{publisher}{Springer}, \bibinfo{pages}{376--402}.
\newblock
\urldef\tempurl%
\url{https://doi.org/10.1007/978-3-030-99336-8\_14}
\showDOI{\tempurl}


\bibitem[Zhu and Xi(2005)]%
        {ats}
\bibfield{author}{\bibinfo{person}{Dengping Zhu} {and} \bibinfo{person}{Hongwei
  Xi}.} \bibinfo{year}{2005}\natexlab{}.
\newblock \showarticletitle{Safe Programming with Pointers Through Stateful
  Views}. In \bibinfo{booktitle}{\emph{Practical Aspects of Declarative
  Languages}}, \bibfield{editor}{\bibinfo{person}{Manuel~V. Hermenegildo} {and}
  \bibinfo{person}{Daniel Cabeza}} (Eds.). \bibinfo{publisher}{Springer Berlin
  Heidelberg}, \bibinfo{address}{Berlin, Heidelberg}, \bibinfo{pages}{83--97}.
\newblock
\showISBNx{978-3-540-30557-6}


\end{thebibliography}

\end{document}